\documentclass[12pt]{iopart}
\usepackage{amssymb}
\usepackage{graphicx}
\usepackage{adjustbox}

\usepackage{color}
\usepackage[usenames,dvipsnames,svgnames,table]{xcolor}

\begin{document}

\title[Real Space Analysis of Colloidal Gels]{Real Space Analysis of Colloidal Gels: Triumphs, Challenges and Future Directions}

\author{C. Patrick Royall}
\address{Gulliver UMR CNRS 7083, ESPCI Paris, Universit\' e PSL, 75005 Paris, France.}
\address{HH Wills Physics Laboratory, Tyndall Avenue, Bristol, BS8 1TL, UK.}
\address{School of Chemistry, University of Bristol, Cantock Close, Bristol, BS8 1TS, UK.}
\address{Centre for Nanoscience and Quantum Information, Tyndall Avenue, Bristol, BS8 1FD, UK.}

\author{Malcolm A. Faers}
\address{Bayer AG, Crop Science Division, Formulation Technology, Alfred Nobel Str. 50, 40789 Monheim, Germany}

\author{Sian L. Fussell}
\address{School of Chemistry, University of Bristol, Cantock Close, Bristol, BS8 1TS, UK.}
\address{Bristol Centre for Functional Nanomaterials,  University of Bristol, Tyndall Avenue, Bristol, BS8 1TL, UK.}

\author{James E. Hallett}
\address{Physical and Theoretical Chemistry Laboratory, South Parks Road, University of Oxford, OX1 3QZ, UK.}


\vspace{10pt}

\begin{abstract}
Colloidal gels constitute an important class of materials found in many contexts and with a wide range of applications. Yet as matter far from equilibrium, gels exhibit a variety of time--dependent behaviours, which can be perplexing, such as an increase in strength prior to catastrophic failure. Remarkably, such complex phenomena are faithfully captured by an extremely simple model -- ``sticky spheres''. Here we review progress in our understanding of colloidal gels made through the use of real space analysis and particle resolved studies. We consider the challenges of obtaining a suitable experimental system where the refractive index and density of the colloidal particles is matched to that of the solvent. We review work to obtain a particle--level mechanism for rigidity in gels and the evolution of our understanding of time--dependent behaviour, from early--time aggregation to ageing, before considering the response of colloidal gels to deformation and then move on to more complex systems of anisotropic particles and mixtures. Finally we note some more exotic materials with similar properties.
\end{abstract}

\maketitle

\section{Background and Motivation}
\label{sectionBackground}

The formation of a network of arrested material with high but finite zero-shear viscosity upon slight quenching is among the most striking features of condensed matter \cite{tanaka2000,poon2002,cipelletti2005,coniglio2004,zaccarelli2007}. Such \emph{gels} are absolutely the stuff of everyday life, for example gels comprise our tissues, numerous foods, cosmetics, coatings, crop protection suspension formulations, pharmaceutical suspension formulations, pigment printing inks, dispersions for 3D printing, ceramic preparations, food preparations, detergent formulations and numerous home care products \cite{drury2003,rose2014,ubbink2012,liang2001}. A wide range of materials also exhibit gelation including proteins \cite{cardinaux2007,leocmach2014,mcmanus2016,fusco2016,riosdeanda2019}, granular matter  \cite{ulrich2009,li2014}, phase-demixing oxides  \cite{bouttes2014}, and metallic glassformers \cite{baumer2013}. Indeed so wide is the range of materials that are commonly termed gels -- from polymer gels, hydrogels, and colloidal gels --  that these materials in fact obey very different physics. Some materials, notably polymers and some colloidal systems \cite{ruzicka2011}, can undergo \emph{equilibrium} gelation, and from a thermodynamic perspective can be regarded as being rather similar to supercooled liquids \cite{saikavoivod2011}.  Such systems form a network stablised by their disinclination to condense, effectively being an \emph{empty liquid} \cite{bianchi2006}.

Unlike equilibrium gels, many (but not all \cite{ruzicka2011,saikavoivod2011,jabbarifarouji2007}) colloidal gels form via the process of \emph{arrested phase separation}. The system begins to phase separate, typically in a spinodal--like fashion \cite{onuki} and often with a bicontinuous texture (Figs. \ref{figSpinodalCP} and \ref{figPhaseGlassGel}). After demixing is initiated, what then happens is that one (or, in principle, both) demixing phases undergo dynamical arrest which results in a solid--like material, a gel. This is an example of \emph{viscoelastic phase separation}, which pertains to the case where the demixed phases have significantly different viscosities \cite{tanaka2000}.  These gels then represent an example of \emph{matter far from equilibrium} as the thermodynamic equilibrium would be phase separated colloid--rich and colloid--poor states \cite{poon2002}. The colloid-rich phase can exist as colloidal crystals for colloids of low polydispersity in shallow quenches \cite{poon2002,lekkerkerker1992} and amorphous networks for deeper quenches \cite{teece2011,klotsa2011,whitelam2015}. This has two interesting consequences, first the properties of the network are not constant but age over time, and second they depend on the shear and processing history, which is of particular importance for formulated products and their manufacturing process.

To obtain such ``spinodal'' gels, a few systems are suitable candidates. Typically (though not always \cite{ferreirocordova2020}), an attraction between the colloidal particles is present. Such an attraction may result from van der Waals attractions \cite{russel,weitz1984}, or from polymer--induced depletion \cite{poon2002,asakura1954,long1973,lekkerkerker1992}, bridging by telechelic (triblock) copolymers \cite{appell1998,gao2015,fussell2019}. or critical Casimir interactions \cite{veen2012,hertlein2008,bonn2009,rouwhorst2020ncomms,rouwhorst2020pre}.  Some evidence has been found for gelation--like behaviour, in the form of a percolating network, in \emph{active colloidal systems} with microtubules \cite{decamp2015} and colloids with dipolar interactions \cite{sakai2020}. Of these, the most common and most well--studied in real space, are depletion gels of spherical colloids and non--absorbing polymer, and these form our primary focus \cite{poon2002}. Indeed so dominant are studies of ``spinodal gels'' and even of one particular  system (see section \ref{sectionEarly}) that we may argue in favour of studying a wider range of systems with a view to exploring other gelation mechanisms for example. We consider work that has been carried out on other systems towards the end of the review in section \ref{sectionExotic}. Often these also exhibit ``spinodal gelation'', so the mechanism can be quite similar, even if the interactions have a different origin (such as critical Casimir interactions \cite{rouwhorst2020ncomms}).

Here, we consider what has been learnt about the nature of colloidal gelation using the technique of \emph{real space analysis}, and in particular imaging of individual colloidal particles via confocal microscopy -- \emph{particle--resolved studies} \cite{ivlev}. We enquire as to the utility of this technique to address the challenges presented by colloidal gels. We note that a salient strength of the technique is its ability to investigate local phenomena, and thus to address the perceived wisdom of materials science that \emph{the microscopic structure determines the dynamics and macroscopic behaviour of the material} \cite{peterquote}.

\emph{The case for ``sticky spheres''. --- } It is worth noting that if hard spheres may be taken as being surprisingly rich in their behaviour, given that the system is driven only by entropy, then, since the seminal work of Baxter \cite{baxter1968} the addition of a short--range attraction, i.e. ``sticky spheres'', opens up a plethora of new phenomena. These include (metastable) liquid-gas phase coexistence and accompanying criticality \cite{miller2003}, in addition to gelation and a multitude of arrest mechanisms \cite{poon2002,zaccarelli2007,pham2002,royall2018jcp}. We suggest that such sticky spheres may be taken as a minimal model for certain classes of matter far from equilibrium and emphasise that approximations to sticky spheres concern much of the material presented here.

\subsection*{Pressing Challenges in Colloidal Gelation}
\label{sectionPressingChallenges}

We have seen from the discussion above that gels sit at the overlap between phase separation via spinodal decomposition (section \ref{sectionSpinodal}) and the glass transition (section \ref{sectionGlass}). If this is how colloidal gels come about, what can we say about their properties?

\begin{itemize}
\item{As non--equilibrium systems, gels age.}
\item{A consequence of ageing is that the properties of gels change over time, and one such property is an increase in mechanical strength.}
\item{Another consequence is their low reproducibility since the early stage network structure formed depends on its quenching history and mechanical processing history.}
\item{Paradoxically, although they become stronger over time, gels fail under their own weight, catastrophically, after a waiting time, which may extend to months or years, when little seems to happen.}
\end{itemize}

Non-equilibrium arrested colloidal gels are ubiquitous amongst everyday formulated products and a key requirement is that they maintain their physical stability and performance during their service life, which can be up to several years. This is imparted by the mechanical strength of the arrested gel network and limited by its ageing rate (which is often temperature dependent), external mechanical disturbances (for example during transport and handling) and gravitational stresses experienced during storage. Notably, in the material response to these, the size and shape of the container plays a surprisingly important role. There is therefore an important need for improved knowledge on ageing of colloidal gels and a better ability to reliably predict their shelf life stability.

\subsection*{Scope and Aims of this Review}

Gels constitute a wide and somewhat ill--defined range of materials. Here we consider \emph{colloidal} gels, which themselves have been subjected to a very wide range of analysis, including theory \cite{lekkerkerker1992,foffi2002,bergenholtz2003,chen2004,shah2003,buscall1987,zaccone2009} and computer simulation \cite{puertas2004,padmanabhan2018,delgado2003,berthier2011}. Experimental techniques include rheology \cite{manley2005time,krishnareddy2012,faers2006}, direct observation and video imaging \cite{starrs2002,bartlett2012} and scattering \cite{poon2002,carpineti1992,cipelletti2000,verhaegh1999}. This has led to great insight, and has been reviewed previously \cite{cipelletti2005,zaccarelli2007}. However we focus here on real space analysis using 3D confocal microscopy with an emphasis on particle--resolved studies. To our knowledge, while particle resolved techniques for colloids have been reviewed  \cite{ivlev}  which briefly covers some early work on gels, no review focussing on gels exists, and that forms the purpose of this work. Where possible, we have referenced relevant review papers, but in any case humbly ask for patience on the part of readers regarding those papers we have missed, or where our opinion seems at odds with theirs. We nevertheless hope to convey the exciting progress made by using real space analysis to study colloidal gels.

This review is organised as follows. Given the wide variety of materials which undergo colloidal gelation, and the broad interest from the basic physics of model far--from--equilibrium systems to a large number of applications, those working with colloidal gels come from a number of disciplines. We have therefore made some attempt to provide sufficient material to cover the underlying physics. Those familiar with spinodal decomposition and/or dynamical arrest may wish to skip the overviews in sections \ref{sectionSpinodal} and \ref{sectionGlass} respectively. In section \ref{sectionAttractionMechanisms} we detail mechanisms of attraction between colloids that lead to gelation. Similarly, readers familiar with these may wish to skip this section. In section \ref{sectionEarly} we review early work on real space analysis of colloidal gels and discuss the consequences of the unexpected degree of electrostatic charging present in the refractive index matched and density matched systems used. In section \ref{sectionEvidence}, we review key developments that provided compelling evidence for spinodal decomposition as the mechanism for colloidal gelation i the context of other proposed mechanisms. In section \ref{sectionMechanism}, we consider the use of real space analysis to provide local mechanisms for dynamical arrest in colloidal gels. Section \ref{sectionDynamical} details the dynamical properties of colloidal gels, and these fall into the following categories. Short--time dynamics with little change in the overall properties are discussed in Section \ref{sectionShortTime}, while changes in the nature of the gel are reviewed in section \ref{sectionAgeing}. Gravitational collapse is considered in section \ref{sectionCollapse} and this leads us to emphasise connections between model systems and real world products in section \ref{sectionBridging}. The response to deformation is considered in section \ref{sectionDeformation}. We then move on to consider the introduction of complexity in depletion systems in section \ref{sectionComplex} before discussing more exotic forms of (colloidal) gels in section \ref{sectionExotic}. We conclude with an outlook in section \ref{sectionOutlook}.

\section{The mechanism of gelation via arrested spinodal decomposition: matter far--from--equilibrium meets the glass transition}
\label{sectionMechanismArrestedSpinodal}

Understanding gelation -- and the mechanism by which gels form invokes two grand challenges of modern physics: matter far from equilibrium and dynamical arrest. Firstly, at least in the context of gelation, the matter far--from--equilibrium aspect pertains to spinodal decomposition, which is reasonably well understood, beginning with the seminal work of 
Cahn and 
Hilliard \cite{onuki,chaikin,cahn1959}.

The second, the glass transition was famously described as ``the deepest problem in condensed matter physics'' by Philip Anderson in 1996 and since then great progress has been made, see e.g. \cite{berthier2011,cavagna2009,royall2015physrep,royall2018jpcm} for reviews and \cite{berthier2016,biroli2013,debenedetti2001,royall2020} for shorter perspectives. However it is fair to say that the problem of the glass transition may be regarded as being solved only for hyperspheres in high dimension \cite{charbonneau2017}.

\subsection{Brief Overview of Spinodal Decomposition}
\label{sectionSpinodal}

\begin{figure}[tb]
\centering
\includegraphics[width=120 mm]{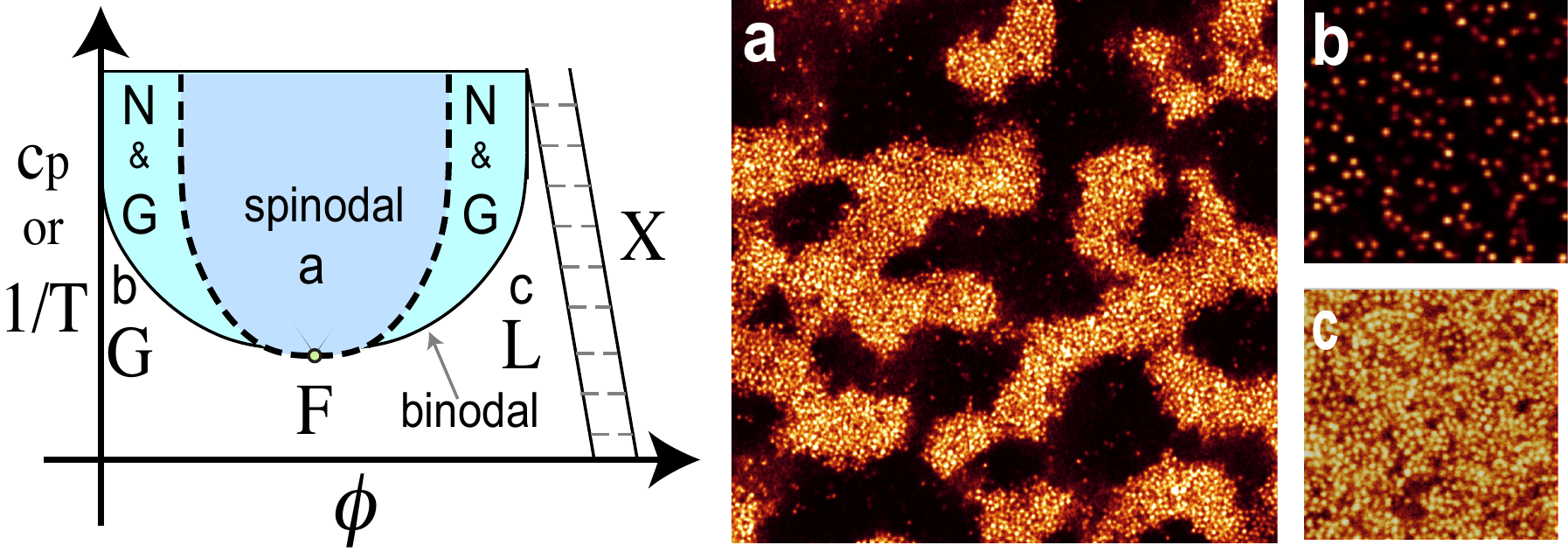} 
\caption{
Schematic of colloid--polymer mixture phase diagram in the case that the effective interaction is \emph{long-ranged}, polymer colloid size ratio $q>0.3$. Here G is a (colloidal) gas, L a liquid, X a crystal and F a (supercritical) fluid. The binodal line marks the limit of the gas--liquid phase coexistence (shaded region). Nucleation and growth is denoted as N\&G, the spinodal line is the thick dashed line. 
(a) A colloidal system undergoing spinodal decomposition, state point is located in the main panel as ``a''.
(b) A colloidal gas,  state point is located as ``b''.
(c) A colloidal liquid,  state point is located as ``c''.
}
\label{figSpinodalCP} 
\end{figure}

We now 
outline spinodal decomposition in the context of colloidal gelation. Figure \ref{figSpinodalCP} is a schematic of a phase diagram for a colloid--polymer mixture 
whose effective colloid--colloid interaction is long--ranged, the polymer--colloid size ratio is $q>0.3$. For such a system, the colloidal liquid (L in Fig.  \ref{figSpinodalCP}) is thermodynamically stable (see section \ref{sectionAttractionMechanisms}) \cite{lekkerkerker1992}. At the binodal line in a colloid--polymer mixture, the chemical potential of the demixed colloidal gas -- colloidal liquid system is the same as the one--phase fluid. However, for sufficient quenching (or, in the case of colloid--polymer mixtures, polymer concentration) the system is unstable even to small fluctuations in composition -- \emph{spinodal decomposition} -- as shown by the dashed line in Fig. \ref{figSpinodalCP}. Here we shall denote composition by $c$ noting that for colloid--polymer mixtures this means the colloid volume fraction $\phi_c$ and polymer concentration $c_p$ which are coupled \cite{lekkerkerker1992}. The canonical treatment is Cahn--Hilliard theory \cite{cahn1959,chaikin}.

We start with a Landau--Ginzburg free energy, which amounts to the observation that because any gradients in concentration $c$ are vectors, and the free energy itself is a scalar, then the lowest--order term will be $(\nabla c)^2$. The free energy of the system of volume $V$ is then

\begin{equation}
F=\int_V f+\kappa (\nabla c)^2 dV
\label{cahnHilliardBegin}
\end{equation}

\noindent where $f$ is the bulk free energy density and $\kappa$ is a constant. We expect that the system will be unstable to small fluctuations in concentration with respect to the mean concentration $c'$, $(c-c')=\delta c=a \sin(\mathbf{k}.\mathbf{r})$. Here the wavevector $k=2 \pi / \lambda_\mathrm{spin}$ with $\lambda_\mathrm{spin}$ a characteristic lengthscale. 

Expanding the free energy density about $c'$ we have

\begin{equation}
f(c)\approx f(c') + (c-c') \left.\frac{\partial f}{\partial c}\right|_{c=c'} + \frac{1}{2}(c-c')^2 \left.\frac{\partial^2 f}{\partial c^2}\right|_{c=c'} + \dots
\label{cahnHilliardMore}
\end{equation}

The integrand in Eq. \ref{cahnHilliardBegin} is then

\begin{equation}
f(c) + \kappa (\nabla c)^2 \approx f(c')  + a \sin(\mathbf{k}.\mathbf{r}) \left.\frac{\partial f}{\partial c}\right|_{c=c'}  + \frac{1}{2} a^2 \sin^2(\mathbf{k}.\mathbf{r}) \left.\frac{\partial^2 f}{\partial c^2}\right|_{c=c'} + a^2 \kappa^2 \sin(\mathbf{k}.\mathbf{r}) + \dots
\label{cahnHilliardSomeMore}
\end{equation}

If we now integrate to obtain the change in free energy over the whole system, we have 

\begin{equation}
\frac{\delta F}{V} = \frac{a^2}{4}\left[ \left.\frac{\partial^2 f}{\partial c^2}\right|_{c=c'} + 2\kappa k^2 \right]
\label{cahnHilliardEvenMore}
\end{equation}

\begin{figure}[tb]
\centering
\includegraphics[width=120 mm]{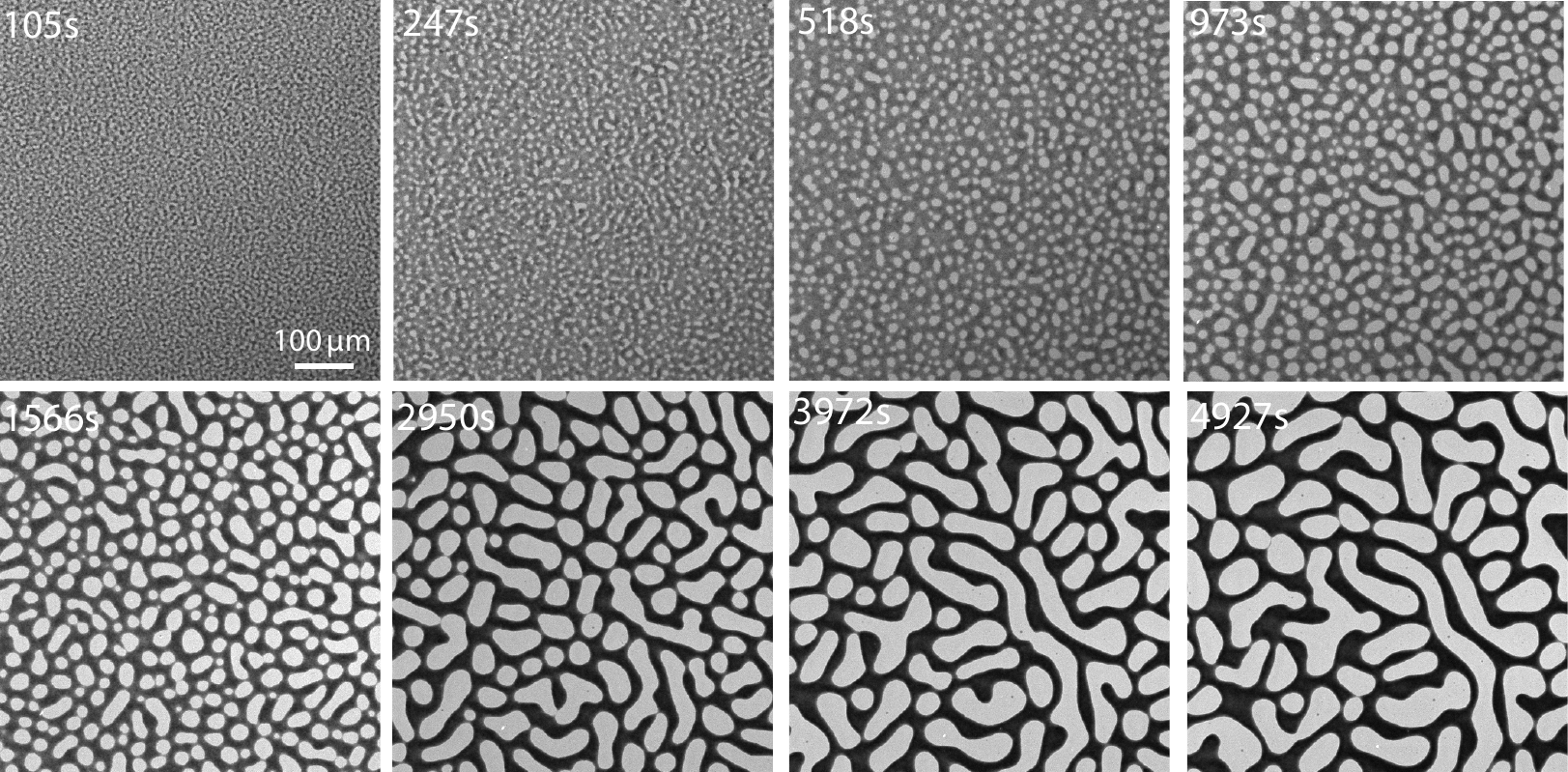} 
\caption{
Confocal microscopy data of a colloid--polymer mixture undergoing spinodal decomposition without arrest. Here the polymer--colloid size ratio is around $q=2.5$, i.e. the polymers are larger than the colloids. The colloidal liquid does not reach volume fractions at which arrest sets in and phase separation proceeds. The early stage of spinodal decomposition (as described by Eqs. \ref{cahnHilliardBegin} -- \ref{cahnHilliardYetMore} is approached in the first two panels. Later stages with coarsening are shown panels corresponding to longer times \cite{jamie2012}.}
\label{figLizJamie} 
\end{figure}

\noindent Thermodynamic stability requires that the term inside the square brackets is positive. So for instability, i.e. spinodal decomposition, we identify a critical (maximum) wavevector

\begin{equation}
k^c = \sqrt{- \frac{1}{2\kappa}\left.\frac{\partial^2 f}{\partial c^2}\right|_{c=c'}  }   
\label{cahnHilliardYetMore}
\end{equation}

\noindent and corresponding lengthscale $\lambda^c = 2\pi/k^c$. What this tells us is that for lengthscales above $\lambda^c$, fluctuations are \emph{unstable}. Now if we consider diffusion we can obtain a fastest growing lengthscale (see Fig. \ref{figLizJamie}). We can also obtain the rate of change of concentration as a function of time which corresponds to that lengthscale. The diffusion equation $\partial c / \partial t= M \nabla^2 \mu $ where $M$ is a constant and the chemical potential in Cahn--Hiliard theory

\begin{equation}
\mu=\frac{\partial F}{\partial c}=\left.\frac{\partial f}{\partial c}\right|_{c=c'} - 2 \kappa \nabla^2 c  
\label{cahnHilliardAndYetMore}
\end{equation}

\noindent We can then relate the free energy density to the rate of change of concentration

\begin{equation}
\frac{\partial c}{\partial t}=M \left[  \left.\frac{\partial f}{\partial c}\right|_{c=c'} \nabla^2 c + 2 \kappa \nabla^4 c \right]
\label{cahnHilliardMoreMore}
\end{equation}

\noindent and as before consider a concentration fluctuation now with a time--dependent component  $\delta c=a \exp(\omega t) \sin(\mathbf{k}.\mathbf{r})$. 
From Eq. \ref{cahnHilliardMoreMore} and dividing by $\delta c$ we have

\begin{equation}
\omega =M \left[  - \left.\frac{\partial f}{\partial c}\right|_{c=c'} k^2 c + \kappa k^4 \right] 
\label{cahnHilliardMoreMoreMore}
\end{equation}

We arrive at a wavevector for the fastest growing fluctuations as

\begin{equation}
k^\mathrm{spin} = \sqrt{-\frac{1}{8k}\left.\frac{\partial^2 f}{\partial c^2}\right|_{c=c'}  }   
\label{cahnHilliardFinal}
\end{equation}

\noindent and corresponding lengthscale $\lambda^\mathrm{spin} = 2\pi/k^\mathrm{spin}$. Therefore, at short timescales, domains of $\lambda^\mathrm{spin}$ increase in colloid volume fraction with growth rate related to $\partial c/\partial t = \omega.$ For \emph{arrested} spinodal decomposition, once the volume fraction in the colloid--rich regime is $\phi_c \approx 0.6$, the system undergoes dynamical arrest [see section \ref{sectionGlass} and Fig. \ref{figJamesPaddyAngell}(a)] \cite{verhaegh1999,manley2005spinodal,lu2008,zaccarelli2008}. This need not be the case, as larger polymer--colloid size ratios lead to colloidal liquids for which $\phi_c < 0.6$, and phase separation proceeds to completion (Fig \ref{figLizJamie}). In this case, at longer timescales, the network coarsens following a power--law $\lambda \sim t^n$ where the power $n$ depends on the particular dynamics of the system.

\begin{figure}[tb]
\centering
\includegraphics[width=150 mm]{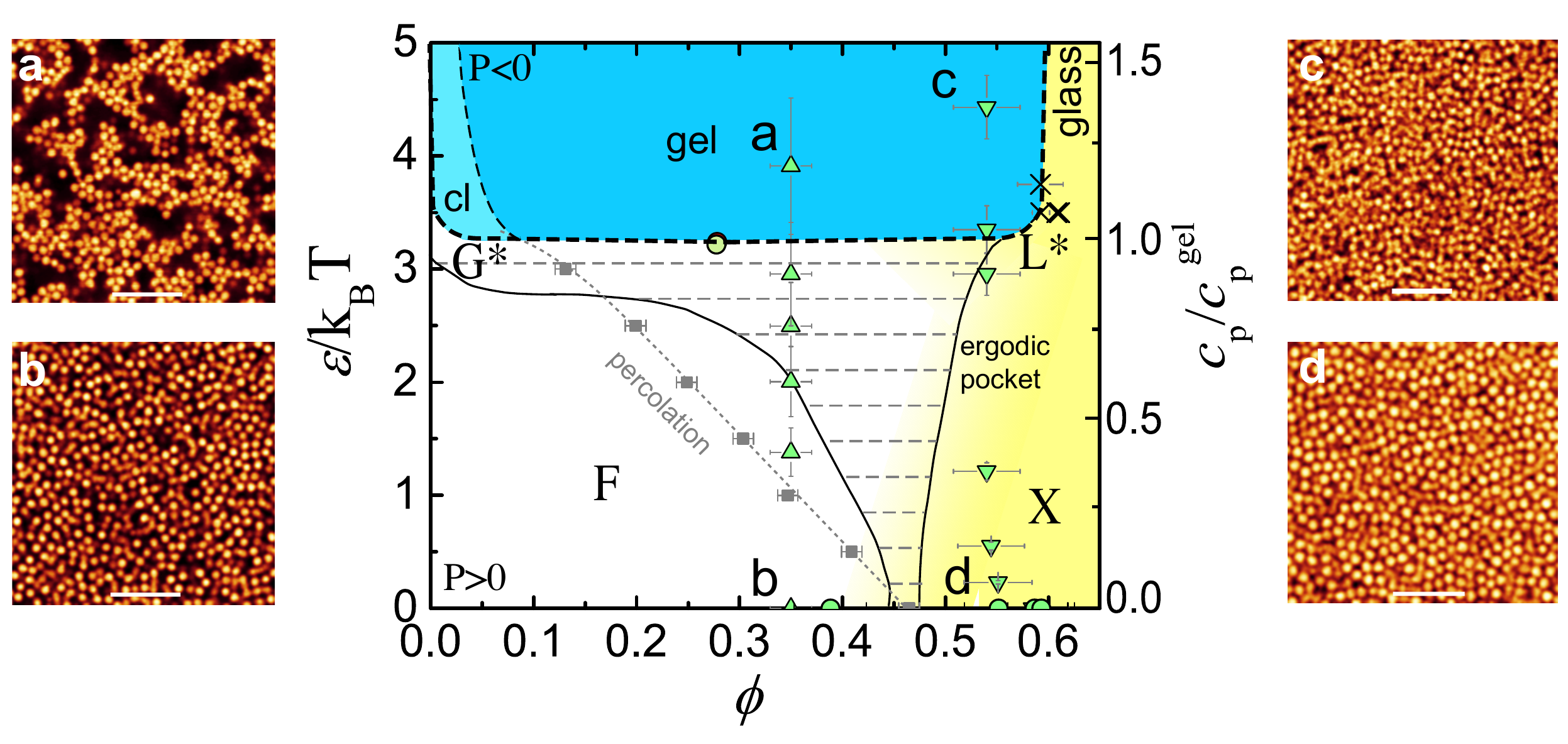} 
\caption{Phase diagram for spinodal decomposition for spheres with a short--ranged attraction (square well with range $0.03 \sigma$). Blue shaded region bounded by thick black dashed line denotes region of spinodal decomposition. Dynamical arrest due to vitrification is shown in shaded yellow regions. Pale shaded region (cl) pertains to cluster--growth driven gelation reflecting the fact that at low volume fraction, there are insufficient colloids to form a percolating network at short times. However the system as still thermodynamically unstable and undergoes spinodal decomposition (section \ref{sectionSpinodal}). Here gelation is time--dependent with a percolating network emerging after some time \cite{manley2005spinodal,griffiths2017}. Experimental data for a colloid--polymer mixture with a short--ranged interaction, $q\approx0.18$ is mapped onto this phase diagram (green data points). 
State points (a-d) are located as shown in the phase diagram. Percolation is indicated by the grey line and is distinct from gelation. 
Bars=20$\mu$m. Adapted from \cite{royall2018jcp}.}
\label{figPhaseGlassGel} 
\end{figure}

\subsection{The Glass Transition in the Context of Spinodal Gelation}
\label{sectionGlass}

Central to the emergence of solidity in colloidal gels is the glass transition that occurs upon densification of the ``colloidal liquid'' phase of the demixing system. We have noted that the colloidal liquid undergoes densification under spinodal decomposition. In colloid--polymer mixtures, partition of the colloids and polymers is such that the colloid--rich phase is largely devoid of polymers. Of course the polymer chemical potential (or, osmotic pressure) may be taken to be equal in both phases \cite{lekkerkerker1992}. However it is nevertheless tempting to view the arrest of the colloid--rich phase as being similar to the glass transition in hard spheres \footnote{While the discovery of the ``attractive glass'' \cite{pham2002} has received much attention, in fact it has more recently been shown that adding attraction to hard spheres does initially accelerate a supercooled hard sphere fluid, but arrest under further attraction is interrupted by gelation for volume fractions at least up to $\phi \approx 0.6$ \cite{royall2018jcp,fullerton2020}.}, and indeed it is hard to avoid the conclusion that the arrest is driven by the steric effects of the colloids.  However it is important to note that in a gel of course there are interfaces between colloid--rich and colloid--poor phases and this is likely where the majority of the particle motion takes place (see section \ref{sectionShortTime}) \cite{puertas2004,gao2007,dibble2008,royall2008naturemater,zhang2013}.

\begin{figure}[tb]
\centering
\includegraphics[width=90 mm]{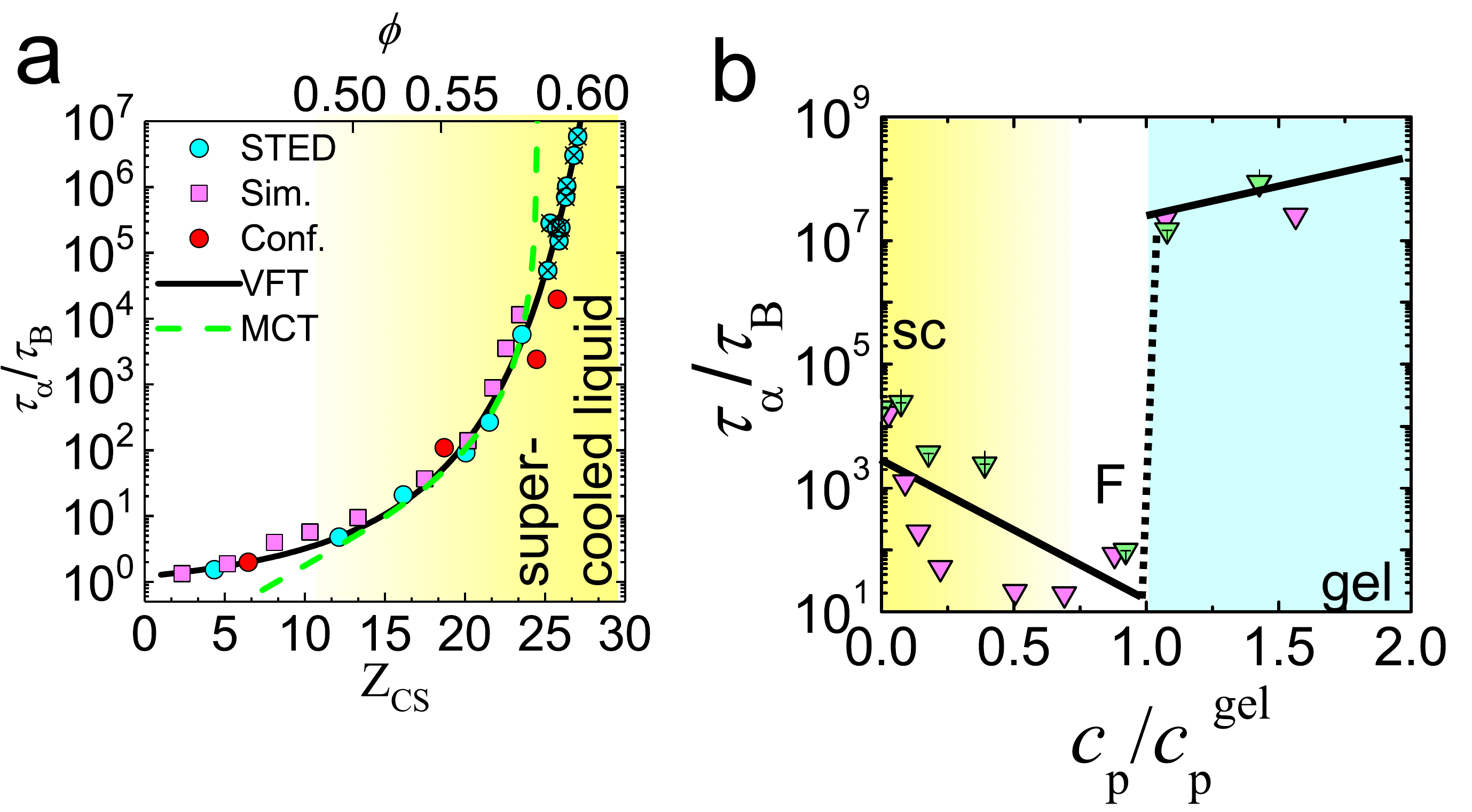} 
\caption{
Differences in dynamical arrest between vitrification and gelation.
(a) The glass transition in hard--sphere colloids. Here $Z_\mathrm{CS}$ is the compressibility factor according to the Carnahan--Starling equation of state. Increasing the volume fraction leads to a \emph{continuous} increase in relaxation time from the colloidal fluid (F) to the supercooled liquid (SC). Here VFT is a fit with the Vogel--Fulcher--Tamman equation (Eq. \ref{eqVFTZ}), red and blue data are from particle--resolved experiments and pink are computer simulation \cite{hallett2018}. 
(b) Gelation. The fluid to gel transition is abrupt, as the system passes through the spinodal line to form the gel (see Fig. \ref{figPhaseGlassGel}). Note that there is also some acceleration for small polymer concentration, this is thought to be due to cage--opening \cite{pham2002}. Pink data are computer simulation, green are particle--resolved experiments \cite{royall2018jcp}.
}
\label{figJamesPaddyAngell} 
\end{figure}

As noted above, the glass transition is a challenge, but one which real space analysis of colloidal hard spheres is amenable to \cite{royall2018jcp,vanblaaderen1995,weeks2000,hallett2020}. Before reviewing some contributions, we give the briefest of outlines of what we consider to be the state of the art. The glass transition is a \emph{scientific revolution} as introduced by Thomas Kuhn \cite{kuhn}. That is to say, there are interpretations (or theories, depending on one's position), which start from \emph{fundamentally different standpoints}, and yet which describe the available data equally well. It is far beyond the remit of this review of real space analysis of colloidal gels to address the glass transition, other than to give reference to some of the many reviews \cite{berthier2011,cavagna2009,royall2018jpcm} and shorter perspectives \cite{berthier2016,biroli2013,debenedetti2001,royall2020} of the glass transition. The best--known first--principles theory, mode--coupling theory (MCT) describes the first 4-5 decades of increase in relaxation time (or viscosity) \cite{goetze,reichman2005}. However, at deeper supercooling, (higher volume fraction in the case of colloids), it has been shown e.g. in [Fig. \ref{figJamesPaddyAngell}(a)] that although mode--coupling theory in its conventional form predicts a transition to a solid with infinite relaxation time, in 
practise relaxation \emph{does} occur, through mechanisms not captured by MCT \cite{hallett2020,brambilla2009,hallett2018}. MCT can be improved, to remove some of the approximations which may serve to neglect higher--order correlations  \cite{janssen2015}, but we are some way from viewing it as a full description.

Mechanisms for the relaxation at deeper supercooling (or higher volume fraction in the case of colloidal hard sphere--like systems) postulated by theories such as Adam--Gibbs \cite{adam1965} random first-order theory (RFOT) \cite{lubchenko2007}, replica theory \cite{charbonneau2017,parisi2010}, and geometric frustration \cite{tarjus2005} are (like MCT) based on thermodynamic principles but are highly approximate. Other interpretations such as dynamic facilitation postulate that it is the dynamical behaviour of the system (not the thermodynamics) which drives arrest \cite{chandler2010}. It has been suggested that these seemingly disparate approaches built on either thermodynamics \cite{charbonneau2017,goetze,reichman2005,adam1965,lubchenko2007,parisi2010,tarjus2005} or dynamics  \cite{chandler2010} may in fact be reconciled \cite{royall2020,turci2017}.

To summarise, it is fair to say that for the purposes of colloidal gelation, colloids at high density may be treated as a viscoelastic system, that is to say, a material that is solid-like on short timescales but ultimately flowing at sufficiently long timescales. It is worth noting that the structural relaxation time $\tau_\alpha$ or viscosity of hard--sphere--like colloidal systems increases \emph{continuously} in response to increasing the volume fraction. This, along with the deviation from MCT, is plotted in Fig. \ref{figJamesPaddyAngell}(a). These data are well--fitted by the celebrated Volgel--Fulcher--Tamman (VFT) ``law''
\begin{equation}
\tau_{\alpha}\left( \phi \right)=\tau_{\infty}\mathrm{exp}\left[ \frac{A}{\left(\phi_{0} - \phi \right) ^{\delta}}\right]
\label{eqVFTZ}
\end{equation}
\noindent 
where $\tau_{\infty} $ is the relaxation time in a dilute system, $\phi_{0}$ is the point at which the relaxation time would diverge, $A$ is a measure of the fragility and $\delta$ is an exponent typically set to one to recover the conventional VFT form. The VFT form has \emph{some} theoretical justification \cite{cavagna2009}, but is far from the only choice, see, \cite{hecksher2008,ozawa2019}, although the VFT fit in Fig. \ref{figJamesPaddyAngell} is clearly better than MCT.

At the end of this section, we identify four characteristics
of spinodal gelation \cite{tanaka2000,zaccarelli2007,ferreirocordova2020}. 
\emph{(i)} The system must undergo spinodal decomposition;
\emph{(ii)} There must be \emph{dynamic asymmetry} between the phases (that is to say, one phase is substantially more viscous that the other);  
\emph{(iii)} The more viscous phase must percolate \cite{cates2004};
\emph{(iv)} The non-equilibrium nature of the gel leads to ageing, in particular coarsening.

\subsection{The Quest for a Theory of Gelation}
\label{sectionTheoryGelation}

This continuous increase in viscosity in vitrification is very different from that of gelation, where the viscosity of the material increases abruptly as the gelation line (the spinodal) is traversed [Fig. \ref{figJamesPaddyAngell}(b)] following the path indicated in Fig. \ref{figPhaseGlassGel} \cite{royall2018jcp}. This is significant, as it emphasises that the glass transition and spinodal gelation are fundamentally different phenomena. One would therefore not expect the same theory to describe both and indeed while it is possible to use advanced liquid state theories to predict the structure of colloidal fluid approaching gelation \cite{shah2003} and mode coupling theory to predict dynamical arrest upon increasing attraction between colloids \cite{bergenholtz2003}, it is unclear that this is really referring to gelation, as the phase separation is not included by default \cite{royall2018jcp}.

We emphasise that gelation via spinodal decomposition is far from the only class of gelation \cite{cipelletti2005,zaccarelli2007}. It forms our focus here because the vast majority of real space studies of colloidal gelation pertain to spinodal gelation. Other standpoints have been put forward including cluster growth \cite{ball1987},  a glass transition of ramified clusters \cite{kroy2004} and an emphasis has been placed on percolation \cite{eberle2011}. In the case of the latter, dynamical arrest was associated with an inferred percolation line in a rheological study \cite{eberle2011}. The relation between percolation and gelation was subsequently probed in real space (Fig. \ref{figPhaseGlassGel}) and also in computer simulation, where no change in dynamics at percolation was found  \cite{royall2018jcp}. This is consistent with fluids of, e.g. square well particles whose second virial coefficient is insufficient for gelation (see below \cite{lu2008}), which can be prepared a volume fractions up to $\phi\lesssim0.58$ without undergoing arrest which of course percolate if one considers the interactions between the particles to constitute a bond. In other words, percolation is a necessary but insufficient condition for gelation. It is possible that the system studied previously \cite{eberle2011} may perhaps have had some additional interactions which might have complicated the situation.

An emphasis has been laid on cluster growth in the approach to gelation \cite{laurati2009} and analogies with the sol--gel transition have been made, lengthscales intermediate between single--particle (microscopic) and macroscopic have been considered. Very recently, gelation has been associated with a non--equilibrium percolation transition \cite{rouwhorst2020ncomms,rouwhorst2020pre}. This is an interesting perspective, playing as it does to particle--level analysis, and is likely to be broadly compatible with the spinodal decomposition mechanism (see section \ref{sectionEarly}, Fig. \ref{figPhaseGlassGel}).

As it stands, while a number of approaches are being pursued \cite{buscall1987,gopalakrishnan2006,chen2004,bergenholtz2003,zaccone2009,laurati2009} theoretical understanding of gelation beyond that outlined here is rather limited, with motivation for future work. Furthermore, this discussion pertains to the process of gelation, the emergence of solidity. The failure of gels under gravity, or external loading is an exciting and challenging research area. Concerning the response of soft materials to external loading, we direct the reader to some excellent reviews \cite{mewis2009,fielding2014,bonn2017,nicolas2018}.

\section{Brief Overview of (Effective) Attractions Between Colloidal Particles}
\label{sectionAttractionMechanisms}

Here we mention some interactions specific to the formation of colloidal gels, and in particular we discuss certain factors pertinent to the work reviewed in the later sections. Interactions between colloids are detailed in a number of textbooks, e.g. \cite{ivlev}.

\textit{Van der Waals interactions. --- }
The van der Waals interactions between colloids result from the sum of these interactions between the constituent molecules, and 
may be taken to have a $1/r^6$ form,

\begin{equation}
u_\mathrm{vdW}(r)=-\frac{C\rho_1\rho_2}{r^6}
\label{eqVDW1}
\end{equation}

\noindent
where $C$ is a constant, and $\rho_{1.2}$ is the number density of molecules in colloids 1 and 2. Integrating over the colloids of volumes $V_{1,2}$ then yields

\begin{equation}
u_\mathrm{vdW}(r)=-\int_{V1} dV_1 \int_{V2} dV_2 \frac{C\rho_1\rho_2}{r^6} 
\label{eqVDW2}
\end{equation}

\noindent
In the case of two spherical colloids of radii $a_1$ and $a_2$, we have

\begin{equation}
u_\mathrm{vdW}(r)=-\frac{A}{3}  \left\{  \frac{a_1a_2}{r^2-(a_1+a_2)^2}  + \frac{a_1a_2}{r^2-(a_1-a_2)^2}+\frac{1}{2}\ln\left(\frac{r^2-(a_1+a_2)^2}{r^2-(a_1-a_2)^2}  \right)  \right\}
\label{eqVDW3}
\end{equation}

\noindent
where $A=\pi^2C\rho_1\rho_2$ is the Hammaker constant. For colloids close together, such that the separation of their surfaces $h=r-(a_1+a_2) \ll \mathrm{min}(a_1,a_2)$

\begin{equation}
u_\mathrm{vdW}(h)\approx-\frac{A}{6h}\frac{a_1a_2}{a_1+a_2}
\label{eqVDW4}
\end{equation}

\textit{Refractive index matching. --- }
Typically the van der Waals interactions between two colloids at contact are thousands of times the thermal energy and result in irreversible aggregation of colloids, if the electrostatic repulsions between the particles are sufficiently screened by the addition of salt. However, refractive index matching of colloids and solvent changes all that. The constant $C$ is related to the (mean) polarisability $\alpha$ of the constituent molecules, which is related to the refractive index $n$ through the Lorentz-Lorenz equation

\begin{equation}
\frac{n^2-1}{n^2+2}=\frac{4\pi}{3}\rho \alpha
\label{eqLorentzLorenz}
\end{equation}

\noindent
where $\rho$ is now averaged. Thus, if the colloids and the solvent have the same refractive index, the net force on the particles will sum to zero and the van der Waals contribution will vanish. In practice, refractive index matching of solvent and colloids is sufficient that the residual van der Waals interactions are a small fraction of $k_BT$ and can often be neglected.

\begin{figure}[tb]
\centering
\includegraphics[width=100 mm]{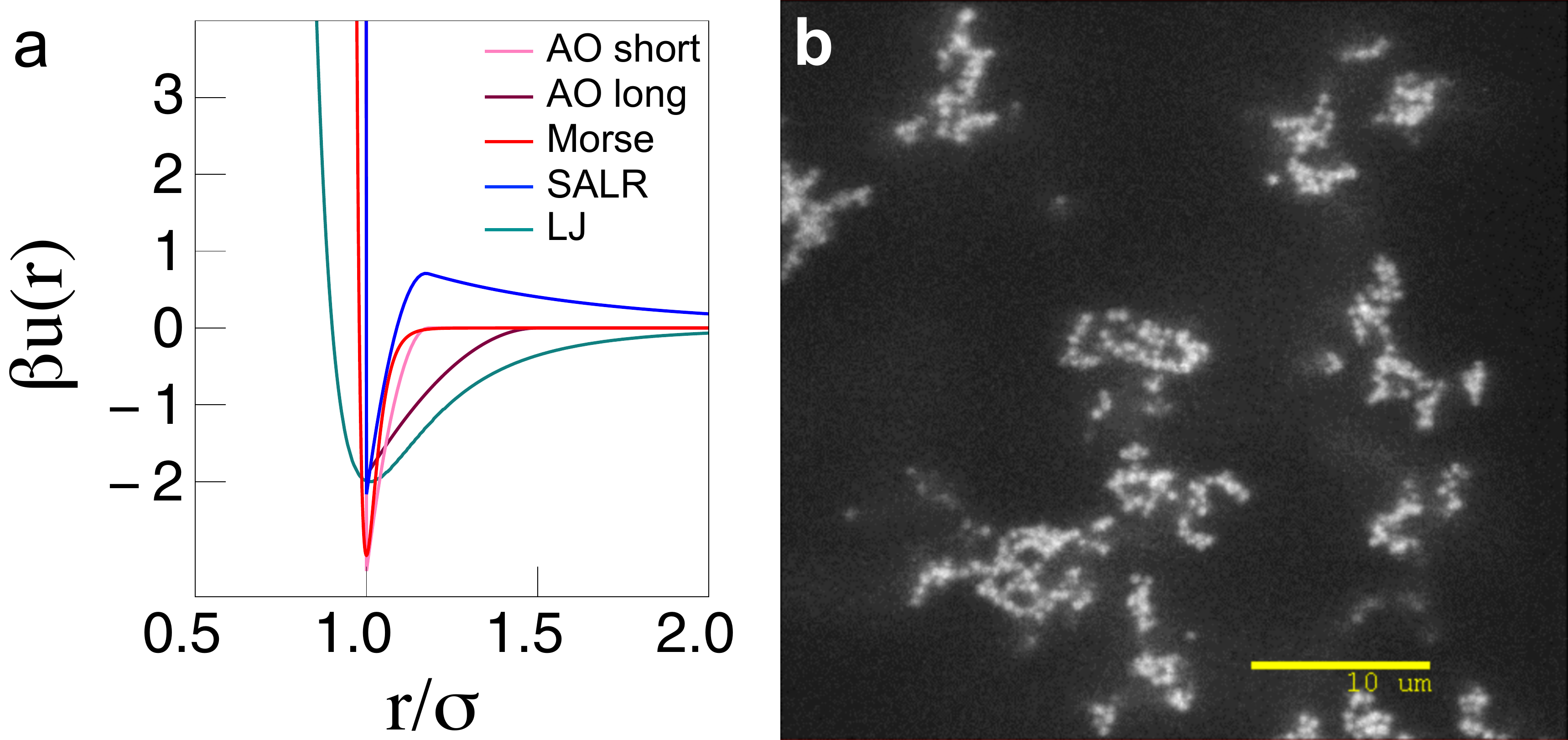} 
\caption{
(a) Spherically symmetric interactions between colloids pertinent to gelation. The Asakura--Oosawa interaction is shown for a short range $(q=0.18)$ system where the colloidal liquid is not thermodynamically stable (pink) and a longer range $(q=0.5)$ where the liquid is stable (maroon). Also shown is the Morse interaction for range parameter $\rho_0=33$ (red) and Lennard--Jones interaction (teal). Finally a competing interaction which is the sum of a short--ranged attraction and long--ranged repulsion (SALR, blue), as postulated for gelation in the density matched PMMA system (see section \ref{sectionEarly}).
(b) Early particle--resolved confocal microscopy image of a colloidal gel. Density matched PMMA colloids with diameter $\sigma=1500$ nm and polystyrene polymer with size ratio $q=0.11$ and colloid volume fraction $\phi_c=0.03$. Bar=10$\mu$m \cite{dinsmore2002}. }
\label{figInteractionsDinsmore} 
\end{figure}

\textit{The depletion interaction. --- }
In addition to the 
short-ranged van der Waals interaction mentioned above, in multispecies colloidal mixtures attraction can be caused by the depletion interaction. This was first introduced by Asakura and Oosawa (AO) \cite{asakura1954} and rediscovered by Brian Vincent and coworkers \cite{long1973}. This depletion interaction is driven by the entropy of the smaller species which can either be colloids or polymers. In the latter case a reasonable approximation is found, so that the polymers are treated as ideal and one can formally integrate them out \cite{dijkstra1999}. This AO model leads to an effective pair interaction between two hard colloidal spheres in a solution of ideal polymers, which is illustrated in Fig. \ref{figInteractionsDinsmore}(a).

\begin{equation}
\beta u_\mathrm{AO}(r)=\cases{\infty&for $r < \sigma$\\
\frac{\pi (2R_G)^3 z_p^r}{6} \frac{(1+q)^3}{q^3}  & \\
\times \{1-\frac{3r}{2(1+q)\sigma}+\frac{r^3}{2(1+q)^3\sigma^3} \} & for $r \ge \sigma < \sigma_C+(2R_G)$ \\
0 & for $r \ge \sigma+(2R_G)$ \\}
\label{eqAO}
\end{equation}

\noindent
where $\beta=1/k_BT$, $R_G$ is the polymer radius of gyration and the polymer--colloid size ratio $q=2R_G/\sigma$, the polymer \emph{fugacity} $z_p^r$ is equal to the number density $\rho_p^r$ of ideal polymers in a reservoir at the same chemical potential as the colloid-polymer mixture. Thus, because the prefactor is proportional to the polymer reservoir concentration, the effective temperature is inversely proportional to the polymer reservoir concentration. The result is an effective interaction between the colloids of range $q\sigma$ and well-depth $u_{\rm AO}^{\rm min}$. For $q \leq 0.1547$ it is formally accurate, in the sense that many--body effects due to integrating out the polymer degrees of freedom are absent  \cite{dijkstra1999,dijkstra2000}. The Morse interaction is a useful, variable ranged attractive interaction which reproduces very well the the AO interaction as shown in Fig. \ref{figInteractionsDinsmore}(a).

\begin{equation}
\beta u_\mathrm{Morse}(r)= \beta \varepsilon_\mathrm{Morse} \exp\left[\rho_{0}   \left(1-r/\sigma \right) \right]    \left(  \exp\left[\rho_{0} \left( 1- r/\sigma   \right) \right] -2  \right) 
\label{eqMorse}
\end{equation}
\noindent
In fact, for some parameters, the Morse interaction appears to reproduce the higher order structure of colloidal fluids better than the one--component AO interaction \cite{taffs2010jpcm}.

\section{Real Space Analysis}
\label{sectionReal}

Here we take real space analysis to pertain to microscope--based studies of colloidal systems. \emph{Particle--resolved studies} is then a subset of real space analysis where the individual colloidal particles are visible \cite{ivlev} and in which the coordinates of colloidal particles may be obtained \cite{ivlev,hunter2012,lu2013,yunker2014}. Particle--resolved studies have had a major impact in a number of fields since 
their inception in the pioneering and far--sighted work of van Blaaderen and coworkers \cite{vanblaaderen1995,vanblaaderen1992}. Such areas include the glass transition \cite{vanblaaderen1995,weeks2000} and crystal nucleation \cite{gasser2001,taffs2013}. Crucial to this breakthrough was the use of a confocal microscope, to enable 3D optical imaging and a colloidal system whose particles were sufficiently large and which were refractive index and density matched to the solvent. The advent of coordinate tracking algorithms \cite{crocker1996,leocmach2013,gao2009,bierbaum2017} meant that the 3D coordinates of the particles could be extracted. In some cases, it is even possible for the \emph{diameter} of every particle to be obtained, which is useful in the case of polydisperse systems \cite{kurita2012}. The technique of particle--resolved studies has been extensively reviewed previously \cite{ivlev,lu2013,hunter2012}. We refer the reader to those reviews and now focus the use of real space analysis in colloidal gelation.

\subsection{Early Work and the Mysteries of Electrostatics in Density--Refractive Index Matched Colloids}
\label{sectionEarly}

The first use of 3D real space analysis for colloidal crystallisation driven by depletion attractions appears to be that of Koenderink \emph{et al.}, who used silica rods as the depleting agent \cite{koenderink1999}. Note that silica is typically not density matched, due to its high density. Indeed to our knowledge, this has only been achieved with a highly volatile solvent \cite{tolpekin2004}. De Hoog \emph{et al.} carried out pioneering particle--resolved studies in a refractive index and (nearly) density matched system with depletion interactions, revealing the interplay between self--assembly and geometric frustration \cite{dehoog2001}. This was followed by Dinsmore and Weitz \cite{dinsmore2002}, which we understand to be the first particle--level  structural characterisation of a colloidal gel. The system considered was the poly--methyl methacrylate (PMMA) colloids with polystyrene polymer in a refractive index and (mass) density--matching solvent. This system, with some variants, has been used extensively since, and some comments are in order.

PMMA colloids were used to mimic hard spheres in the seminal work of Pusey and van Megen \cite{pusey1986}. The thin ($\sim5-10$ nm \cite{poon2012,royall2013myth}) layer of steric stabilisation afforded a degree of softness small in comparison to the diameter of the particles (typically around 400 nm). That and later work which used light scattering for structural analysis of the system was able to use smaller colloids, where the effect of gravity was relatively small. The effect of gravity may be quantified via the gravitational length, which is the height that a colloid may be raised such that the gravitational potential energy is equal to the thermal energy $k_BT$, $\lambda_g=k_BT/(\delta\rho\pi\sigma^3g/6)$ where $\delta\rho$ is the mass density difference between the solvent and the colloids and $\sigma$ is the colloid diameter. Substituting typical values in (assuming that the solvent 
is the commonly used \emph{cis}--decalin, whose density is less than that of PMMA), we find a gravitational length of around $100\sigma$ for 400 nm diameter particles.

However, particle resolved studies requires larger colloids of 2--3 $\mu$m if the coordinates are to be tracked \cite{ivlev}. In the \emph{same solvent} considering the gravitational length \emph{in units of the colloid diameter}, we have a gravitational length of just 0.03$\sigma$, putting such 3 $\mu$m particles firmly in the regime of granular matter where the effects of gravity dominate the thermal energy. Clearly, something must be done to reduce the effects of gravity if these larger particles are to behave as colloids and be dominated by the thermal energy. The solution was to match the density of the solvent and particles, by mixing two solvents together, for example adding cyclo--hexyl bromide (whose density is greater than that of PMMA, and whose refractive index is, like that of \emph{cis}--decalin, conveniently close to that of PMMA), then the $\delta\rho$ term becomes small and colloidal behaviour is regained with $\lambda_g \gg \sigma$. It is worth noting that \emph{perfect density matching is not possible}, due for example to the different thermal expansivities of the PMMA colloids and the solvent mixture, and the temperature fluctuations that occur during experiments. Good density matching might correspond to a gravitational length of (say) $\sim100\sigma$ \cite{royall2005sedimentation}.

Although the sedimentation issues of the larger PMMA colloids were resolved by using the cyclo--hexyl bromide (CHB) --  \emph{cis}--decalin solvent mixture, the larger particles behaved in a manner very unlike the earlier work with smaller sub--micron colloids. The reason for this was the fact that the electrostatic charge borne by the larger PMMA colloids was sufficient to massively change the phase behaviour. The reason is two--fold: firstly the solvent is different: the dielectric constant of the density matching CHB --  \emph{cis}--decalin mixture is 5.37, while that of \emph{cis}--decalin alone is 2.4 \cite{royall2003}. Although ion dissociation is weak in the density matching mixture, it is very much more than in pure \emph{cis}--decalin, and indeed the solvents used in the early work with light--scattering of small particles had dielectric constants around 2, in which there is less ion dissociation with respect to the CHB--\emph{cis}--decalin density matching mixture. Secondly, the particles are much larger. If we make the not entirely unreasonable assumption that the surface charge per unit area is fixed, then an increase in particle size from 400 nm to 3 $\mu$m corresponds to an increase in surface area by a factor of 56. Given that the electrostatic interaction scales with the \emph{square} of the charge it is clear that the electrostatic interactions would be very much stronger. For example, if we consider the case of a 400 nm colloid system with 7 electronic charges per colloid and Debye length of 1 micron, this would have a contact potential of $k_BT$ with little effect upon phase behaviour. A 3 micron colloid, in the same solvent with the same Debye length with a similar surface charge density would have a carry a charge of 415, corresponding to a contact potential of around 100 $k_BT$ -- which is substantially larger than any attraction driven by depletion, so there \emph{should} be no gelation at all \cite{klix2010,royall2018mermaid}. Moreover in the lower dielectric constant solvents used previously with the smaller colloids, the charge carried on each colloid would be even less than 7 elementary charges and often safely neglected. While some experiments on colloids in micro--gravity have been carried out (using techniques other than real space analysis)\cite{manley2005spinodal}, it is hard to see how a confocal microscope would survive a launch in order to be operated in space, so for now, particle--resolved studies remains earthbound. This need not mark the end of the story. For example it has been shown that by simulating microgravity on earth through slow rotation, that gelation can be suppressed \cite{elmasri2012}.

\begin{figure}[tb]
\centering
\includegraphics[width=100 mm]{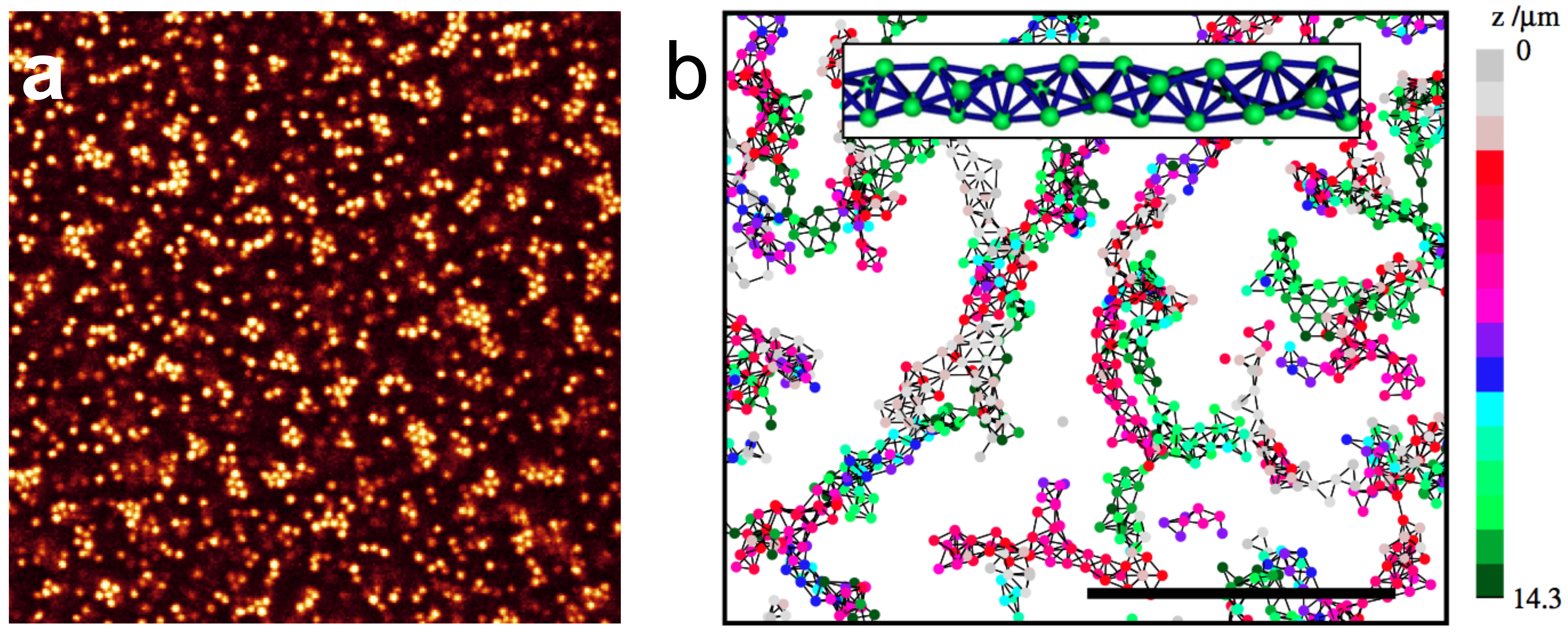} 
\caption{Indications of ``Mermaid-like'' interactions in clusters and gels. 
(a) A confocal micrograph of the clusters in a sample with volume fraction $\phi= 0.086$ and polymer concentration $c_p = 3$ mg cm$^{3}$ $q=0.021$. Here the colloids had a diameter of 1320 nm.
Note the spacing between the monomers, indicating a significant strength and range of the repulsive interactions. Modified with permission from \cite{sedgwick2004}.
(b) A two-dimensional projection of the particle centers within a slab of gel in a colloidal system with competing interactions.
Particles are coloured as a function of their depth within the sample and drawn 40\% of their actual size for clarity. Inset: A spiral chain formed from tetrahedra of particles sharing faces. From \cite{campbell2005}.
}
\label{figHelenWilsonPaul} 
\end{figure}

In retrospect, then, we can see that while Pusey and van Megen found excellent agreement with hard sphere like behaviour with 400 nm colloids in a solvent with dielectric constant around 2 \cite{pusey1986}, Yethiraj and van Blaaderen \cite{yethiraj2003} found that the larger PMMA crystallised at a volume fraction $\phi \lesssim 0.01$, some fifty times less than hard spheres \cite{royall2003,riosdeanda2015}, and Wigner glasses (dynamical arrest driven by electrostatics) have been found at similar volume fractions $\sim 0.01$ \cite{klix2010}. The effects of such electrostatic behaviour have been reviewed in some detail \cite{royall2013myth,royall2018mermaid}, but before leaving this topic we note that the interaction of these nonaqueous systems with water can be very complex \cite{leunissen2007}, and depends on the amount of water the system comes into contact with. What this means is that even the same system may not exhibit the same behaviour, if environmental factors, like the humidity, change. For now we note that electrostatic interactions can have a profound, complex and sometimes uncontrollable, effect upon real space analysis of gelation in colloids.

In applying the then newly--developed density--matched, refractive index--matched PMMA--CHB--\emph{cis}--decalin system to gelation, the work of Dinsmore and Weitz \cite{dinsmore2002} demonstrated the power of particle--resolved studies. The characteristics of colloidal gels, the fractal dimension, mass distribution of clusters and neighbour distributions were obtained. Also, local measures, inaccessible to bulk techniques such as scattering, like the chain topology and the number of pivots for each particle in an ``arm'' of a gel were found. Soon after, the same lab exploited the effect of the range of the depletion interaction: small polymers ($R_G\approx6$) nm lead to short--ranged attractions (Eq. \ref{eqAO}) which are of the same order as the stabiliser layer on the PMMA colloids. This suppresses the rotation of the colloids around one another, presumably as the stabiliser ``hairs'' would 
 interdigitate. Longer ranged interactions with ($R_G\approx35$) nm led to structures consistent with the particles rotating around each other. Thus it is possible to include an angular dependence into a depletion interaction through a judicious choice of polymer size \cite{dinsmore2006}. We note that surface roughness has been explored to considerable effect in the elegant work of Mason and coworkers \cite{zhao2007}.

The work of Dinsmore and Weitz was followed from the same lab, by Lu \emph{et al.}'s work on clusters \cite{lu2006}. This demonstrated a truly exceptional degree of density matching, as clusters of hundreds or even thousands of particles were studied. Given the discussion above, to density match such large assembles (of already rather large colloids), represented a truly Herculean task of exquisite control over the sample preparation and experimental conditions. Lu \emph{et al.} used a large amount of salt to screen the electrostatics. While Yethiraj and van Blaaderen \cite{yethiraj2003} had demonstrated the use of salt to screen the electrostatic interactions, they used relatively small quantities, and indeed it was later determined by Leunissen and coworkers that the saturation concentration of the tetrabutyl ammonium bromide (TBAB) salt often used is a mere 260 nM \cite{royall2003,leunissenthesis}. At this salt concentration, the Debye length ($\approx 100$nm) is similar to the range of the depletion interaction and the electrostatics strongly influence gelation \cite{royall2005competing}. Lu \emph{et al.} used 4mM of the same salt, and it is possible that, due to the low ionic strength and slow approach to equilibrium in these systems \cite{royall2003} some non--equilibrium salt concentrations above the 260 nM were possible. While ion--dependent non-equilibrium phenomena might seem surprising, given that small ions ``should'' relax in the timescale of picoseconds, over a timescale of hours to days, a variety of \emph{peculiar, time--dependent phenomena} have been observed: the low density crystals noted above melt \cite{royall2003,riosdeanda2015} and clusters in the ``mermaid--like'' systems undergo electrostatically--induced fission \cite{klix2010}. In any case, Lu \emph{et al.}'s work found good agreement with expectations from theory (e.g. \cite{lekkerkerker1992,dijkstra1999}), except that the clusters observed would be expected to be a metastable state, with the equilibrium being a (colloidal) gas--crystal phase coexistence. It is worth noting that the approach to equilibrium for such a system can be extremely slow, in principle running to years or more \cite{ruzicka2011}.

An important advance around this time was to explicitly investigate the interplay of solvent composition (and density matching) and electrostatics, by Sedgwick \emph{et al.} \cite{sedgwick2004}. It is clear from the discussion above that this is an important issue, and three scenarios were investigated. The first used a pure \emph{cis}--decalin solvent, which was not density matched but the behaviour was otherwise consistent with expectations. The second case was to consider a density--matched solvent but with added salt, which was also found to be consistent with expectations. The third case where no salt was added was more intriguing as although gels were formed, more polymer was needed than in the case where the electrostatics were screened. This was not surprising, because now the stronger, unscreened electrostatic repulsions needed to be overcome for aggregation, but what was intriguing was the observation of \emph{elongated} clusters [Fig. \ref{figHelenWilsonPaul}(a)]. These were interpreted as being a result of the electrostatic repulsions \cite{groenewold2001}, and indeed similar behaviour was observed in Lyzozyme protein solutions, where the attractions came from van der Waals interactions between the protein molecules \cite{stradner2004}. For proteins, it is important to emphasise that while some observations are possible with real space analysis using optical microscopy (see section \ref{sectionExotic} \cite{riosdeanda2019}), x--ray scattering can probe the lengthscales appropriate to interactions between protein molecules \cite{stradner2020,Bucciarelli2015,bucciarelli2016,myung2018}. Indeed, x--ray scattering was used to infer the protein clusters in question, however the results were later disputed \cite{shukla2008}. It is worth noting that while such spherically symmetric short--range attraction -- long--range repulsion (or ``mermaid'') interactions [Fig. \ref{figInteractionsDinsmore}(a)] are a simple model for protein--protein interactions, in reality the situation is far more complex, due not least to the relatively small number of discrete charged sites on the protein and side--chains \cite{riosdeanda2019,mcmanus2016,zhang2008}.

Unlike the protein clusters, due to the larger size of the particles, the elongated colloidal clusters were clearly visible in the confocal images in Fig. \ref{figHelenWilsonPaul}(a). Gels of such particles with these ``mermaid'' interactions were even shown to be resistant to gravitational collapse (section \ref{sectionCollapse}) \cite{vanschooneveld2009}. However the behaviour of the electrostatic interactions in this and other work \cite{royall2005competing,klix2010,campbell2005,klix2013} have been analysed in some detail. In the case that the electrostatics are strongly screened through the addition of salt \cite{royall2005competing,capellmann2016}, behaviour consistent with a superposition of attractive (depletion) and repulsive (electrostatic) contributions can be realised. However in this case the range of the electrostatic interactions is comparable to, or less than, the depletion-induced attractions. In the case that the screening is weak, such that the electrostatic repulsions are longer--ranged than the attractions a breakdown in spherical symmetry of the interactions has been identified \cite{royall2018mermaid,klix2013}, not to mention large inconsistencies in the calculated strengths of the attractive and repulsive components \cite{klix2010,royall2018mermaid}. Even in the case that the dielectric constant is around 2 through a suitable choice of solvents, so that electrostatic interactions are small, surprises can still occur \cite{klix2013}. In the multicomponent index and density matching solvents used, one or other component can be absorbed by the colloids and this breaks the refractive index matching, leading (in addition to poor imaging) to significant van der Waals attractions between the colloids \cite{royall2013myth,ohtsuka2008}.

Other work showed fascinating Bernal spirals [Fig. \ref{figHelenWilsonPaul}(b)] \cite{campbell2005} which were attributed to the interplay between long--ranged electrostatics and polymer--induced depletion attractions. These spirals were subsequently reproduced by computer simulation \cite{sciortino2005}, although the experimental system likely exhibited the breakdown in spherical symmetry of the interactions noted above.

The interplay of electrostatics and depletion in the PMMA system continues to arouse interest. Recently, Kohl \emph{et al.} suggested that \emph{directed} percolation, in which percolation occurs through steps through the network only in a preferred direction could be identified in gelation of the charged PMMA system \cite{kohl2016}. This was curious, as there is no obvious reason for the symmetry breaking implicit in a preferred direction for directed percolation. Percolation (that is to say, un--directed percolation) had previously been linked to gelation \cite{kim2013}, but subsequent work has shown that while percolation is a necessary condition for gelation, no change in dynamical behaviour could be identified upon percolation \cite{royall2018jcp}. The same holds for directed percolation which occurs at higher colloid volume fraction than does (undirected) percolation, but there was found to be no change in dynamical behaviour at the directed percolation transition \cite{richard2018jcp}.

\begin{figure}[tb]
\centering
\includegraphics[width=100 mm]{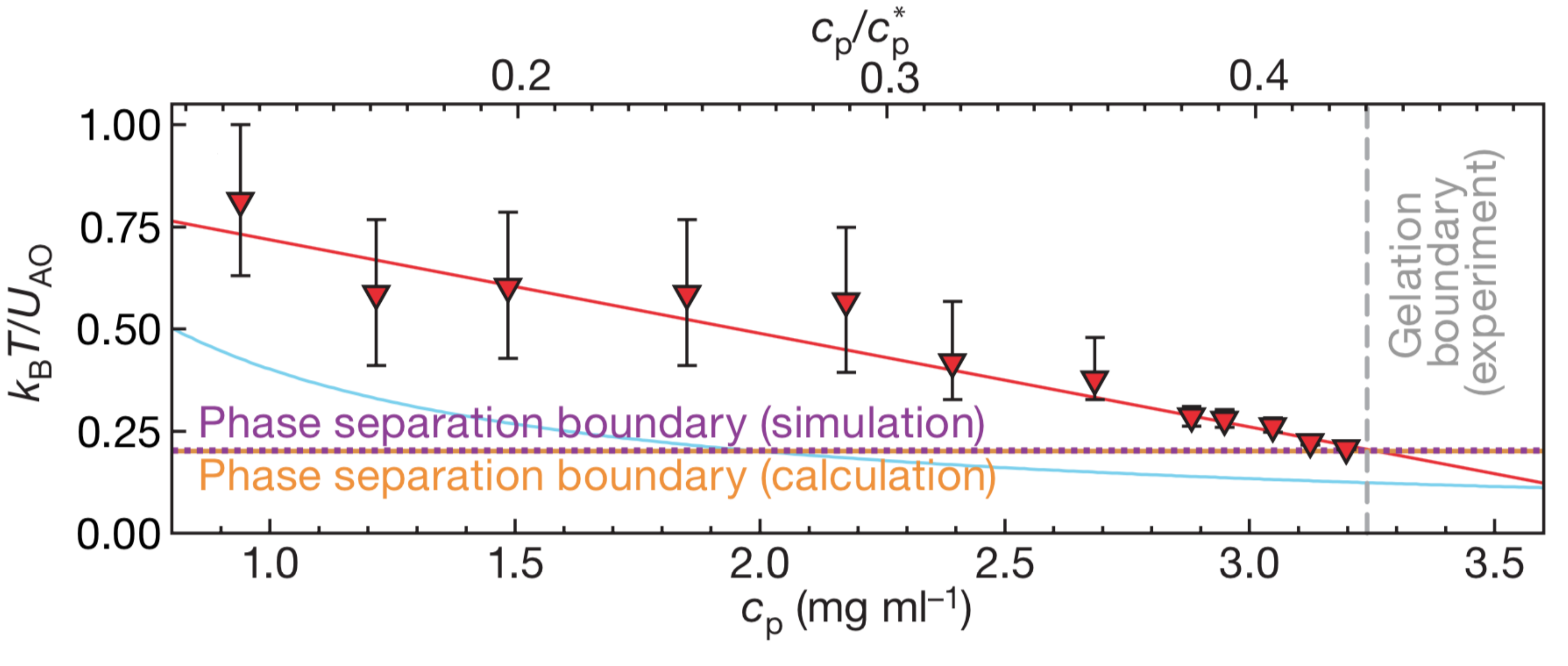} 
\caption{
Mapping interactions in particle--resolved experiments to computer simulation using cluster size distributions. Note that the measured gel point coincides precisely with that predicted for the system to undergo phase separation.
 \cite{lu2008,zaccarelli2008}. }
\label{figPeterLu} 
\end{figure}

\subsection{Evidence for spinodal decomposition as a mechanism for gelation}
\label{sectionEvidence}

While we have assumed that spinodal decomposition is the mechanism by which a colloid--rich network is formed (the beginnings of phase separation), and local dynamical \emph{arrest} prevents gels of ``sticky spheres'' from fully phase separating, in fact we have provided little hard evidence in that direction. Spinodal decomposition had been proposed for some time \cite{carpineti1992,cipelletti2000,verhaegh1999,manley2005spinodal}, but was not the only proposed mechanism, others included MCT \cite{bergenholtz2003} and arrest of fractal clusters \cite{kroy2004}, along with cluster aggregation \cite{ball1987,rouwhorst2020ncomms}. It is fair to say that the \emph{tour de force} of Lu \emph{et al.} in 2008 \cite{lu2008,zaccarelli2008} provided, if not unequivocal proof then very compelling evidence in support of spinodal decomposition. In an ingenious approach, they mapped the experimental PMMA with salt system onto the square well interaction potential, whose phase diagram is well known. Now there are various ways to map colloidal systems onto well--known models \cite{poon2012,royall2013myth,royall2003}, but it is often challenging and confounded by poorly controlled phenomena like electrostatics (see above) \cite{royall2018mermaid}. The method employed here was novel and elegant: the size distribution of clusters in the ergodic fluid phase was compared between simulation and experiment and from this equivalent state points were identified as shown in Fig. \ref{figPeterLu}. This mapping of equivalent state points enabled the reduced second viral coefficient $B_2^*$ to be determined in the experiment (it can readily be calculated for the simulation). It is known that we expect spinodal decomposition for the square well system for $B_2^*\approx3/2$ \cite{noro2000} \footnote{Strictly this holds only at criticality, however the spinodal of the square well system for these short ranges is so flat, that for practical purposes the same interaction strength induces gelation for off-critical colloid volume fractions also \cite{royall2018jcp}.}, and this is precisely what Lu \emph{et al.} found. They were also able to use their particle--resolved data to measure the volume fraction of the dense phase for the first time and arrived at a value of $\phi_c=0.60$, certainly sufficient for dynamical arrest [see Fig. \ref{figJamesPaddyAngell}(a)]. The strength in Lu \emph{et al.}'s work lay at least in part in its precision: at the very attraction strength at which phase separation and spinodal decomposition is expected, is when gelation is observed. Later work showed that the dynamics change abruptly at the point of gelation \ref{figJamesPaddyAngell}(b) \cite{royall2018jcp}. This does not mean that other mechanisms do not contribute and particularly at low volume fractions, there is strong evidence for the role of cluster growth \cite{manley2005spinodal,griffiths2017} (see also the discussion in section \ref{sectionShortTime}). However such clusters almost certainly lie within the region of the phase diagram where spinodal demixing occurs, with percolation taking longer to occur at these low volume fractions (see the pale blue region where cluster growth is important in Fig. \ref{figPhaseGlassGel}). Percolation is of course  a necessary (but insufficient) condition for gelation) \cite{cates2004}.

\begin{figure}[tb]
\centering
\includegraphics[width=60 mm]{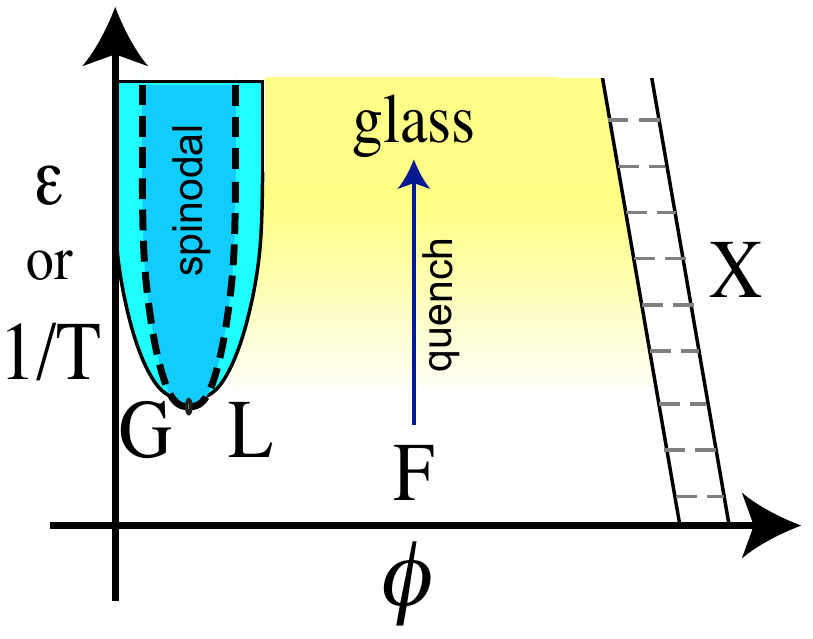} 
\caption{
Schematic of inducing dynamical arrest in ``empty liquids'' via increasing the attraction strength. An ergodic fluid with weak interactions will exhibit progressively slower dynamics (yellow shaded region) upon an increase in the attraction strength (blue arrow maked ``quench''). By passing to the high--density side of the liquid--gas phase separation (which occurs at very low volume fraction in ``empty liquids'') spinodal decompositon is bypassed. The dynamics are expected slow in a manner than is Arrhenius (or close to Arrhenius)  \cite{saikavoivod2011}, consistent with strong glassforming systems like silica \cite{royall2015physrep}. This is quite unlike the drastic change upon crossing the spinodal (Fig. \ref{figJamesPaddyAngell}(b)). Such a continuous increase in the structural relaxation time is indistinguishable form vitrification from a thermodynamic point of view \cite{royall2018jcp}.}
\label{figPhaseEmptyArrest} 
\end{figure}

One may enquire as to the nature of gelation (if any) of systems in which spinodal decomposition does not occur. In a most elegant work, Bianchi \emph{et al.} showed that by reducing the ``valency'' of colloidal particles, the liquid--gas transition can be suppressed to low volume fraction \cite{bianchi2006}. Such \emph{empty liquids} thus enable attraction--driven arrest without spinodal decomposition intervening. As shown schematically in Fig. \ref{figPhaseEmptyArrest} one may ``quench'' the system by increasing the attraction strength higher volume fraction than the liquid--gas binodal and spinodal decomposition is avoided. It has been shown that under such conditions, the increase in structural relaxation time is continuous and gradual \cite{saikavoivod2011}, quite unlike the drastic increase in structural relaxation time that occurs in the case of spinodal gelation (Fig. \ref{figJamesPaddyAngell}(b)). In fact, such systems are thermodynamically indistinguishable form strong glasses formed from network glassformers like silica \cite{saikavoivod2011,royall2015physrep,berthier2011}, and thus in a sense, are not gels at all -- unless we recast most famous glass of all as a gel!

\subsection{A Local Mechanism for Arrested Spinodal Decomposition}
\label{sectionMechanism}

The mechanism of dynamical arrest in glasses, and, by implication, colloidal gels, has long confounded our understanding, as discussed in section \ref{sectionGlass}. Among the key challenges is the noted perceived wisdom that \emph{the microscopic structure underlies the macroscopic behaviour} \cite{peterquote}. The near--absence of any change in the two--point structure (for example the static structure factor)  in atomic and molecular  \cite{royall2015physrep} or colloidal \cite{vanmegen1998} systems approaching the glass transition thus presented a challenge. Higher--order correlation functions, for example three--body correlations or geometric motifs such as the icosahedra originally proposed by Sir Charles Frank, due to their low potential energy with respect to crystalline structures \cite{frank1952} and later incorporated in the theory of geometric frustration by Tarjus and coworkers \cite{tarjus2005} are hard, though not impossible \cite{royall2015physrep} to identify in atomic and molecular systems.

\begin{figure}[tb]
\centering
\includegraphics[width=150 mm]{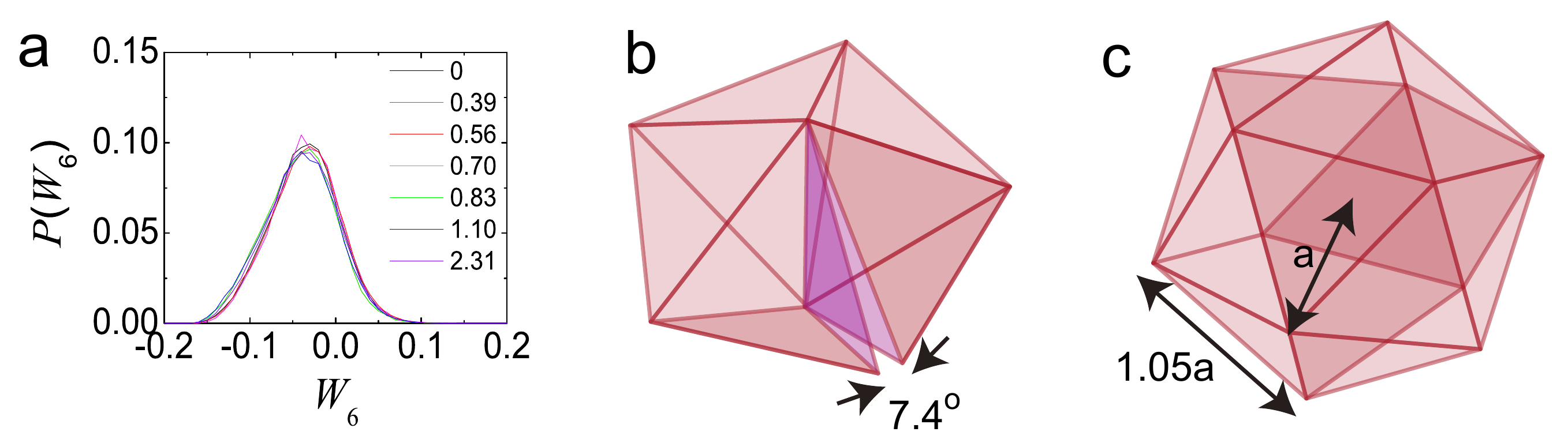} 
\caption{Geometric frustration in fivefold symmetric polyhedra. 
(a) $W_{6}$ bond-order parameter distribution. Colours correspond to polymer weight fractions $\times10^{-4}$. Here gelation occurred at $c_{PG}=7.7\pm0.7\times10^{-5}$ \cite{royall2008naturemater}. 
(b) The pentagonal bipyramid is constructed with five tetrahedra which leads to a gap of $7.4^\circ$. 
(c) 13-membered icosahedron. Icosahedra have bonds between particles in the shell stretched by 5\% with respect to the bonds between the central particle and the shell particles.
\label{figBakaFivesW6}
}
\end{figure}

However, in particle--resolved studies access to higher--order correlation functions is afforded by access to the particle coordinates, which was recognised by van Blaaderen and Wiltzius \cite{vanblaaderen1995}, in their determination of the Steinhardt bond--orientational order parameter $W_6$ \cite{steinhardt1983}, whose negative value was interpreted as evidence for fivefold symmetry, consistent with the conjecture of Frank 
\cite{frank1952}. Some of us  built on Frank's ideas and applied them to colloidal gelation. Frank's conjecture was that because the icosahedron was the minimum energy structure for thirteen Lennard--Jones atoms \emph{and} because a common coordination number in bulk liquids is 12, one should expect such geometric motifs in supercooled liquids. It was therefore tempting in the case of colloidal gels, to test this idea. This was done 
and the $W_6$ distribution in a colloidal system undergoing gelation is shown in Fig. \ref{figBakaFivesW6}(a). The results were \emph{extremely} disappointing, with hardly any change in the $W_6$ distribution upon gelation \cite{royall2008naturemater}.

Now colloidal gels, as Figs. \ref{figPhaseGlassGel}(a) and \ref{figInteractionsDinsmore}(b) make plain, are not homogenous materials like liquids, rather they are highly inhomogenous. Moreover the Asakura--Oosawa interaction between the colloids is rather different to the Lennard--Jones interaction [Fig. \ref{figInteractionsDinsmore}(a)]. Notably it exhibits a hard core, and hard spheres and ``sticky spheres'' have a distinct and more complex energy landscape than does the longer--ranged Lennard--Jones interaction \cite{robinson2019,meng2010,malins2009,miller1999}. Indeed it turns out that only in the limit of infinite interaction range, i.e. infinitely large polymers, does the icosahedron acquire the same binding energy as the crystal structures, suggesting that Frank's hypothesis might not apply to colloid--polymer mixtures at all. This is due to the hard core in the AO model, and packing of spheres in an icosahedron requires either a small compression of the inner sphere or that the outer spheres do not touch one another, see Fig. \ref{figBakaFivesW6}(b,c).

Doye \emph{et al.} explored the effect of the interaction range on the minimum energy clusters \cite{doye1995}, by considering the variable--ranged Morse potential (Eq. \ref{eqMorse}). The Morse potential is controlled by its range parameter, $\rho_0$, and for $\rho_0=6$, it is rather similar to the Lennard--Jones interaction, and indeed the minimum energy cluster for $m=13$ particles is indeed the icosahedron. However, as shown in Fig. \ref{figInteractionsDinsmore}(a), for a range parameter $\rho_0=33$, the interaction is very similar to an AO interaction for a polymer--colloid size ratio $q=0.18$ and in this case, the icosahedron is not the minimum energy structure.

There is further consideration which suggests that minimum energy clusters may be appropriate when considering the local structure of colloidal gels. Loosely speaking, clusters may be thought of as ``zero--dimensional'' objects. Bulk liquids, of course, are three--dimensional. What this means is that while, for example the icosahedron is the minimum energy cluster of 13 particles in isolation, this is not \emph{a priori} valid in a bulk liquid. In fact, taking a mean--field treatment of the surrounding liquid, 
Mossa and 
Tarjus showed that indeed the icosahedron is still the minimum energy structure in a bulk Lennard--Jones liquid, albeit with a significantly reduced energy benefit with respect to other structures, compared to 13 particles in isolation \cite{mossa2003}. More troubling is the observation than in the Lennard--Jones liquid at the triple point, only one particle in a thousand is part of an icosahedron \cite{taffs2010jcp}, although upon supercooling (in the absence of crystallisation) the population increases markedly in a variety of systems \cite{hallett2020,hallett2018,royall2015jnonxtalsol,malins2013jcp,malins2013fara}. Consideration of colloidal gels, as shown in Figs. \ref{figInteractionsDinsmore}(b) and \ref{figHelenWilsonPaul}(b), suggests that thinking of the ``arms'' of the gel to be quasi--one dimensional, we may expect that the clusters might be a reasonable description of the local structure of a colloidal gel.

\begin{figure}[tb]
\centering
\includegraphics[width=140 mm]{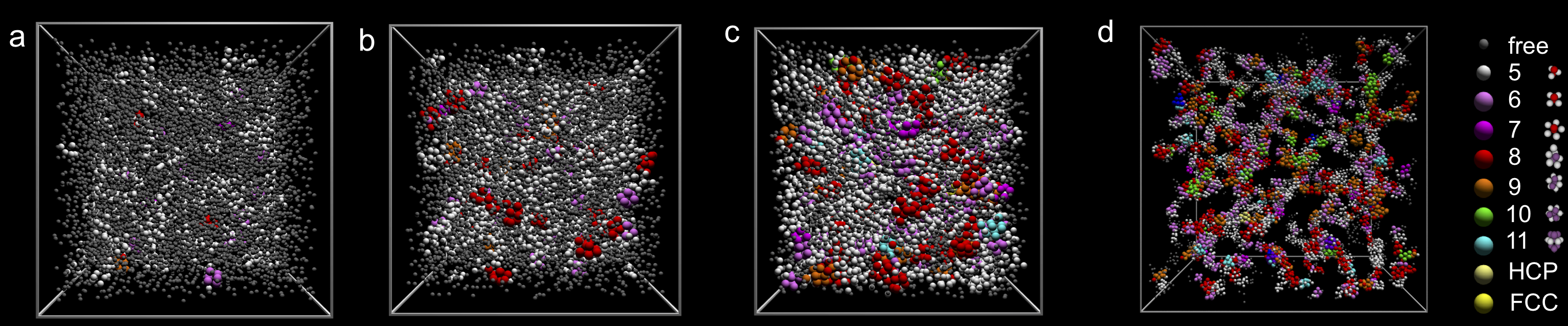} 
\caption{\textbf{Coordinates identified as belonging to different clusters}. 
(a) colloidal fluid, polymer mass fraction $c_{p}/c_{pg}=0.73$. 
(b) (ergodic) liquid close to gelation, $c_{p}/c_{pg}=0.92$. 
(c) colloidal gel, $c_{p}/c_{pg}=1.4$. 
(d) dilute gel $\phi=0.05$, $c_{p}/c_{pg}=2.28$ showing percolating cluster structure. Particles are colour-coded according as follows: grey, free (not in any cluster)
shown 0.4 actual size, white, $m=5$, shown 0.6 actual size. Yellow, crystalline, shown 0.8 actual size. Other particles are members of cluster of size $m$ given by the colours, shown 0.8 actual size.
Here $c_{pg}$ is the polymer concentration required for gelation \cite{royall2008naturemater}.
\label{figFivesTCC}
}
\end{figure}

The idea of Frank, of 13--particle icosahedra, could therefore be extended in two ways. Firstly, minimum energy clusters appropriate to colloid--polymer mixtures should be considered, and here we appeal to the work of Doye \emph{et al.} \cite{doye1995}. Secondly, given the inhomogenous nature of the material, we expect that cluster sizes other than $m=13$ would be appropriate. To address this, the \emph{topological cluster classification} (TCC) was developed \cite{royall2008naturemater,malins2013tcc,royall2008aip}. This algorithm detects groups of particles whose bond network is identical to the minimum energy clusters depicted in Fig. \ref{figFivesTCC}. The transition in cluster population from the ergodic fluid (dominated by particles not in a cluster, rendered in grey in Fig. \ref{figFivesTCC}) to the gel which is dominated by particles in clusters, such the five--membered triangular bipyramid rendered in white and five--fold symmetric 8--membered cluster rendered in red is dramatic, especially given the failure of the bond--orientational order parameter $W_6$ [Fig. \ref{figBakaFivesW6}(a)].

The importance of clusters was later emphasised in work which related long--lived and stable clusters to the mechanical properties of gels, and also explored the role of isostaticity \cite{hsiao2012}. The role of isostaticity gained further traction in an ingenious series of experiments performed by Tsurusawa \emph{et al.} \cite{tsurusawa2018}. Here the process of gelation was observed \emph{in--situ}. The emergent rigidity of the gel network was identified with a strong increase in the population of 6--fold coordinated (isostatic) particles, and in particular a network of these. Related ideas have been expressed 
recently, this time focussing on contacts between clusters \cite{whitaker2019}. And indeed, clusters have been postulated as a mechanism for gelation due to their packing \cite{kroy2004}. Given that many of these papers use one or other method to analyse their data, it is hard to escape the impression that the different analyses are different facets of the same phenomenon: 
there is a clear motivation to apply a range of analysis techniques to a single set of data.

\begin{figure}[tb]
\centering
\includegraphics[width=100 mm]{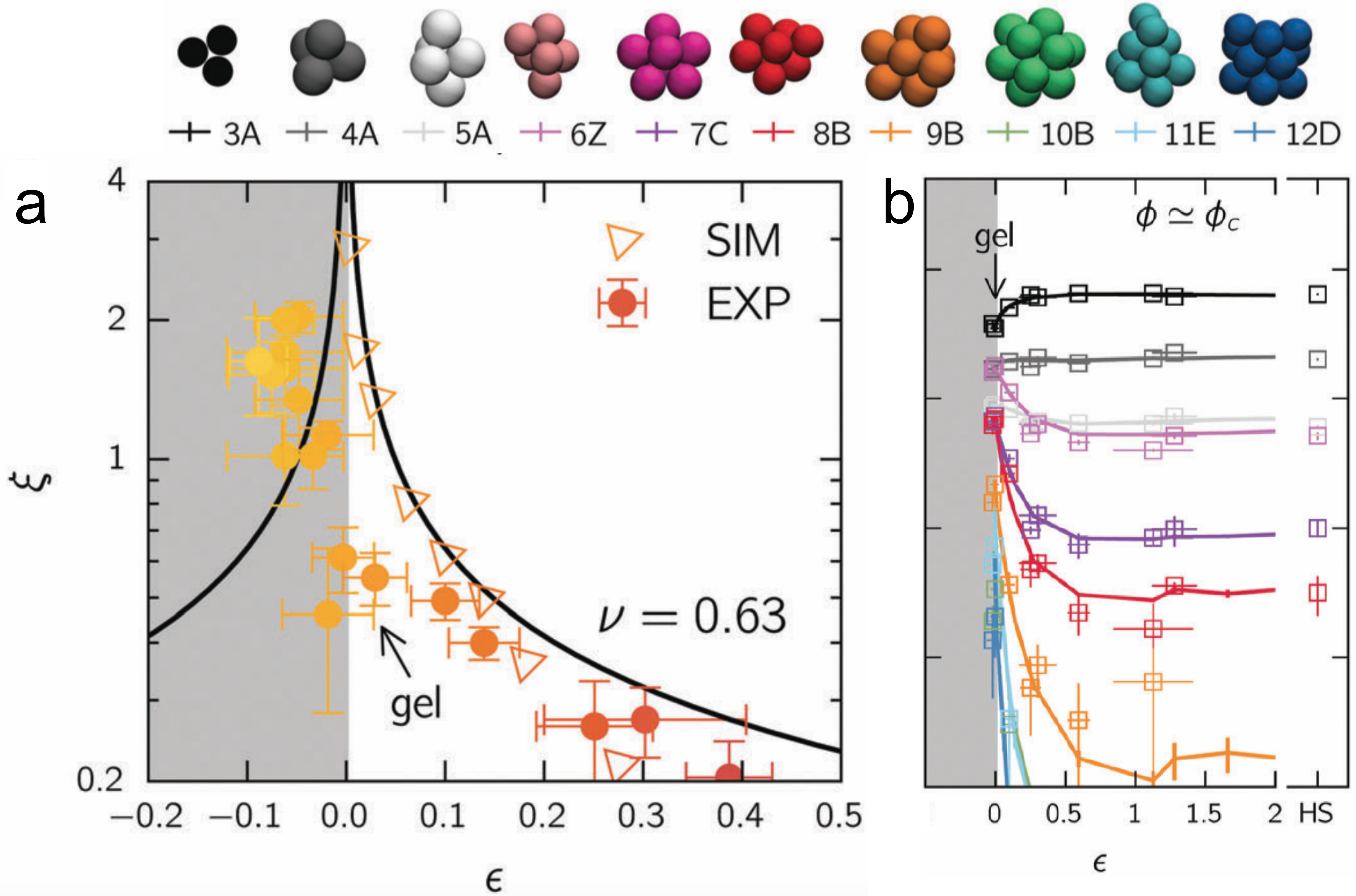} 
\caption{\textbf{Critical behaviour in colloidal gels}. 
(a) correlation length of density--density fluctuations $\xi$ as a function of the reduced effective temperature $\varepsilon=(c_p^c-c_p)/c_p$ for the volume fraction corresponding to the critical isochore $\phi_c^c$.
(b) population of TCC clusters (indicated above) as a function of the reduced effective temperature $\varepsilon$
\cite{richard2018softmatter}.}
\label{figDavidRichard}
\end{figure}

\textit{Criticality. -- } Spinodal gelation in sticky sphere--like systems with short--ranged attractions is intimately related to liquid--gas demixing (Fig. \ref{figPhaseGlassGel}) \cite{royall2018jcp}. One may enquire what happens around the critical point. Naively one imagines the fluid side is similar to a system with long--ranged interactions where liquid--gas demixing is thermodynamically stable. It is as if the system does not ``know'' that it will undergo gelation upon demixing. Thus one can determine the correlation length $\xi$ of density--density fluctuation approaching criticality. As shown in Fig. \ref{figDavidRichard}(a) it is consistent with the onset of critical scaling \cite{richard2018jcp}, which is known to apply rather far from criticality, indeed until $\xi \approx \sigma/2$ \cite{royall2007naturephys}. On the gel side, the correlation length does not obey critical scaling, as the system is not in equilibrium. These density fluctuations lead to a substantial increase in the cluster population [Fig. \ref{figDavidRichard}(b)], though little evidence of crystalline clusters is found even close to criticality (which might have been expected from the increase in nucleation rate in the vicinity of the critical point see section \ref{sectionAgeing}).

\section{Dynamical Properties of Colloidal Gels}
\label{sectionDynamical}

We have discussed above mechanisms for solidification in colloidal gels. We now turn our attention to dynamical behaviour. This can be interpreted at three levels, pertaining to different timescales:

\begin{itemize}
\item{Timescales where the properties of the gel as a whole remain largely unchanged. In this regime, most results concern single--particle motion and the timescales are short, on the order of the Brownian time $\tau_B$.}
\item{Timescales, over which there is significant ageing or even failure of the gel. Here the properties of the gel change measurably over time. Depending on the age of the gel, these can be short ($\sim\tau_B$) or long ($\gg\tau_B$).}
\item{Timescales related to (periodic) deformation such as through shear or flow. In this instance the dynamical mechanisms at play are strongly related to the choice of deformation rate.}
\end{itemize}

Before discussing each case, we emphasise that for particle--resolved studies, an important consideration is particle--tracking errors \cite{ivlev,royall2007jcp}. In the case that particles move rather little, say $\lesssim 0.1\sigma$, it is likely that particle tracking errors can contribute significantly to the measured displacements, It is often hard to distinguish what is a ``real'' movement and what emerges from the limitations of the tracking analysis, and this should be borne in mind when interpreting data.

\subsection{Short--Time Dynamics}
\label{sectionShortTime}

In the case of short--time dynamics, it is natural to consider dynamic heterogeneity. This has been extensively studied in supercooled liquids and glasses since its discovery by 
Harrowell and coworkers in the early 1990s \cite{hurley1995}. In the case of glasses and supercooled liquids, which are rather homogenous, dynamic heterogeneity pertains to regions where the particles relax at different rates, but the structure is often rather similar between slow and fast regions \cite{royall2015physrep}.

\begin{figure}[tb]
\centering
\includegraphics[width=140 mm]{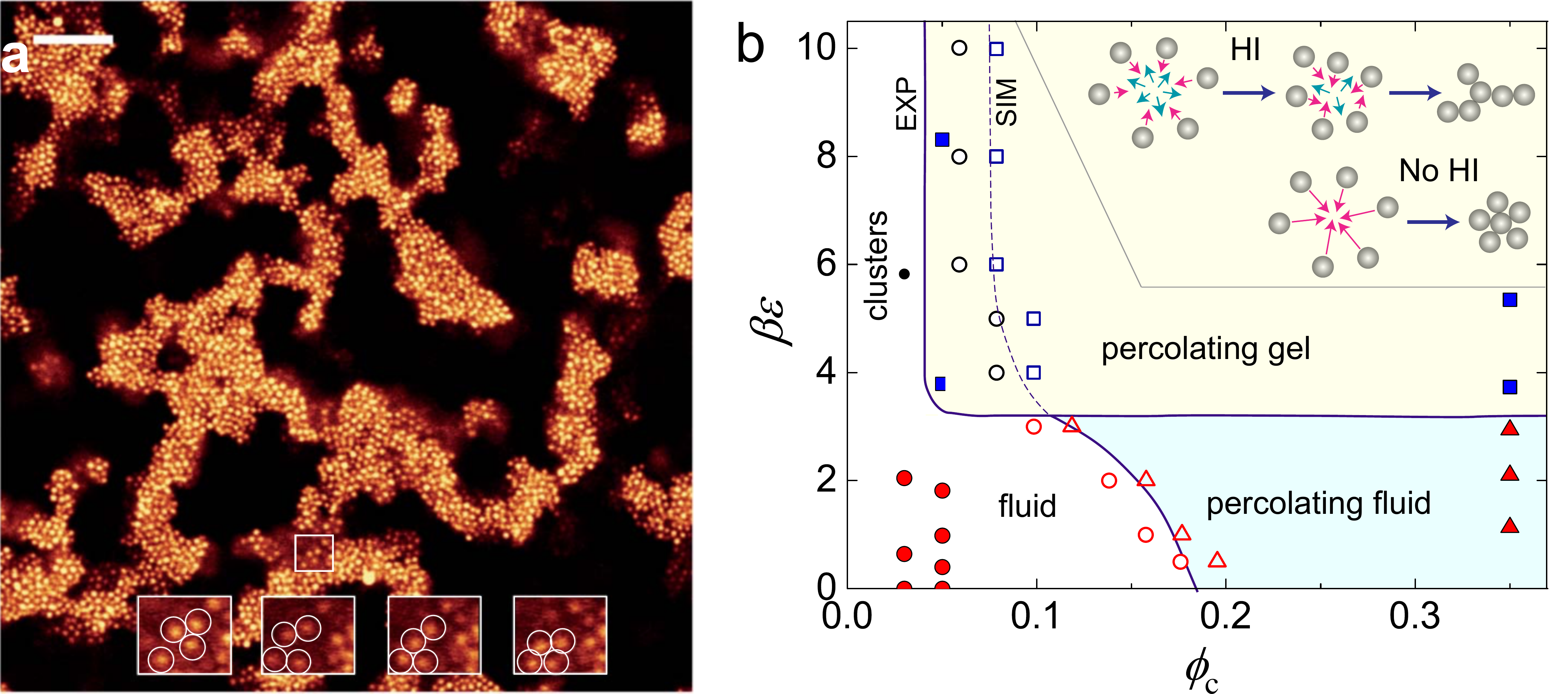} 
\caption{
(a) Localised particle-hopping on the surface of an arrested gel network. Magnified snapshots of an area (small box) are shown in the four larger boxes as a time-sequence. The particles undergo a local re-arrangement over $\sim 500$ s, as indicated by the outlined particles. The snapshots are taken at $t_w=320,410,420$ and $460$ s after imaging began, which correspond to $460, 590, 600$ and $660\tau_B$ respectively. Here the polymer concentration  with respect to that required for gelation $c_p/c_p^\mathrm{gel}=3.90,$ polymer--colloid size ratio $q=0.49$, and volume fraction $\phi=0.2$. Scale bar in top left-hand corner indicates 10 $\mu$m \cite{zhang2013}.
(b) State diagram showing gelation in an experimental and simulated system in the $\beta \epsilon$ (potential well depth)-$\phi$ (colloid volume fraction) plane with $q=0.18$. Phase separation is observed for $\beta \varepsilon \gtrsim 3$. Filled symbols are experimental data, Brownian dynamics simulation data are unfilled symbols. Red symbols are one-phase fluids, circles lie below the percolation threshold, triangles above. Black circles are (non-percolating) cluster fluids. Blue squares are gels. 
Inset shows effects of hydrodynamics on colloidal aggregation. 
Top line: hydrodynamic interactions lead to solvent flow (cyan arrows) which influences the motion of the aggregating colloids. The incompressibility of the fluid allows only transverse (rotational) flow, resulting in the formation of an elongated structure rather than a closed one. 
Bottom line: no hydrodynamic interactions. Their attractive interactions lead the colloids to aggregate (pink arrows) to form a compact structure. Adapted from \cite{royall2015prl}.
\label{figHydrodynamicsIsla}
}
\end{figure}

Gels, with their inhomogeneous nature present many interfaces between the ``arms'' of the gel and voids in between the arms and indeed here the particles have fewer neighbours and can move more easily \cite{gao2007,dibble2008,royall2008naturemater,ohtsuka2008,prasad2003}. Thus we expect that the most mobile particles may be located at these interfaces, and indeed the first study of dynamic heterogeneity in gels, using computer simulation, found precisely that \cite{puertas2004}. Particle--resolved studies of dynamic heterogeneity in gels began with Gao and Kilfoil who found distinct populations of slow and fast particles \cite{gao2007}. Later work confirmed this \cite{royall2008naturemater,zhang2013,vandoorn2018,verweij2019} as shown in Fig. \ref{figHydrodynamicsIsla}(a). It is important to note that in this sense, dynamic heterogeneity in gels is profoundly different to that in the more homogenous supercooled liquids and glasses where there are no interfaces. Furthermore gels are far--from--equilibrium and the ageing is much more significant: supercooled liquids are typically metastable to crystallisation, but their properties are usually stable over many relaxation times, unlike the case of colloidal gels formed via spinodal decomposition. It is possible that, deep inside the ``arms'' of a gel, there may be dynamic heterogeneity of a similar nature to that in supercooled liquids \cite{hurley1995}, but to our knowledge, no such investigation has been made.

Curiously, in the density--matched PMMA system (section \ref{sectionEarly}), Solomon and coworkers found a peculiar kind of dynamic heterogeneity \cite{dibble2008}.  The observation was that the mobility depended on the number of neighbours that a particle had in a weakly quenched gel (with a small amount of polymer) and not in the case that more polymer was added such that the gel could be regarded as more deeply quenched. In the former case, the gel had a cluster--like structure while the latter case the structure was described as ``stringlike''. It is worth noting that this was inferred from 2D measurement of the dynamics, and revisiting this phenomenon with 3D tracking and/or computer simulation (subject to the discussion above in section \ref{sectionEarly} and ref. \cite{royall2018mermaid}) would be interesting. Some work has been carried out in this direction: in a study with a solvent of dielectric constant $\sim 2$ so that electrostatics were very weak, Ohtsuka \emph{et al.} \cite{ohtsuka2008} found that close to gelation, the colloidal fluid exhibits transient but rather long--lived clusters where again mobility depended on the number of neighbours.

\begin{figure}[tb]
\centering
\includegraphics[width=120 mm]{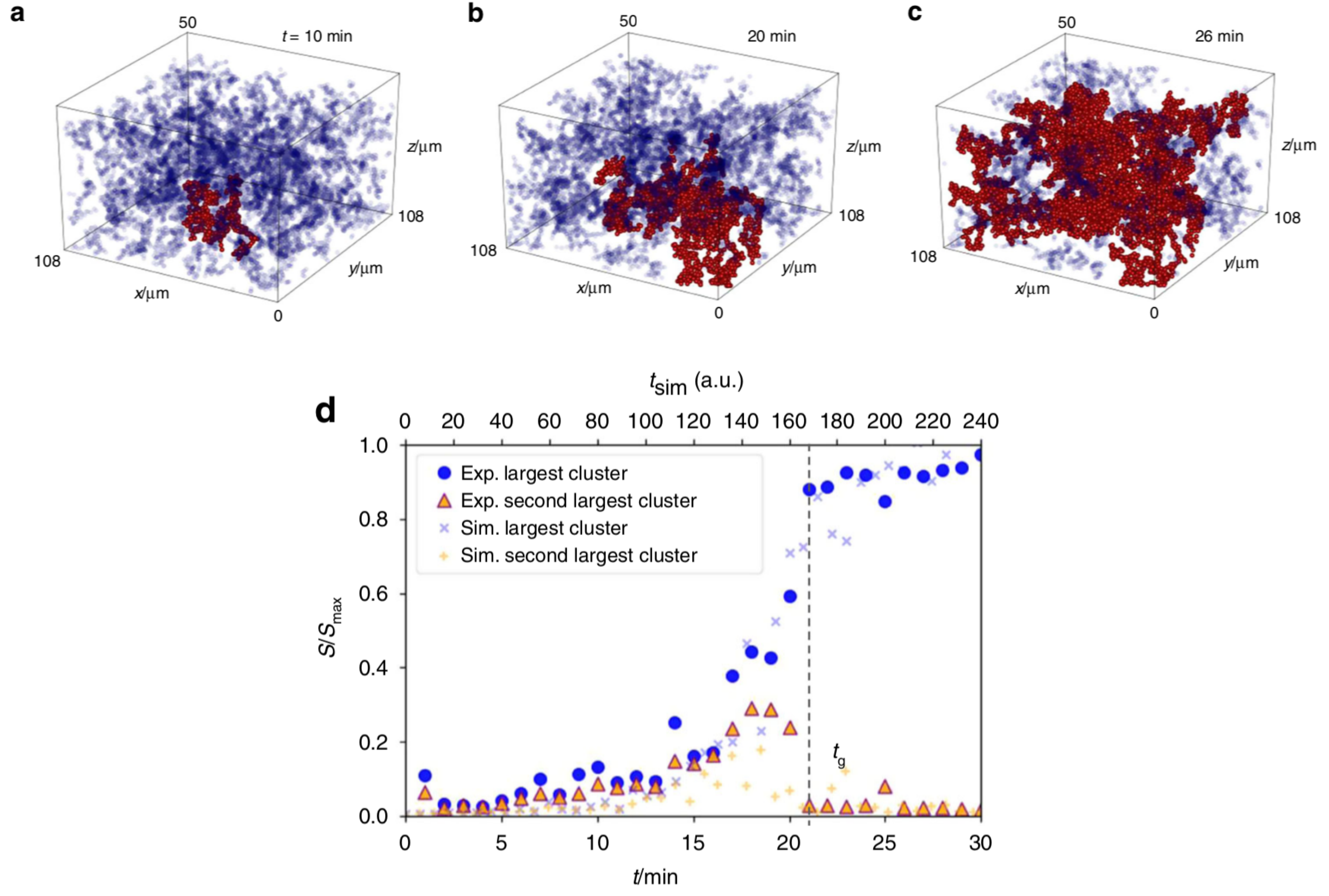} 
\caption{
Experimental and computer simulation observation of aggregating colloidal particles interacting with critical Casimir forces at the times noted. 
(a--c) Largest connected cluster is marked in red. 
(d) Size of the largest, and second-largest cluster normalized by the maximum cluster size as a function of time. Reproduced from \cite{rouwhorst2020ncomms}.
\label{figPeterSchall} }
\end{figure}

\subsection{Evolution: Aggregation, ageing and Crystallisation}
\label{sectionAgeing}

Ageing of quiescent (unperturbed) colloidal gels has been studied on two timescales. Short timescales ($\sim\tau_B$), \emph{prior to arrest} under which the gel assembles and follows spinodal decomposition and long timescales ($\gg\tau_B$) of post--arrest coarsening. A major effect on the former, short timescale case, is the role of hydrodynamic interactions induced by the solvent. This effect was first emphasised by Furukawa and Tanaka using computer simulation with the hydrodynamics included \cite{furukawa2010}. It was later shown that including hydrodynamics in simulation gives a much more accurate reproduction of the experimental formation of gels than does Brownian dynamics simulations, which do not consider the effect of hydrodynamic interactions between particles \cite{royall2015prl}. The principle effect appears to be that the outflow of solvent in condensing clusters of colloids in the very initial stage when density inhomogeneities emerge leads to much ``stringier'' clusters of colloids than those found in Brownian dynamics simulations, though nevertheless the particles in the experiments have been found to be isostatic \cite{tsurusawa2019}. As shown in Fig \ref{figHydrodynamicsIsla}(b), these then percolate at a volume fraction \emph{a factor of two} less than is the case for Brownian dynamics simulations under the same conditions. It is important to note, however, that any link between volume fraction and percolation must have a timescale associated with it. This is because the clusters formed by aggregating colloids have a fractal dimension less than three (typically 1.8--2) so that in the limit of long timescales of aggregation, the volume fraction required for percolation vanishes. This was first emphasised by Weitz and coworkers \cite{manley2004} and later demonstrated with computer simulation \cite{griffiths2017}.  Very recently, novel particle--resolved experiments where the cluster aggregation could be initiated \emph{in--situ} have been carried out. The cluster size size as a function of time can be followed directly, as shown in Fig. \ref{figPeterSchall}. This has been made possible by the use of critical Casimir forces \cite{rouwhorst2020ncomms,rouwhorst2020pre}.

\begin{figure}[tb]
\centering
\includegraphics[width=110 mm]{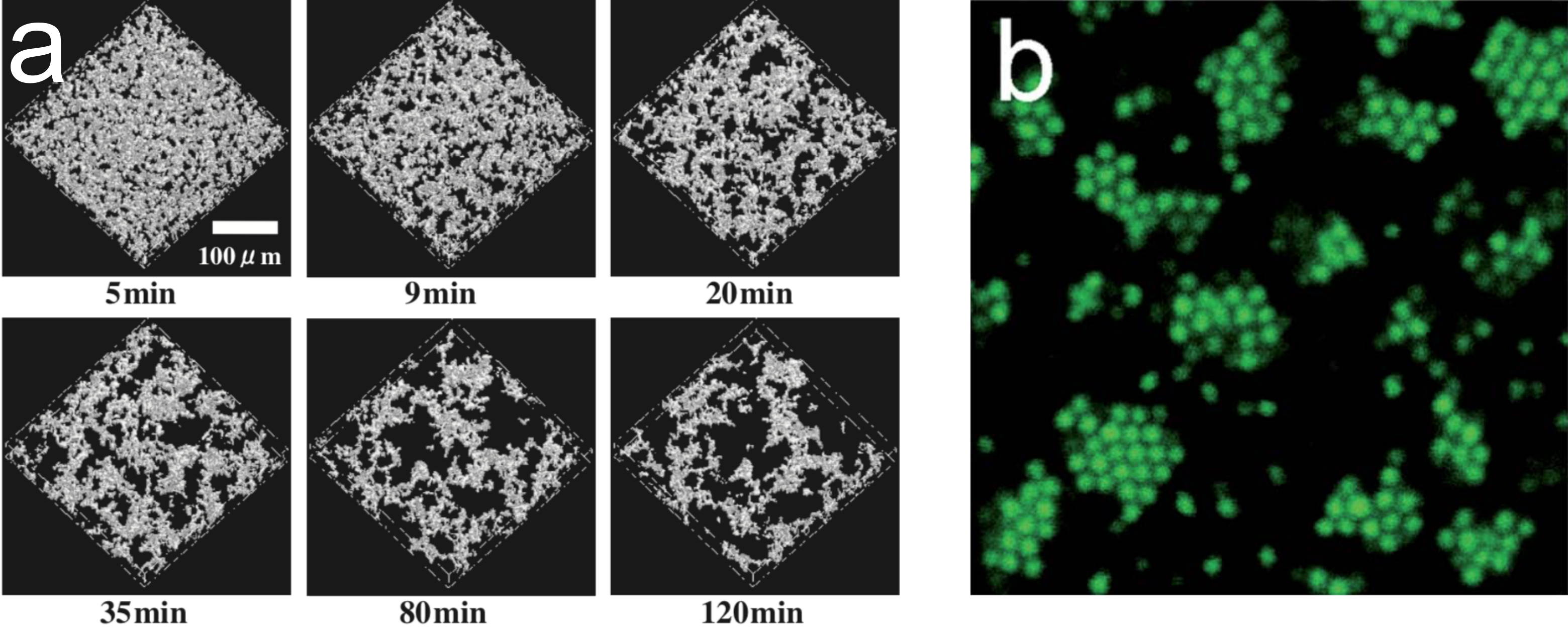} 
\caption{
(a) Temporal change in 3D phase-separation patterns (240 $\mu$m $\times$ 240 $\mu$m $\times$  80 $\mu$m) during phase separation. The system was 50 nm diameter polystyrene particles in water \cite{tanaka2005}.
(b) Crystallisation in the ``arms'' of a colloidal gel with $u_{AO}\approx 5.5k_BT$ and $\phi_c=0.12$.
\cite{zhang2012}.
\label{figNishikawaWillem}
}
\end{figure}

At longer timescales, where dynamical arrest of the colloid--rich phase occurs, ageing, the irreversible change of the properties of the gel is a significant feature. This is essentially due to the fact that the colloid dynamics is not totally arrested, it is rather \emph{slowed}. Moreover, relative to the homogeneous supercooled liquid the interfaces in the gels mean that there is another mechanism of relaxation. Thus on long timescales the gel network will exhibit the coarsening dictated by the spinodal decomposition (section \ref{sectionSpinodal}). Such a scenario is shown in Fig. \ref{figNishikawaWillem}(a), an early 3D analysis of a very dilute (volume fraction $\sim0.01$) gel of 50 nm polystyrene particles formed through van der Waals attractions. It is possible that the strong van der Waals interactions in this system acted to suppress creaming (sedimentation where gravity ``points in an unusual direction'') due to the polystyrene particles being lighter than the water solvent \cite{tanaka2005}.

Another means to investigate the coarsening dynamics is to controllably accelerate the time--evolution of the system. Such an approach was used by Isla Zhang \emph{et al.} \cite{zhang2013} who reduced the volume fraction of the colloid--rich phase such that the dynamics accelerated [recall Fig. \ref{figJamesPaddyAngell}(a)]. Zhang \emph{et al.} were thus able to observe coarsening prior to sedimentation even in non--density matched depletion gels. This reduction in volume fraction was achieved in two ways. Firstly, a larger polymer--colloid size ratio $q$ means that, for a given degree of effective ``cooling'' with respect to the critical point, by adding more polymer, corresponds to a colloidal liquid of lower volume fraction \cite{lekkerkerker1992}. That is to say, if one adds (say) 10\% more polymer than that corresponding to criticality, the volume fraction of the colloid--rich phase is lower in the case of larger size ratio $q$. Secondly, if one reduces the concentration of polymer (approaching criticality), the volume fraction of the colloidal liquid decreases and, in principle, even very short--ranged effective attractions can result in colloidal liquids without arrest \cite{royall2018molphys}.

\begin{figure}[tb]
\centering
\includegraphics[width=110 mm]{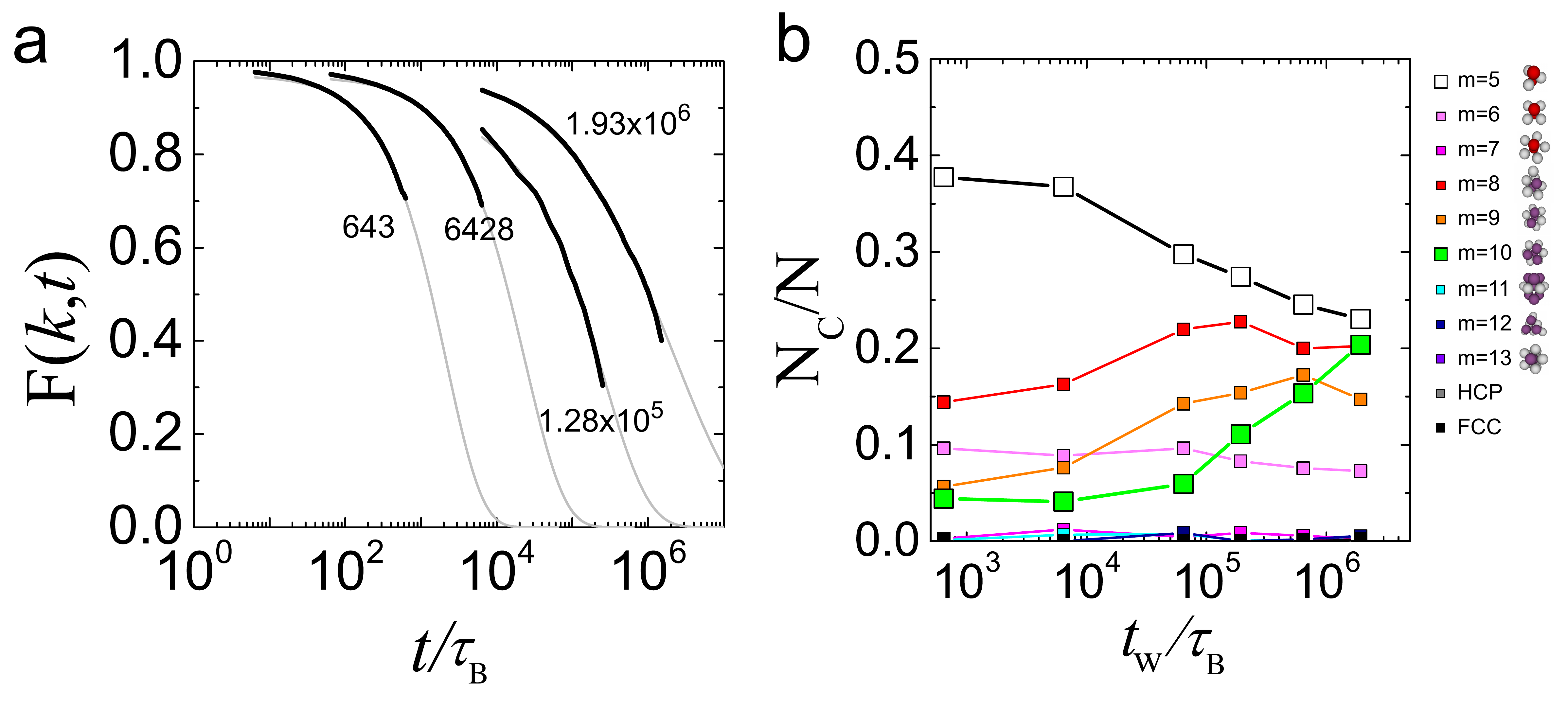} 
\caption{
Stiffening (increase of relaxation time) and change in local structure in aging gels.
(a) Intermediate scattering functions from simulation data for a square well with range $0.03\sigma$, volume fraction $\phi_{\rm e}=0.35$ and $c_{\rm p}/c_{\rm p}^{\rm gel}=1.864$. Different waiting times are expressed in units of the Brownian time $\tau_B$. Here the results from simulation (black lines) are fitted with a stretched exponential (grey lines). 
(b) Topological cluster classification analysis (see section \ref{sectionMechanism} \cite{malins2013tcc}) of the structural evolution as a function of waiting time $t_w$. TCC structures are illustrated in the legend right 
\cite{royall2018jcp}.
\label{figAgingMisplaced}}
\end{figure}

The local structural analysis underlying the process by which gels become rigid (section \ref{sectionMechanism} and e.g. Fig. \ref{figFivesTCC}) can also be used to probe the change in local structure during ageing. It is seen here in Fig. \ref{figAgingMisplaced} that for the system studied (with $8$\% polydispersity to suppress crystallization), aging is consistent with an increase in local fivefold symmetry, via in the increase in population of the 10--membered defective icosahedron. This structure also happens to be the locally favoured structure in hard spheres upon supercooling \cite{royall2015jnonxtalsol,hallett2018,hallett2020}. In Fig.  \ref{figAgingMisplaced}(a) we also see strong evidence for stiffening in the form of the increase in the structural relaxation time.

A further key aspect of ageing is crystallisation. For suitably monodisperse systems, gels are metastable to colloidal fluid--crystal phase coexistence \cite{lekkerkerker1992}, as shown in the early work on particle resolved studies of crystallisation in colloid--polymer mixtures by Else de Hoog \emph{et al.} \cite{dehoog2001}. Simulations found that a slow increase in attraction strength gave a higher degree of crystallinity in monodisperse systems interacting via the Morse potential (Eq. \ref{eqMorse}). In a system with just 4\% polydispersity system the effect was similar, although the degree of crystallisation was rather reduced  \cite{royall2012}. Even highly polydisperse systems have been predicted to fractionate and crystallise for hard spheres \cite{sollich2010} and there seems little reason to suppose that the same mechanism would not apply here. Around criticality, fluctuations have been shown to accelerate crystallisation \cite{tenwolde1997} and this has been reproduced in real space analysis of colloid--polymer mixtures \cite{savage2009,taylor2012}. Alternatively, crystallisation can occur in the ``arms'' of the gels, as shown in Fig. \ref{figNishikawaWillem}(b) \cite{zhang2012}. Interestingly it has been shown that this is an unexpected example of the Bergeron process which underlies atmospheric ice crystallisation, where particles ``evaporate'' from the amorphous arms of the gel and then condense onto a crystalline region \cite{tsurusawa2017}.

\subsection{Failure: Collapse Under Gravity}
\label{sectionCollapse}

\begin{figure}[tb]
\centering
\includegraphics[width=110 mm]{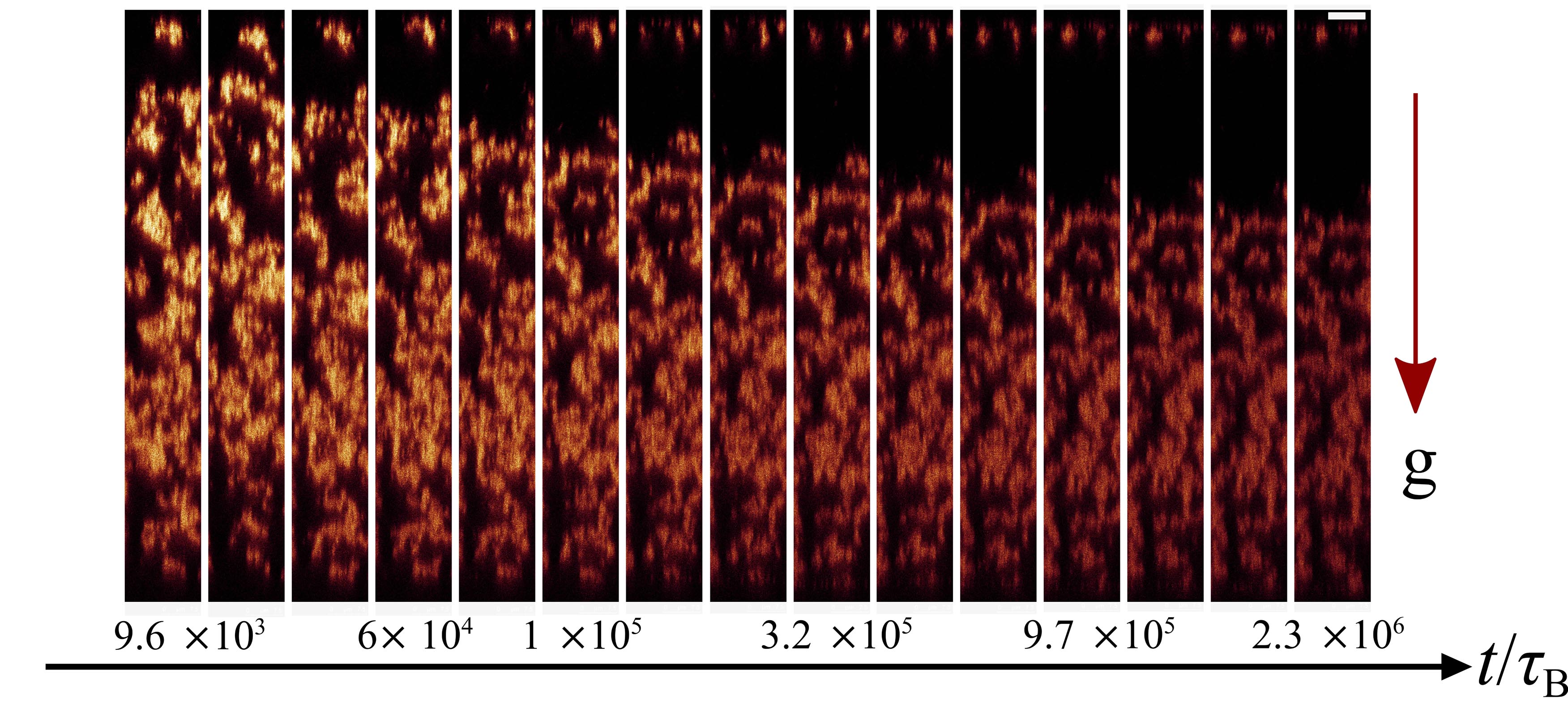} 
\caption{
Time-sequence of sedimenting gels captured from a PMMA system in \emph{cis}--decalin with a contact potential of around 7.0 $k_\mathrm{B}T$. The scale bar in corresponds to 7.5 $\mu$m, the system is around 100 $\mu$m in height.}
\label{figSedimentProfileAzaima}
\end{figure}

A particularly intriguing and challenging problem in colloidal gels is collapse under gravity \cite{buscall1987,poon1999,buscall2009,teece2014}. While we have noted above in section \ref{sectionEarly} that considerable efforts have been made to density match colloids and solvent to mitigate the effects of gravity, firstly, perfect density matching does not exist, secondly and more importantly, colloidal gelation is exploited to suppress sedimentation in a wide range of products such as coatings, crop protection suspension formulations, pharmaceutical suspension formulations, various cosmetic formulations, pigment printing inks, dispersions for 3D printing, ceramic preparations, food preparations, detergent formulations and home care products amongst many others for example. In equilibrium, or at least, in well--aged systems, many intriguing phenomena of hard and sticky spheres have been explored in the elegant work of Piazza and coworkers, who have largely focussed on rather smaller colloids than are amenable to real space imaging. From this and related work, at least equilibrium sedimentation is quite well understood \cite{piazza2012,piazza2014}.

Collapse falls into two categories. \emph{Rapid collapse,} in which the system starts to collapse as soon as it is prepared \cite{razali2017}  and \emph{delayed collapse} in which little happens to the height of the top of the gel for some considerable time until, rather suddenly, it fails catastrophically and collapses in a timescale much shorter than the waiting time \cite{starrs2002,poon1999}. Intriguingly, rheological work does suggest a coupling between collapse times, interaction strength and mechanical properties \cite{kamp2009}. Overall, however, gel collapse is a poorly understood phenomenon, combining as it does highly non-equilibrium behaviour and two very distinct time-evolutions. It has been observed that often, larger systems (say $>1$ cm in sample height) tend to exhibit delayed collapse while smaller systems can undergo rapid collapse. Some of us looked at the behaviour of a small system undergoing rapid collapse, see Fig. \ref{figSedimentProfileAzaima} \cite{razali2017}. Notably, these small, rapidly collapsing systems can be captured with particle-level computer simulation. It is worth noting that in the small system sizes as studied by Razali \emph{et al.} \cite{razali2017}, even a drop of a few microns is readliy measured, which would likely go un--noticed in larger systems. Moreover, in these small systems, the gravitational length of the colloids may be comparable to the system size. In this way, a hard sphere system would not sediment, but, as we see in Fig. \ref{figSedimentProfileAzaima}, a gelling system does. This shows that, the interplay between gelation and sedimentation is \emph{qualitatively} different between large and small systems. For now, the longer timescales and larger system sizes pertaining to delayed collapse remain beyond the grasp of direct particle-level simulation.

\begin{figure}[tb]
\centering
\includegraphics[width=100 mm]{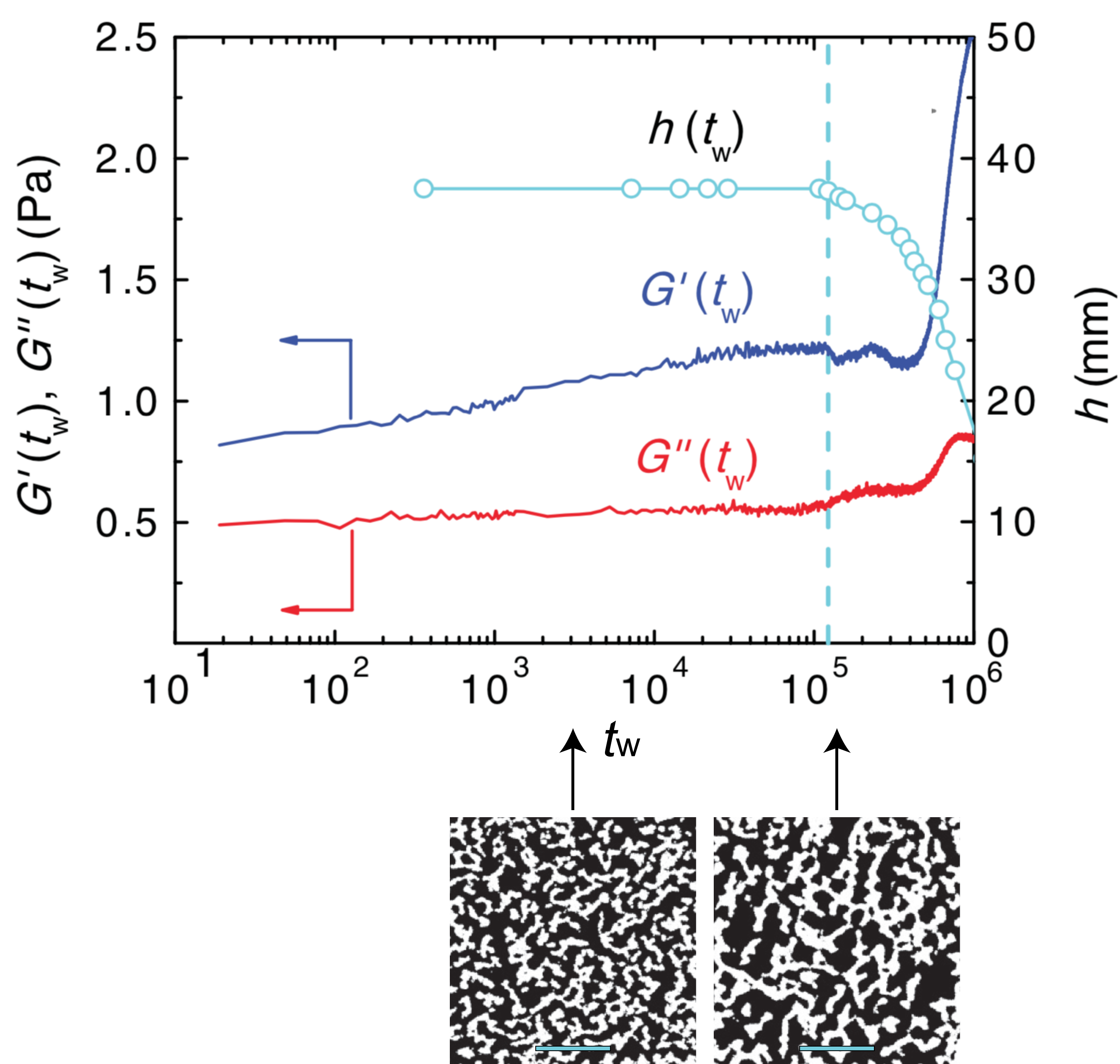} 
\caption{Combining techniques to investigate delayed collapse. Direct observation reveals the macroscopic height of the sample $h(t_w)$. 
Storage and loss moduli determined from rheology show a change in the mechanical properties, particularly the storage modulus $G'(t_w)$. This indicates an increase in stiffness of the material over time. Binarised confocal images (bars$=40$ $\mu m$) show the time--evolution of the structure.
\cite{bartlett2012}
}
\label{figPaulLisaMalcolm}
\end{figure}

An example of the combination of real space analysis and rheology and direct observation is shown in Fig. \ref{figPaulLisaMalcolm}. Here we see that little happens to the height of the gel $h(t_w)$ but the storage and loss shear moduli both show an increase as a function of waiting time. In other words, the gel becomes \emph{stronger} as it ages prior to catastrophic collapse, here after around $10^5$s storage and loss shear moduli both show an increase as a function of waiting time \cite{bartlett2012}. 
Here real-space analysis reveals coarsening of the gel prior to collapse, concurrent with a significant increase in the storage modulus, indicating that the gel is becoming stiffer.

\subsection{Bridging the gap between model systems and real world products}
\label{sectionBridging}

In the real world, the particles comprising colloidal gels are often no longer monodisperse or even spherical, sample sizes can be significantly larger (e.g. 0.1 to 10 litres). Gels have greater strength, stability timescales are much longer, secondary network ageing effects can occur such as Ostwald ripening (droplet/crystal growth) of colloidal emulsion droplets or suspensions of crystalline organic compounds such as active ingredients, many products are water based and importantly nearly all are opaque. Polydisperse silicone oil emulsions index matched in water/glycol/glycerol mixtures gelled with non-adsorbing polymers (hydroxyethyl cellulose, xanthan) which are more representative of real products have been developed and studies exploring how ageing of the network structure influences the rheology and gravitational stability in larger sample sizes \cite{teece2011,bartlett2012,zhang2013,faers2003}.

Interestingly such systems can reproduce features of both academic and industrial systems allowing academic research to be applied to real-world challenges with a key finding coming from \emph{in-situ} vane rheology measurements in relatively large samples (height $\approx 4$cm) that the gel strength increases during ageing until collapse occurs, rather than collapse occurring because the network is becoming weaker, and therefore that the gravitational stress within the sample must also be increasing until it exceeds the strength of the gel (see Fig. \ref{figPaulLisaMalcolm}) \cite{teece2011,bartlett2012}. Such systems are invaluable in exploring the larger scale ageing and gravitational stability of colloidal gels since they can be readily prepared in large volumes (5-10 litres) and help to answer the pressing challenges of colloidal gelation. Another interesting feature of polydisperse systems is that they do not form colloidal crystals on experimental timescales, allowing ageing to be studied in shallow quenches closer to the spinodal line.


\subsection{Deformation}
\label{sectionDeformation}

In many real world applications of gels, they undergo deformation as they are squeezed or scraped out of their containers. Therefore it is pertinent to ask how these materials change when they are sheared or when they flow.  While solids respond elastically and liquids flow in response to deformation, complex fluids such as gels can show a combination of both responses: They respond elastically below a critical strain or stress, while beyond a characteristic yield strain or stress they flow \cite{bonn2017,fielding2014}. Typically the response to shear is measured using rheometry, which yields the elastic (storage) $G'$ and viscous (loss) $G''$ moduli. The transition between liquidlike and solidlike response, such as by progressively increasing polymer concentration to increase the attraction strength between colloidal particles, is evidenced by a crossover between \textit{G'} and $G''$ at lower frequency, while for a gel $G'$ is greater than $G''$ in the elastic regime. One measure of yielding is, at large strains ($ \gamma \geq 10\% $), $G'<G''$. For a gel however, the decay of \textit{G'} and $G''$ with increasing strain provides evidence of residual structure until $ \gamma > 100\% $ \cite{laurati2009,laurati2011}. Indeed, Hsiao \textit{et al.} \cite{hsiao2012} note that for a comparable system of non-interacting particles, \textit{G’} is too small to measure for ($ \gamma \geq 15\% $), yet for a seemingly fluidised gel sample, there is still some contribution to rigidity for strains of at least 60\%. Through simultaneous confocal imaging during shear, they attribute this to hydrodynamic coupling between a sub-population of stress-bearing, rigid particle clusters which imparts some rigidity to the gel - despite breakdown of the quiescent percolating gel network.

When studying 
the flow or shear of colloidal gels it is useful to consider the P\'{e}clet number, $\mathrm{Pe}=\dot{\gamma} \tau_{\mathrm{B}}$, where $\dot{\gamma}$ is the shear rate and $\tau_{\mathrm{B}}$ is the Brownian time. For a gel, large P\'{e}clet numbers correspond to a simple viscous response (stress increasing linearly with shear rate), as the gel structure is rapidly broken up, while small P\'{e}clet numbers result in a soft-solid response (displaying a yield stress plateau).

While the rheological response of gels provides compelling evidence of their time-dependent structure, it is useful to \textit{simultaneously} image gels under shear to directly relate mechanical information to changes in gel configuration. To perform real-space imaging and shear measurements simultaneously, one of two approaches are typically used: A confocal optical setup is used with a (modified) conventional rheometer \cite{koumakis2015}, or a custom shear stage is used that can be mounted directly to a commercial confocal microscope \cite{tolpekin2004,smith2007,lin2014}. Similarly, measurements under applied flow typically use a custom flow cell with a standard microscope \cite{isa2007,pandey2014}.

\subsubsection{Shear}
\begin{figure}
\centering
\includegraphics[width=80mm]{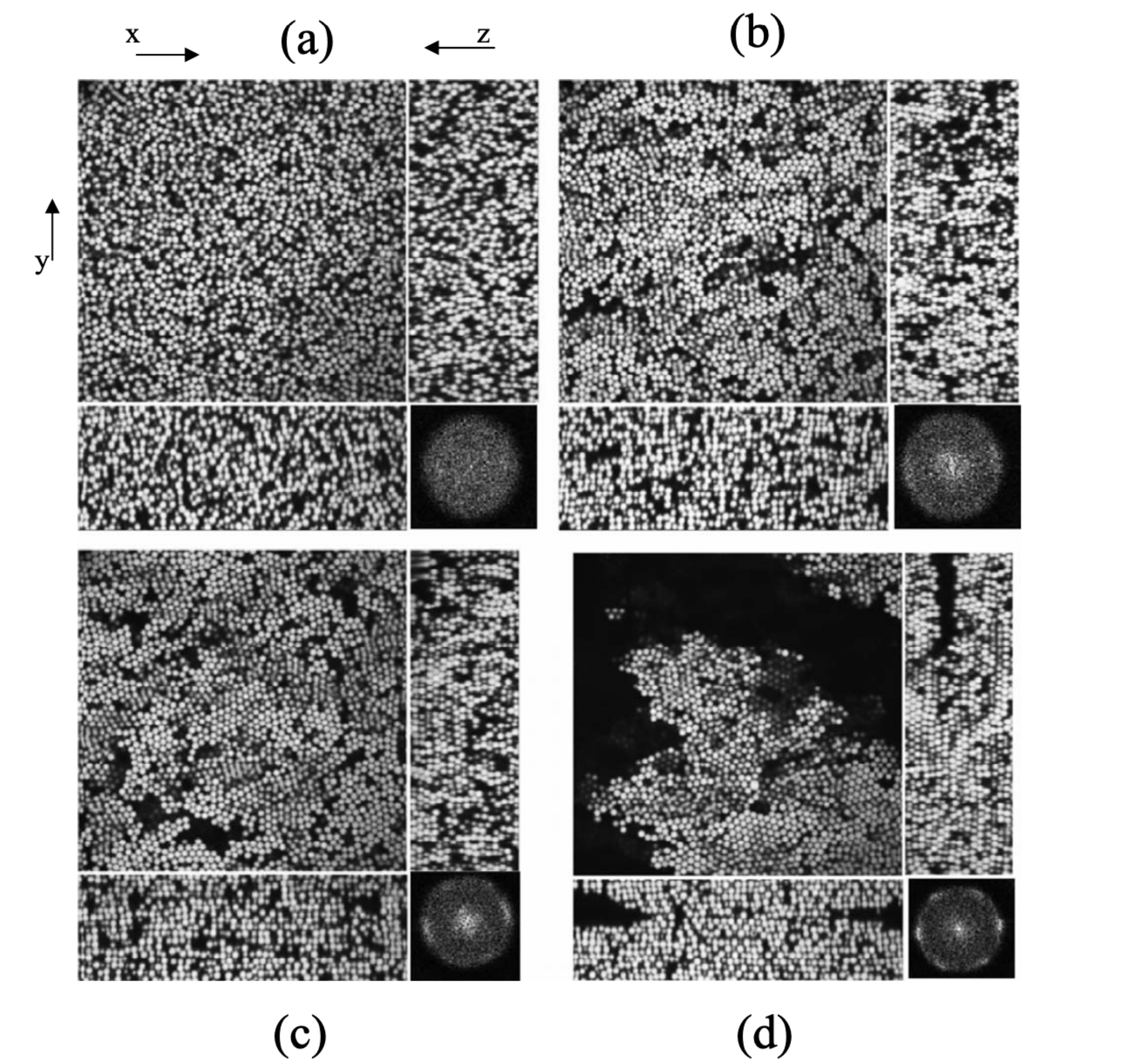}
\caption{\label{figSmith}Confocal microscopy images of a colloid-polymer gel ($\phi = 0.4$) after 31 minutes of oscillatory shear at a frequency of 70 Hz and a strain amplitude of 0.15. (a) Shows the gel in its quiescent state. (b-d) Shows typical, highly ordered and void-rich regions respectively. All regions are shown in xy, xz and yz slices and a representative fast Fourier transform (FFT) is also shown. Reproduced from \cite{smith2007}.}
\end{figure}

As with a polymer gel, shear can act to coarsen or melt the gel structure. Further (detectable) complexity is also accessible in the formation of locally favourable and/or crystalline structures. Smith \textit{et al.} \cite{smith2007} used a custom parallel plate shear cell where both plates were moved in opposite directions, accessing a stationary plane in the centre of the sample (as opposed to a stage with a single mobile plate, for which the stationary plane is the fixed plate) to shear a very stiff ($u_{\mathrm{AO}} / k_{B}T= -46 \pm 16  $) colloid-polymer mixture (PMMA-polystyrene in a density and refractive index matching solvent mixture, section \ref{sectionEarly}). They found that gel yielding and crystallisation occur concurrently, and that the strain required increased with decreasing shear frequency, while for high strain values the crystallites were melted. Figure \ref{figSmith} shows the rich variety of shear-induced structure for a typical sample, including small crystalline, large crystalline domains and large voids.

\begin{figure}
\centering
\includegraphics[width=90mm]{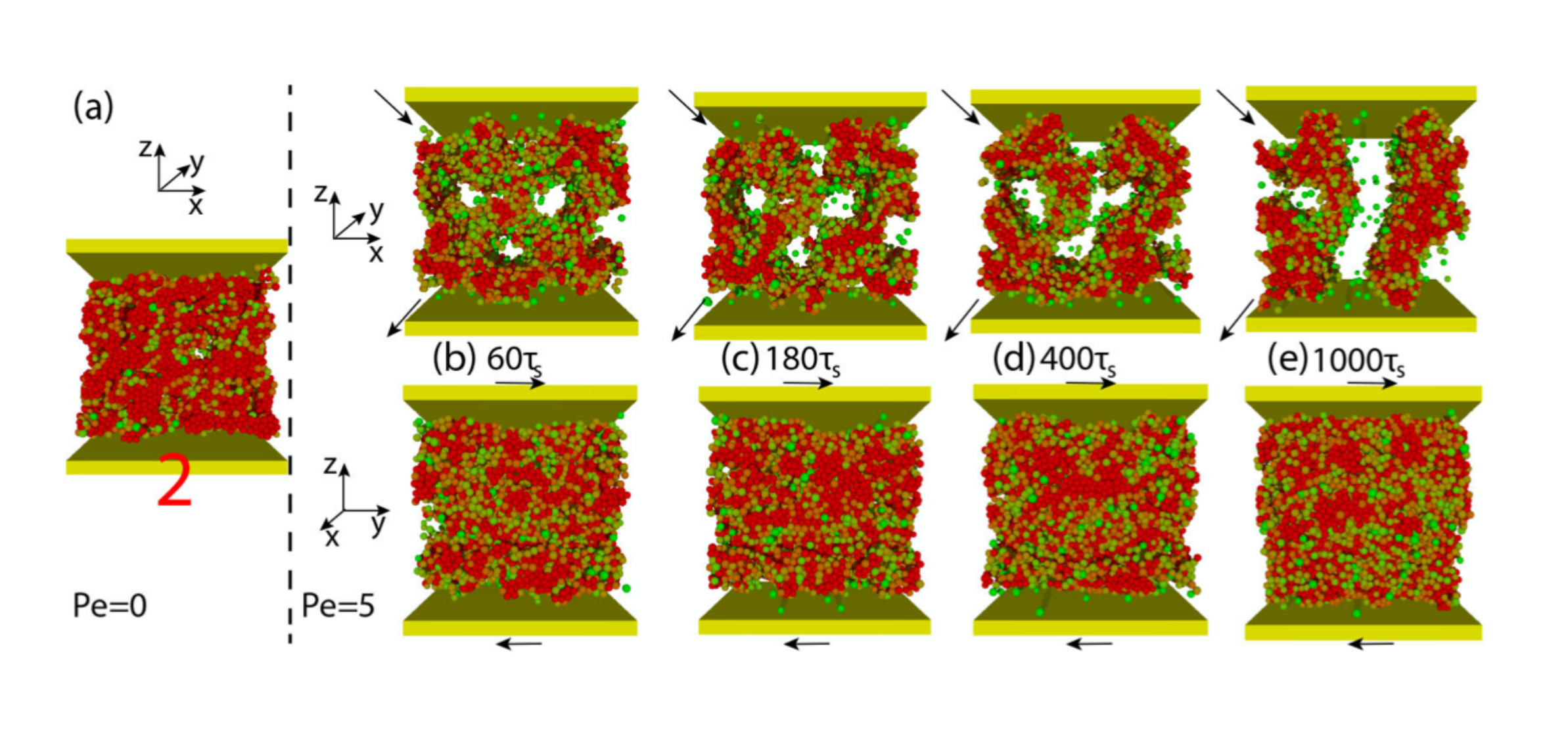}
\caption{\label{figKohl} Brownian dynamics snapshots of gel ($U_{\mathrm{AO}}$ -4.4 $k_{B}T) $ over many shear cycles (indicated b-e) compared to quiescent case (a). While the yz projection shows little change in structure or bond number (where red indicates 5 or more bonds, green 1 or fewer), the $xz$ plane (perpendicular to the shear direction) shows the emergence of pronounced gel-slabs held between the upper and lower plates, separated by voids. Reproduced from \cite{kohl2017}.}
\end{figure}

Koumakis \textit{et al.} \cite{koumakis2015} also applied shear to a colloid polymer gel of similar composition to Smith \textit{et al.} \cite{smith2007} (albeit with slightly less strong attractions of -16 $ k_{B}T$ and -23.2 $ k_{B}T$). Utilising a modified rheometer to apply shear to the gel, imaged \textit{in situ} with a confocal insert, it was possible to study the interplay between microstructure and mechanical properties. Gel microstructure was characterised through the void distribution and the average number of interparticle bonds. For high shear rates complete cluster melting was observed, while for smaller rates cluster compactification occurred, producing significant structural inhomogeneity. By studying the time evolution of gels prepared with high and low pre-shear, it was shown that strong pre-shear resulted in homogeneous, stronger solids (analogous to an instantaneous thermal quench), while weak preshear lead to heterogeneous weaker gels (analogous to low rate thermal quenching, resulting in the formation of more compact clusters than possible by Brownian relaxation alone). Unusual void formation was also reported by Kohl \textit{et al.} \cite{kohl2017} in Brownian dynamics simulations of shear flow of relatively soft gels ($u_{\mathrm{AO}}$ between  -2.2 $k_{B}T $ and -6.1 $k_{B}T$). Rather than randomly distributed, isotropically oriented domains reported previously, they discovered ``slab-like'' domains in sheared gels that emerged over many shear cycles (Fig. \ref{figKohl}). This behaviour persisted to high shear strengths before eventually being shear melted. This unusual structure was compared to the syneresis phenomena that are observed in a variety of gelling systems.

The ability to encode structural or mechanical properties into a gel through specific shear conditions as a form of ``memory'' was recently explored by Schwen \textit{et al.} \cite{schwen2020} using a custom biaxial shear stage \cite{lin2014} on a confocal microscope. By using many shear cycles at a given strain to ``train'' the gel, strain memories were embedded, such that a reversible structural rearrangement (such that the image difference before and after strain was small) in response to further strains \textit{below} the training strain amplitude was observed: In effect, a new yield strain is set for the gel at the training strain. However for large strains this was not the case -- the applied strain would always result in rearrangements. Remarkably the memory was also encoded orthogonally to the shear direction: that is, shear applied along the $x-z$ shear plane also resulted in a new yield strain along the $x-y$ shear plane. The main structural transformation during this process appeared to be an increase in mean contact number for the trained gel. Indeed, particles most likely to move between shear cycles were those with fewer number of contacts.

\emph{Shear banding}, a commonly observed feature in colloidal glasses and yield stress fluids \cite{bonn2017} has been observed using magnetic resonance imaging for a colloidal gel \cite{moller2008}. This real space technique, with less resolution than microscopy, but not requiring special preparation of the material (refractive index matching) serves as a useful complement to confocal microscopy. It is also possible to study shear banding using the set--up noted above, combining a confocal microscope and a rheometer \cite{fall2010}. Bonn and coworkers demonstrated the role of using rods as a bridging mechanism between emulsion droplets to cause gelation and were able to relate the rheological behaviour of thixotropic and so--called simple yield stress fluids \cite{fall2010}. Shear banding is intimately related to so--called viscosity bifurcating fluids, where for certain stresses the material can catastrophically fail, with a massive drop in its viscosity \cite{fielding2014,bonn2017}. A predominantly rheological study which incorporated some real space imaging by Sprakel \emph{et al.} showed intriguing scaling of the increase in strength of the gel prior to yielding which could be accounted for by a simple model \cite{sprakel2011}.

\subsubsection{Flow}

The behaviour of colloidal gels flowing under narrow confinement is widely applicable to many real-world situations, such as in biological systems or soft robotics. In most shear measurements the shear is typically uniform, whereas a real-world system might combine nonuniform shear and hydrodynamic forces as the fluid experiences different levels of constriction, such as in a dispensing nozzle. Various experiments and simulations have attempted to address these differences.

\begin{figure}
\centering
\includegraphics[width=80mm]{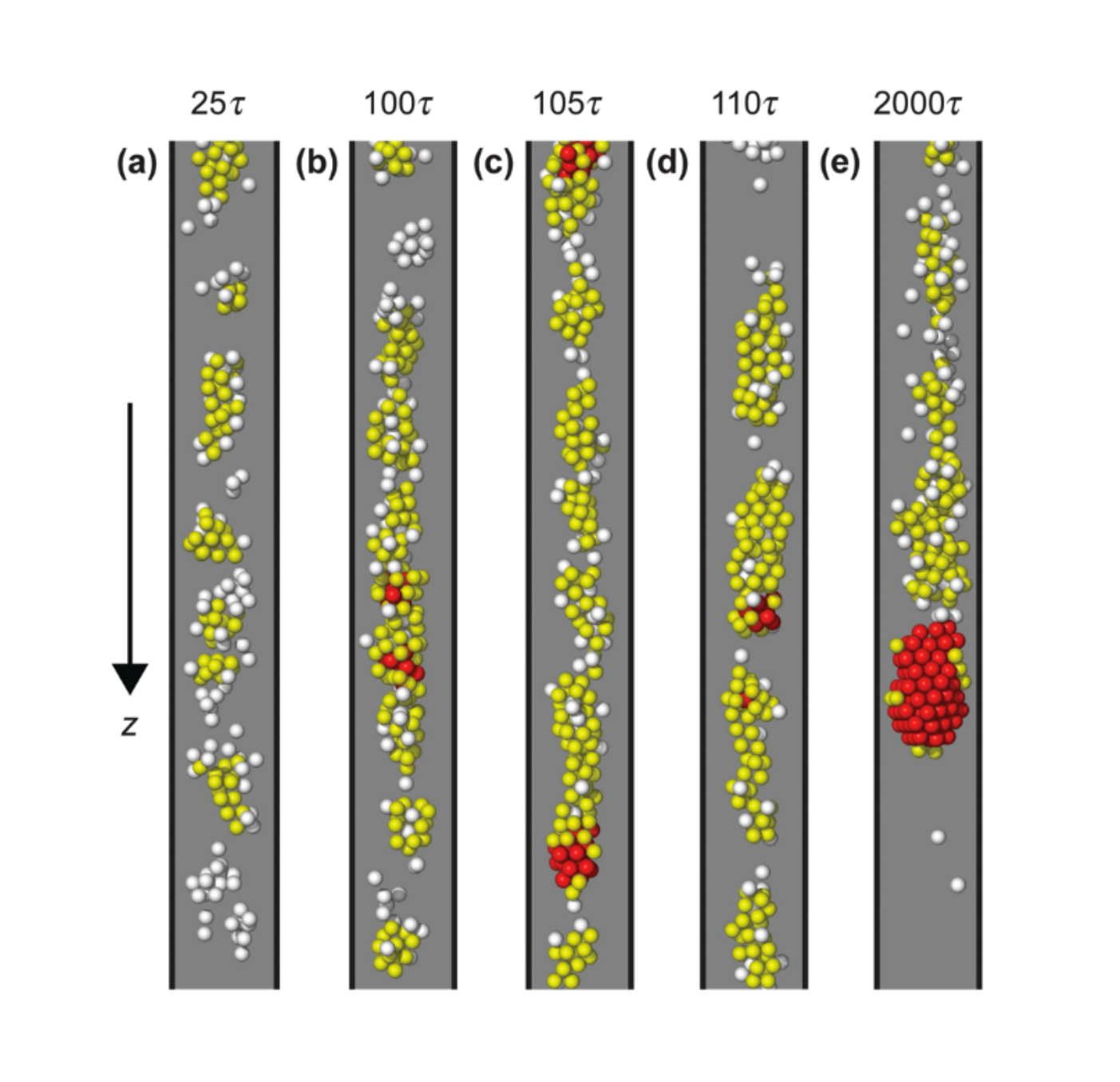}
\caption{\label{figHan} Time dependence of cluster structure in microchannel flow. (a) Shear flow acts to align attractive clusters along channel axis that form connected strands (b). Threads form series of 'packets' (c)  before breaking up into crystalline droplets (d), which become increasingly ordered over many cycles (e). Colour coding indicates colloidal gas (white), liquid (yellow) and crystal (red). Reproduced from \cite{han2019}.}
\end{figure}

Han \textit{et al.} \cite{han2019} used hydrodynamic simulations to explore the behaviour of a colloidal gel subject to microchannel flow. High shear flow acts to disrupt any clustering, while for low shear flow the colloids form gel like clusters and small crystal nuclei. For intermediate shear flows (where the global shear forces approach the same magnitude as the attractive forces between the particles) crystallisation is greatly enhanced. In the regime between crystallisation and melting the non-uniform shear results in crystalline cluster cores localised in the channel centre, but melting at the cluster surfaces. This effect was reduced for wider channels as the flow becomes more uniform. Remarkably they also noted gel-strand formation along the channel axis, that repeatedly break up into increasingly crystalline droplets. This is shown in Fig. \ref{figHan} for average shear forces 1.4 times attractive forces.

Conrad and Lewis used confocal microscopy to study the flow of a silica-polyelectrolyte colloidal gel in a microchannel of width $\sim$ 50 diameters \cite{conrad2008}. High flow rates prevented 3D imaging during measurements (owing to the slow framerate of recording a 3D stack). As with the simulations, they note that the network structure of a quiescent gel is disrupted during shear flow, first into small clusters, then smaller clusters and individual particles with increasing pressure. They also note a transition from pluglike to fluidlike flow beyond a critical shear rate. They also used a similar setup to study the structure of a colloidal gel passing through a channel constriction \cite{conrad2010}, and found evidence of filter pressing as the particle concentration increased in the constriction. They also noted failure of the gel network at cluster boundaries. For small shear pluglike flow was observed, but as for the straight channel, at high shear the flow became fluid-like. They also found that a stagnation zone developed close to the constriction, resulting in transient wall jamming and further disruption of the gel network, leading to fluidization further downstream.


\section{Introducing Complexity: Anisotropy and Mixtures}
\label{sectionComplex}

\begin{figure}
\centering
\includegraphics[width=140mm]{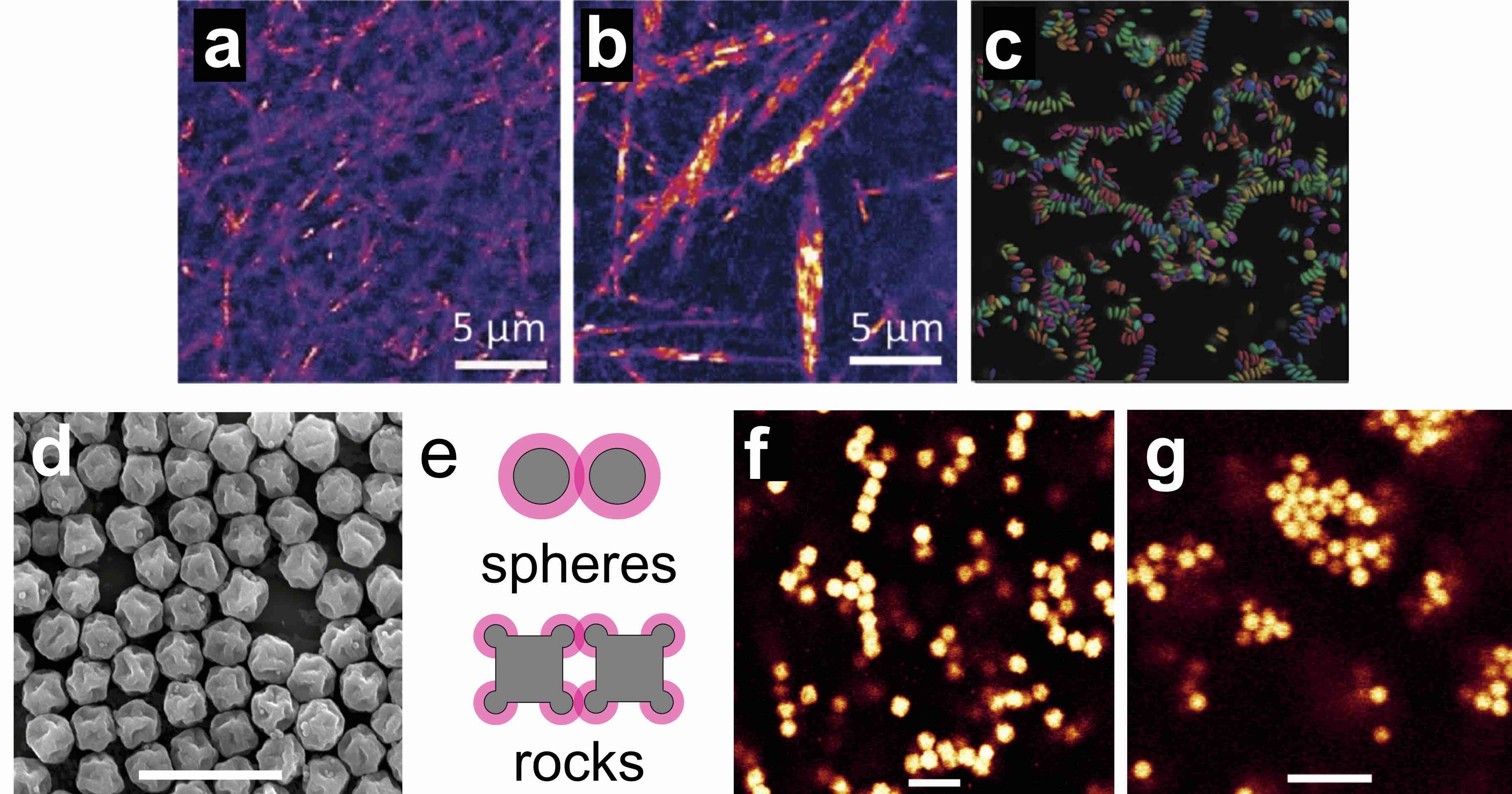}
\caption{
\textbf{Depletion--induced gelation in anisotropic colloids. }
(a, b) Bundling in colloidal rods. 
(a) Confocal image showing a mobile rod network with no added polymer. 
(b) Bundled structures for a rod suspension with $c_p/c_p* = 0.52$ \cite{wilkins2009}.
(c) Chain--like clusters of PMMA discoids \cite{hsiao2015}. 
(d) SEM image of colloidal ``rocks''. 
(e) schematic showing the mechanism of rigid bonding in the rocks contrasted with spheres. 
(f) confocal microscopy image of a gel of rocks $c_{p}/c_{p}^\mathrm{gel}=1.13$.
(g) confocal microscopy image of a gel of spheres $c_{p}/c_{p}^\mathrm{gel}=1.44$
Scale bars in (d,f,g) are 10 $\mu$m. \cite{rice2012}.}
\label{figRoxBundles}
\end{figure}

Gels can be formed from systems other than spherical colloids with an (effective) attraction. Complexity may be introduced via anisotropic particles such as rods and platelets. Depletion gels formed of colloidal rods may form at very low volume fractions for suitably high aspect ratio rods and have been reviewed by Solomon and Spicer, who suggested that the structures assumed by the rods could be interpreted as ranging from fractal clusters to homogeneous rod networks \cite{solomon2010}. An important observation in the case of gels formed of rods is that percolation can occur at very low volume fraction, with implications for materials based on percolation such as photovoltaic cells (which rely on percolation of conducting particles) \cite{schilling2007,kyrylyuk2011}. It is important to note that the equilibrium phase diagrams of rods interacting through depletion interactions are rather more complex than those of spheres, due in no small part to the emergence of a number of liquid crystalline phases \cite{lekkerkerker} and thus one may expect complex mechanisms of gelation. Real space analysis of rods took a step forward with the development of polyamide rods in the group of Solomon, which revealed a novel bundling transition, as shown in Fig. \ref{figRoxBundles}(a,b). The same group managed to form discoids, by compressing PMMA above its glass transition temperature and then redispersing it. The resulting discoids formed chains with their flatter sides oriented towards one another, which would maximise the overlap volume and thus be favoured by the depletion interaction [Fig. \ref{figRoxBundles}(c)] \cite{hsiao2015}.

Another form of anisotropy was explored by Rice \emph{et al.} \cite{rice2012}. As shown in Fig. \ref{figRoxBundles}(d-g), here colloidal ``rocks'' assembled into gels. Notably, the rocks developed highly rigid bonds between the particles, as it was hard for the rocks to roll around one another compared to spheres [Fig \ref{figRoxBundles}(d)] \footnote{The case of very small polymers which lead to stabiliser interdigitation not withstanding, see section \ref{sectionEarly} \cite{dinsmore2006}.} These rigid bonds led to rather linear chains one particle thick, markedly different from the more compact clusters formed of spheres and in fact reminiscent of the ``empty liquids'' mentioned in section \ref{sectionEvidence} \cite{bianchi2006}.

\begin{figure}
\centering
\includegraphics[width=120mm]{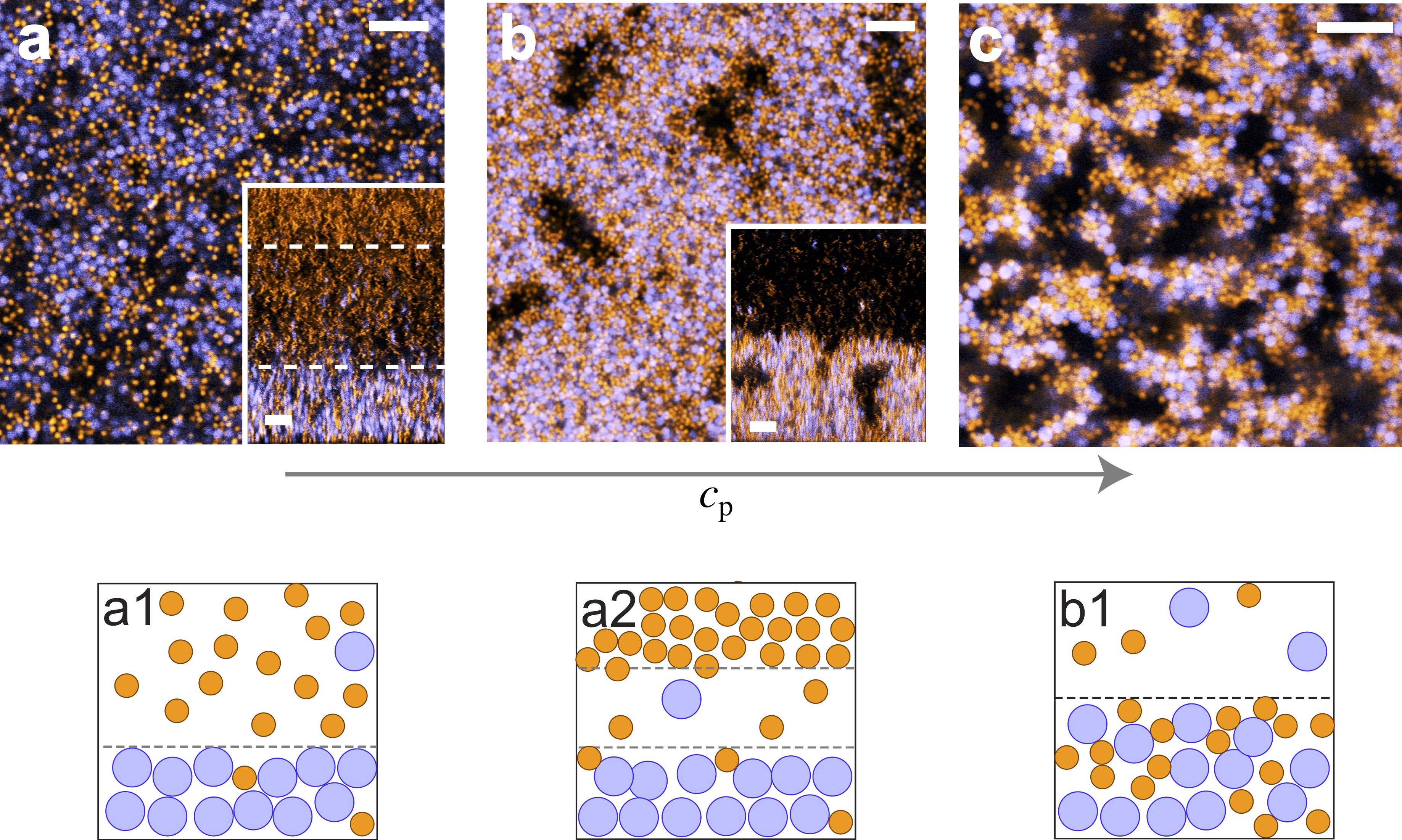}
\caption{Phase behaviour of binary colloid-polymer mixtures.
(a-c) Confocal microscopy images (see text) with larger colloids (blue) and small (orange). Insets show $xz$ profiles which are 100 $\mu$m in height.  
(a) $c_p/c_p^*=0.059$; Possible two- or three-phase demixing as indicated by dashed lines in inset. 
(b) $c_p/c_p^*=0.069$; 
(c) $c_p/c_p^*=0.090$, gelation.  Scale bars denote 10 $\mu$m. 
Possible scenarios by which the experimental data may be interpreted: 
(\emph{a1}) Two-phase coexistence --- the lower (liquid) phase is rich in large particles and the upper (vapor) phase is rich in small particles and polymer;
(\emph{a2}) The colloidal mixture exhibits liquid-liquid demixing. 
(\emph{b1}) Upon deeper quenching more small particles become entrained in the colloid-rich liquid.
Reproduced from \cite{zhang2018}.
}
\label{figCCCP}
\end{figure}

A further means to increase the complexity of the system is to introduce further species. This was done by Isla Zhang and coworkers, where binary colloids and polymer were investigated, as shown in Fig. \ref{figCCCP}  \cite{zhang2018}. The size ratios were approximately 4:2:1 for the large colloids, small colloids and polymer, respectively. A superficial analysis of Eq. \ref{eqAO} might suggest that, upon increasing the polymer concentration, the larger colloids would start to demix prior to the smaller particles as the big--big depletion interaction is much stronger than the small--small depletion interaction, as indicated in Fig. \ref{figCCCP}(a2). However, the reality is more intricate with a single critical point for the entire three--component system, which separates into two mixed phases [Fig. \ref{figCCCP}(a1)]. Further addition of polymer results in a complex composition inversion phenomenon, where upon demixing the colloid--rich phase is initially dominated by the larger colloids but this effect lessens dramatically as further polymer is added, ultimately resulting in arrest of the colloid--rich phase and gelation \cite{zhang2018}.

\section{Exotic Gels in Real Space}
\label{sectionExotic}

\begin{figure}
\centering
\includegraphics[width=80mm]{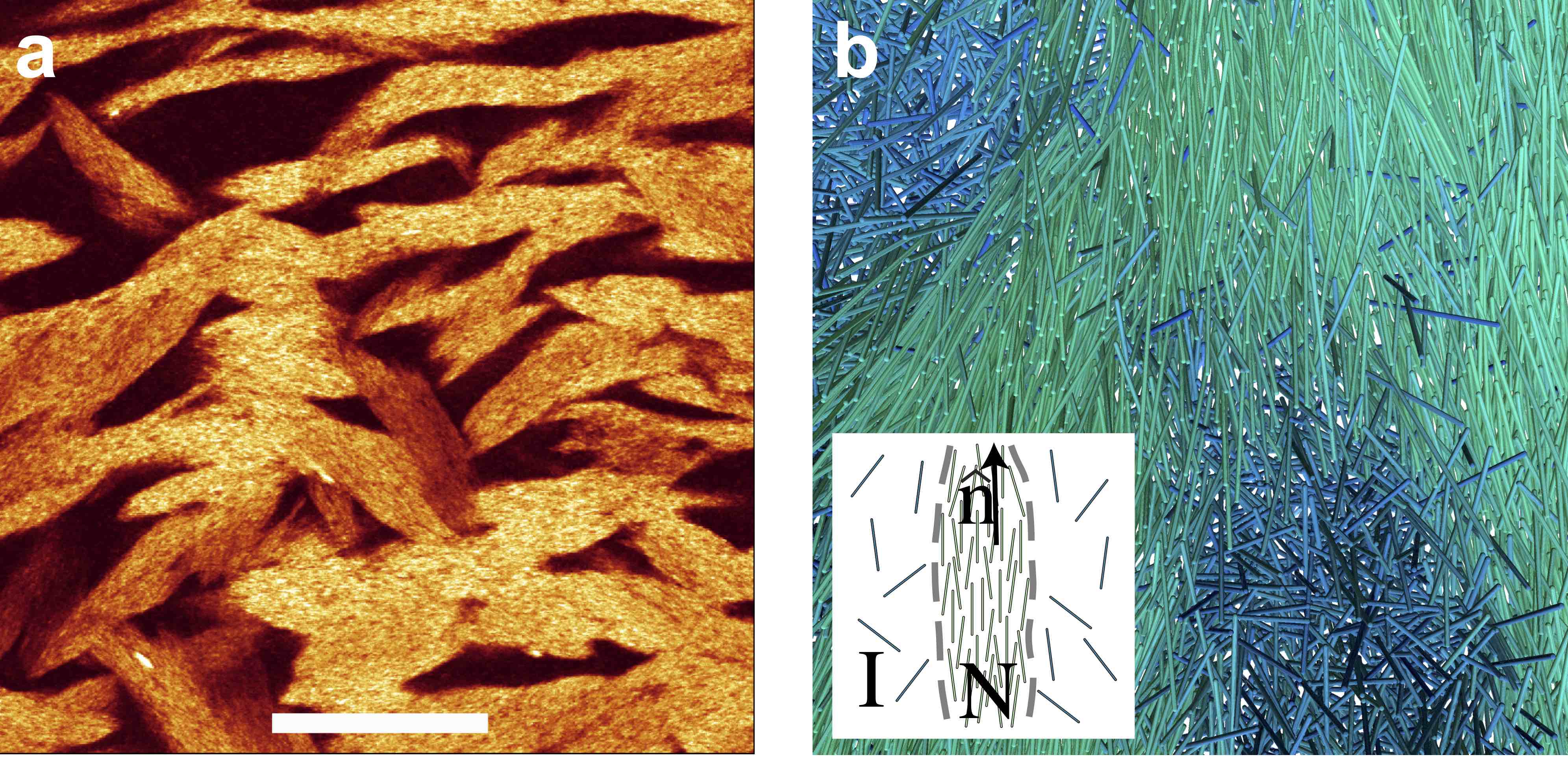}
\caption{\textbf{Non--sticky Spinodal gelation in polydisperse hard rods.}
Spinodal decomposition leads to a bicontinuous network of isotropic (I) and nematic (N). 
(a) confocal image of gel at rod volume fraction $\phi=0.043$, bright regions of nematic phase. Scale bar represents $4 \; \mathrm{\mu m}$. 
(b) Snapshot of a thin slice of a simulation box of rods at $\phi=0.140$. The colours indicate the local order parameter  of each rod and range from dark blue for rods in the isotropic phase to bright green for the nematic.
Inset shows that rods align  parallel to the director, indicated as $\mathbf{\hat{n}}$. We expect that the viscosity parallel to the director is low enough to permit flow, while perpendicular flow is strongly suppressed. Reproduced from \cite{ferreirocordova2020}.}
\label{figClaudia}
\end{figure}

Most of what we have discussed pertains to real space analysis of gels formed in colloid--polymer mixtures. While these have received the lion's share of the attention, gels analysed in real space take many forms. Sometimes the behaviour is similar, sometimes it is profoundly different. Given the nature of spinodal gelation, among the more surprising systems to undergo gelation is polydisperse hard rods (without any appreciable attraction between the particles) \cite{ferreirocordova2020}. Ferreiro-C\'{o}rdova and coworkers used real space analysis of a model system of polydisperse hard rods to demonstrate such \emph{non-sticky gelation}. These sepiolite colloidal rods exhibited two important features. Firstly, any attractions were small so the gelation was not attributed to (effective) attractions. Secondly, the rods are rather polydisperse (with an effective aspect ratio  $\langle L^{\prime}/D^{\prime} \rangle=24.6 \pm 9.5$) leading to a large gap in volume fraction between isotropic and nematic phases at phase coexistence. This large gap in volume fraction meant that the relaxation in the nematic phase was around 1000 times slower than that in the isotropic phase (depending on the orientation of the director). This leads to viscoelastic phase separation and, for suitable state points, spinodal decomposition. The four characteristics of spinodal gelation noted in the introduction are then met. Such a gel of hard particles is shown in Fig. \ref{figClaudia}(a) and computer simulation with similar behaviour revealing local rod orientation in (b).

\begin{figure}
\centering
\includegraphics[width=130mm]{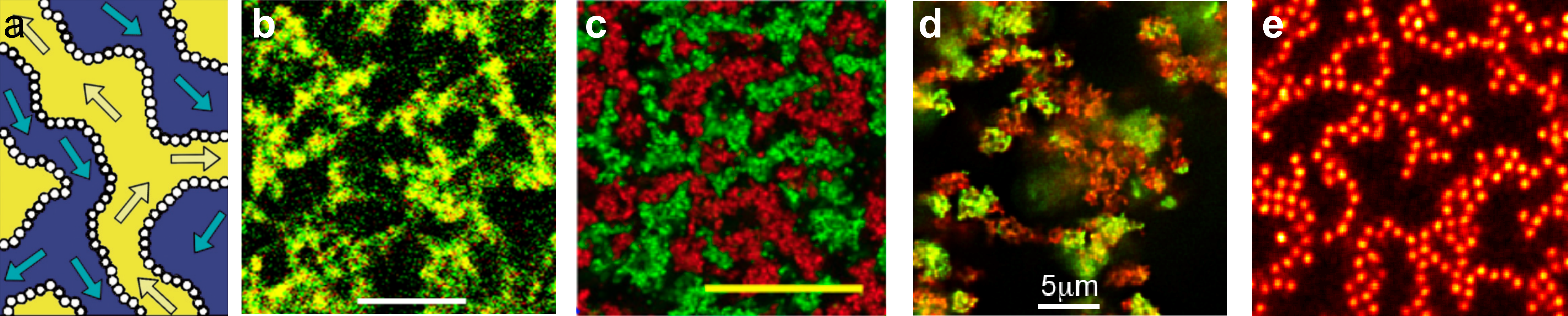}
\caption{\textbf{Exotic Gels and Gel--like systems.}
(a) Schematic geometry of a bijel. The bicontinuous morphology of the bijel allows two immiscible fluids to be passed through the material in opposite directions via a continuous process. Reproduced from \cite{cates2008}.
(b) Confocal images of an example binary gel that forms in mixtures of pNIPAM microgels (3 wt\%) and triblock-copolymer (3 wt\%) at elevated temperatures, the combined signal from the fluorescein labelled microgels and Nile red labelled triblock-copolymer is shown. Images are taken at 50$^{\circ}$C, bar=10  $\mathrm{\mu m}$ \cite{fussell2019}.
(c) Bigel of binary colloidal system with specific DNA--mediated interactions assembled into two independent networks. Bar=25  $\mathrm{\mu m}$ \cite{varrato2012}. 
(d) Binary protein gel with tunable domain sizes  \cite{riosdeanda2019}.
(e) Labyrinth topology in active dipolar gel \cite{sakai2020}.
}
\label{figExotic}
\end{figure}

Bijels are formed for two demixing liquids \cite{herzig2007} . However, a novel mechanism is used to prevent full demixing. The bijel is formed by the interface between the two liquids -- to which colloids are strongly absorbed, as in a pickering emulsion. Since full phase separation leads to a loss of interfacial area, if one can arrest the decrease in interfacial area between the two phases, a gel--like material can be formed. The interfacial area is then fixed by (2d) close packing of colloids which are strongly absorbed to it. Such a \emph{bijel} is shown schematically in Fig. \ref{figExotic}(a) \cite{cates2008}.

Hegelson and coworkers used \emph{polymer bridging} of emulsion droplets as a mechanism to drive colloidal gelation \cite{gao2015}. Here the ``sticky ends'' of the triblock copolymer tend to be absorbed by two neighbouring emulsion droplets, so that the polymer effectively bridges the droplets. Similar mechanisms had been developed by 
Porte and coworkers at a much smaller lengthscale \cite{filali2001}. Binary gels between poly(N-isopropylacrylamide) (pNIPAM) microgels and triblock-copolymer surfactant have been investigated by Fussell and coworkers [Fig.  \ref{figExotic}(b)], these gels only from at temperatures above the collapse temperature of the microgels. This results in the ability to form gels reversibly with heating. The gelation is caused by the temperature responsive association (i.e. bridging) of the microgel colloids and triblock-copolymer, driven at elevated temperatures by hydrophobic interactions \cite{fussell2019,fussell2021}.

An exciting development in the context of designing nano--architectures was the work of Eiser and Foffi and coworkers \cite{varrato2012}. Here a binary colloidal system with controllable \emph{specific} interactions (unlike polymer--induced depletion which ``sticks everything to everything'' -- see section \ref{sectionComplex}) was designed using DNA--coated colloids. The interactions in the binary system were designed such that each species was attracted to itself, but there was no significant attraction between the two species. Thus, the simultaneous formation of two independent networks in the same suspension was possible, as shown in Fig.  \ref{figExotic}(c). Similar behaviour has since been obtained in protein gels \cite{blumlein2015}.

\textit{Protein gels --- } We have noted that proteins can form gels. Often the small lengthscales of proteins have led to scattering studies being carried out \cite{mcmanus2016,stradner2004,stradner2020,Bucciarelli2015,bucciarelli2016,myung2018}. However, fluorescent proteins are amenable to real space analysis \cite{blumlein2015}. It is thought that the mechanism of gelation in proteins is often similar to that discussed here \cite{mcmanus2016}. However, the interactions of protein molecules are far more complex, enabling novel assembly pathways to be designed. In Fig. \ref{figExotic}(d) we show a gel of the fluorescent proteins eGFP and mCherry, whose interactions have been tailored such that the size of domains of each protein can be controlled. Notably, because here the proteins retained their fluorescence, we may infer that they retain their functionality, with potential for the assembly of novel nanomaterials \cite{riosdeanda2019}. Very recently, depletion driven gelation in a protein system has been quantitatively mapped to a colloid--polymer model. Given the usually complex nature fo protein--protein interactions, mapping of eGFP dimers to hard spherocylindwers was thought to be possible due to polymer--induced passivation which led the protein molecules to interact more like hard particles \cite{cheng2021}.

\textit{Active gels --- } Active gels of actin tubules form a fascinatingly rich system \cite{decamp2015}. Our opinion is that the underlying physics is so different to the systems considered here that we refer the reader to the appropriate literature \cite{bechinger2016,marchetti2013}. However, recently, colloidal gel--like assembly has been achieved with active colloids with dipolar interactions Fig.  \ref{figExotic}(e) \cite{sakai2020}. These feature a steady--state ``labyrinthine'' structure which is reminiscent of a transient state of the passive dipolar system \cite{dassanayake2000}.

\section{Outlook}
\label{sectionOutlook}

We close with some conclusions about what has been learned and identify some open questions. We begin with what we understand.
\begin{itemize}
\item{Perhaps surprisingly, given that colloidal  systems are prized as simple models, it proved more challenging than expected to realise the simplest colloidal gelformer, sticky spheres in experimental systems suitable for real space analysis with index and in particular density matching (Section \ref{sectionEarly}). That now seems to be under control, but care should be exercised when preparing experimental systems: it is harder than in seems.}
\item{We understand the mechanisms of how a rigid network emerges from spinodal decomposition, see sections \ref{sectionMechanism} and \ref{sectionShortTime}. Rigidity seems largely geometric and we emphasise the role of hydrodynamics in the early stages of aggregation.}
\end{itemize}
Moving to open questions, loosely speaking it seems that we understand how gels form. Much less clear is what happens next.
\begin{itemize}
\item{How do gels fail? In the quiescent case (section \ref{sectionCollapse}), small systems collapse rapidly, but the delayed collapse phenomenon, important though it is in application, continues to elude our understanding. It is perhaps not the ideal task for real space analysis, suited as the technique is to detailed analysis of \emph{local} phenomena, due to the desire to study the whole sample. While excellent work has been carried out with scattering techniques \cite{aime2018} and computer simulation can yield much insight \cite{puertas2004,padmanabhan2018,harich2016,varga2016}, one cannot help but feel that real space analysis is nevertheless a good way to pinpoint the origins of this phenomenon, perhaps combined with magnetic resonance imaging \cite{moller2008}.}
\item{Although investigations have been made (section \ref{sectionDeformation}), a consensus on how the ``arms'' of a gel fail seems to be lacking. A detailed, particle--resolved analysis of this process, even in perhaps a somewhat artificial set up, would be most useful.}
\item{The response to shear may provide a good starting point to tackle gel failure. Whether failure mechanisms under shear are the same as those in a gravitational field is unclear, but at least shear provides a means to ``cause'' failure in such a way that it can be analysed in a reasonably straightforward manner. It may be possible to build on this and build elastoplastic models \cite{nicolas2018}, or to use theoretical approaches such as soft glassy rheology \cite{fielding2014} or elastoplastic models \cite{nicolas2018} which may access the relevant timescales (hours to years) and system sizes (at least $10^{12}$ particles) pertinent to delayed collapse which currently lie outwith the reach of computer simulation of particles.}
\item{The potential of computer simulation of particle--based for improving our understanding of gelation lies, if not i the systems sizes and timescales, rather in its precision. This is exemplified in the work of van Doorn \emph{et al.} \cite{vandoorn2018} where local mechanism for failure are investigated in an idealised set--up. Other possibilities include the use of advanced sampling techniques to address the challenges of timescales of up to years important for example in gel failure. However, gels age during this time, rather more obviously than do glasses, for example. One expects that, given a suitable order parameter for aging techniques like forward flux sampling may be one way to generate gels which have been ``aged'' in an accelerated manner \cite{allen2009}.} 
\item{What is the nature of the colloidal glass which leads to the arrest of the phase separation? In recent years, evidence has emerged for a \emph{Gardner Transition} in hard sphere glasses between a stable (thermal) glass and a marginal glass or ``Gardner phase'', depending on the compression rate \cite{charbonneau2017,charbonneau2014}. The volume fraction of the ``arms'' of the gel has been determined as $\phi_c\approx0.60$ \cite{lu2008,royall2018jcp}, which may place the system in the Gardner phase, with notable consequences for the mechanical properties \cite{charbonneau2017}.}
\item{We may expect to see more exotic systems. It is notable and perhaps remarkable that the vast majority of work reviewed here has concerned the polymethyl methacrylate--polystyrene system (see section \ref{sectionEvidence}). This is a most amenable system for study, but surely other systems are suitable, and an obvious candidate would seem to be soft colloids, e.g. microgels \cite{fussell2019}. Furthermore, although the colloidal ``rocks'' (Fig. \ref{figExotic}(d-g)) share some characteristics \cite{rice2012}, as yet there is no particle--resolved work on empty liquid systems  \cite{bianchi2006}, with their intriguing dynamic behaviour \cite{saikavoivod2011}. While such systems do not necessarily undergo spinodal gelation and indeed thermodynamically, their behaviour is more akin to that of network glassformers like silica than the systems we have discussed here (see section \ref{sectionEarly} and Fig. \ref{figPhaseEmptyArrest}), it would certainly be interesting to explore empty liquids in real space used ``patchy particle'' systems.}
\item{The age of dynamic arrest of active colloids is in its infancy \cite{klongvessa2019}, and we expect to see more. Meanwhile the protein gels in Fig. \ref{figExotic}(d) raise the following question: if proteins may be naively though of as sticky spheres, then why is that protein gels have ``arms'' on the same lengthscale as colloidal gels when the molecules are a thousand times smaller in diameter. That is to say, a lengthscale of a few microns corresponds to a few ($<10$) colloid diameters while that same lengthscale corresponds to thousands of protein molecules. How is it that if proteins and colloids somehow obey the same physics (here), then why such a massive difference in lengthscale (in units of the diameter)? It is tempting to speculate that because proteins are so much smaller, and the Brownian time $\tau_B \sim \sigma^3$, then in a sense, time runs $10^9$ times faster in the case of proteins. So we are looking at a gel at a much later stage than is the case with the colloids -- for the same laboratory timescale. Such speculations of course warrant careful investigation.}
\end{itemize}
We close by concluding that in the \emph{18 years} since the pioneering work of Dinsmore and Weitz \cite{dinsmore2002} a great deal has been done with real space analysis of colloidal gels, and many open questions remain.

\vspace{2cm}
\ack
In connection with the preparation of this article, and for enlightening discussions pertaining to gelation, the authors would like to thank 
Dirk Aarts,
Ludovic Berthier, 
Paul Bartlett,
Daniel Bonn,
Patrick Charbonneau,
Rui Cheng,
Luca Cipeletti,
Marjolein Dijkstra,
Roel Dullens,
Jeroen van Duijneveldt,
Stefan Egelhaaf,
Bob Evans,
Jens Eggers,
Claudia Ferreiro-C\'{o}rdova,
Guiseppe Foffi,
Sharon Glotzer,
Peter Harrowell,
Rob Jack,
Willem Kegel,
Christian Klix,
Tannie Liverpool,
Henk Lekkerkerker,
Mattjieu Leocmach,
Hartmut L\"{o}wen,
Jennifer Mcmanus,
Kuni Miyazaki,
Takehiro Ohtsuka,
Rattachai Pinchiapat,
Wilson Poon,
Itamar Procaccia, 
Azaima Razali,
Rebecca Rice,
Ioatzin Rios de Anda,
David Richard,
Roland Roth,
John Russo,
Nariaki Saka\"{i},
Francesco Sciortino,
Mike Solomon,
Ken Schweizer,
Thomas Speck,
Grzegorz Szamel,
Hajime Tanaka,
Shelley Taylor,
Francesco Turci,
Brian Vincent,
Eric Weeks,
Nigel Wilding,
Stephen Williams,
Emanuela Zaccarelli,
Isla Zhang.
and
Roseanna Zia.
CPR gratefully acknowledges the Royal Society  and EPSRC grant EP/T031077/1. 
CPR and JH acknowledge European Research Council (ERC Consolidator Grant NANOPRS, project number 617266).
SF is supported by a studentship provided by the Bristol Centre for Functional Nanomaterials (EPSRC grant EP/L016648/1)


\begin{thebibliography}{100}

\bibitem{tanaka2000}
Hajime Tanaka.
\newblock Viscoelastic phase separation.
\newblock {\em J. Phys.: Condens. Matter}, 12(15):R207--R264, April 2000.

\bibitem{poon2002}
W~C~K Poon.
\newblock The physics of a model colloid\textendash polymer mixture.
\newblock {\em J. Phys.: Condens. Matter}, 14:R859, 2002.

\bibitem{cipelletti2005}
Luca Cipelletti and Laurence Ramos.
\newblock Slow dynamics in glassy soft matter.
\newblock {\em J. Phys.: Condens. Matter}, 17(6):R253--R285, February 2005.

\bibitem{coniglio2004}
A~Coniglio, L~De Arcangelis, E~Del Gado, A~Fierro, and N~Sator.
\newblock Percolation, gelation and dynamical behaviour in colloids.
\newblock {\em J. Phys.: Condens. Matter}, 16(42):S4831--S4839, October 2004.

\bibitem{zaccarelli2007}
Emanuela Zaccarelli.
\newblock Colloidal gels: Equilibrium and non-equilibrium routes.
\newblock {\em J. Phys.: Condens. Matter}, 19(32):323101, August 2007.

\bibitem{drury2003}
Jeanie~L. Drury and David~J. Mooney.
\newblock Hydrogels for tissue engineering: Scaffold design variables and
  applications.
\newblock {\em Biomaterials}, 24(24):4337--4351, November 2003.

\bibitem{rose2014}
S{\'e}verine Rose, Alexandre Prevoteau, Paul Elzi{\`e}re, Dominique Hourdet,
  Alba Marcellan, and Ludwik Leibler.
\newblock Nanoparticle solutions as adhesives for gels and biological tissues.
\newblock {\em Nature}, 505(7483):382--385, January 2014.

\bibitem{ubbink2012}
Job Ubbink.
\newblock Soft matter approaches to structured foods: From ``cook-and-look'' to
  rational food design?
\newblock {\em Faraday Discuss.}, 158:9, 2012.

\bibitem{liang2001}
Jingmei Liang, Yue Ma, Yi~Zheng, H.~Ted Davis, Hung-Ta Chang, David Binder,
  Syed Abbas, and F.-L. Hsu.
\newblock Solvent-{{Induced Crystal Morphology Transformation}} in a {{Ternary
  Soap System}}: {{Sodium Stearate Crystalline Fibers}} and {{Platelets}}.
\newblock {\em Langmuir}, 17(21):6447--6454, October 2001.

\bibitem{cardinaux2007}
Fr{\'e}d{\'e}ric Cardinaux, Thomas Gibaud, Anna Stradner, and Peter
  Schurtenberger.
\newblock Interplay between {{Spinodal Decomposition}} and {{Glass Formation}}
  in {{Proteins Exhibiting Short}}-{{Range Attractions}}.
\newblock {\em Phys. Rev. Lett.}, 99(11):118301, September 2007.

\bibitem{leocmach2014}
Mathieu Leocmach, Christophe Perge, Thibaut Divoux, and S{\'e}bastien
  Manneville.
\newblock Creep and {{Fracture}} of a {{Protein Gel}} under {{Stress}}.
\newblock {\em Phys. Rev. Lett.}, 113(3):038303, July 2014.

\bibitem{mcmanus2016}
Jennifer~J. McManus, Patrick Charbonneau, Emanuela Zaccarelli, and Neer
  Asherie.
\newblock The physics of protein self-assembly.
\newblock {\em Current Opinion in Colloid \& Interface Science}, 22:73--79,
  April 2016.

\bibitem{fusco2016}
Diana Fusco and Patrick Charbonneau.
\newblock Soft matter perspective on protein crystal assembly.
\newblock {\em Colloids and Surfaces B: Biointerfaces}, 137:22--31, January
  2016.

\bibitem{riosdeanda2019}
Ioatzin~R{\i}os {de Anda}, Angelique {Coutable-Pennarun}, Chris Brasnett,
  Stephen Whitelam, Annela Seddon, John Russo, J.~L.~Ross Anderson, and
  C.~Patrick Royall1.
\newblock Decorated {{Protein Networks}}: {{Functional Nanomaterials}} with
  {{Tunable Domain Size}}.
\newblock {\em ArXiv191105857 Cond-Mat Physicsphysics Q-Bio}, November 2019.

\bibitem{ulrich2009}
Stephan Ulrich, Timo Aspelmeier, Klaus Roeller, Axel Fingerle, Stephan
  Herminghaus, and Annette Zippelius.
\newblock Cooling and {{Aggregation}} in {{Wet Granulates}}.
\newblock {\em Phys. Rev. Lett.}, 102(14):148002, April 2009.

\bibitem{li2014}
Jindong Li, Yixin Cao, Chengjie Xia, Binquan Kou, Xianghui Xiao, Kamel Fezzaa,
  and Yujie Wang.
\newblock Similarity of wet granular packing to gels.
\newblock {\em Nat Commun}, 5(1):5014, December 2014.

\bibitem{bouttes2014}
David Bouttes, Emmanuelle Gouillart, Elodie Boller, Davy Dalmas, and Damien
  Vandembroucq.
\newblock Fragmentation and {{Limits}} to {{Dynamical Scaling}} in {{Viscous
  Coarsening}}: {{An Interrupted}} {\emph{in situ}} {{X}}-{{Ray Tomographic
  Study}}.
\newblock {\em Phys. Rev. Lett.}, 112(24):245701, June 2014.

\bibitem{baumer2013}
R.~E. Baumer and M.~J. Demkowicz.
\newblock Glass {{Transition}} by {{Gelation}} in a {{Phase Separating Binary
  Alloy}}.
\newblock {\em Phys. Rev. Lett.}, 110(14):145502, April 2013.

\bibitem{ruzicka2011}
Barbara Ruzicka, Emanuela Zaccarelli, Laura Zulian, Roberta Angelini, Michael
  Sztucki, Abdellatif Moussa{\"i}d, Theyencheri Narayanan, and Francesco
  Sciortino.
\newblock Observation of empty liquids and equilibrium gels in a colloidal
  clay.
\newblock {\em Nature Mater}, 10(1):56--60, January 2011.

\bibitem{saikavoivod2011}
Ivan {Saika-Voivod}, Heather~Marie King, Piero Tartaglia, Francesco Sciortino,
  and Emanuela Zaccarelli.
\newblock Silica through the eyes of colloidal models\textemdash when glass is
  a gel.
\newblock {\em J. Phys.: Condens. Matter}, 23(28):285101, July 2011.

\bibitem{bianchi2006}
Emanuela Bianchi, Julio Largo, Piero Tartaglia, Emanuela Zaccarelli, and
  Francesco Sciortino.
\newblock Phase {{Diagram}} of {{Patchy Colloids}}: {{Towards Empty Liquids}}.
\newblock {\em Phys. Rev. Lett.}, 97(16):168301, October 2006.

\bibitem{jabbarifarouji2007}
Sara {Jabbari-Farouji}, Gerard~H. Wegdam, and Daniel Bonn.
\newblock Gels and {{Glasses}} in a {{Single System}}: {{Evidence}} for an
  {{Intricate Free}}-{{Energy Landscape}} of {{Glassy Materials}}.
\newblock {\em Phys. Rev. Lett.}, 99(6):065701, August 2007.

\bibitem{onuki}
Akira Onuki.
\newblock {\em Phase {{Transition Dynamics}}}.
\newblock {Cambridge University Press}, 2002.

\bibitem{lekkerkerker1992}
H~N~W Lekkerkerker, H.N.W., Wilson C~K Poon, P.~N. Pusey, A.~Stroobants, and
  P.~B. Warren.
\newblock Phase-behavior of colloid plus polymer mixtures.
\newblock {\em Europhys. Lett.}, 20:559--564, 1992.

\bibitem{teece2011}
Lisa~J. Teece, Malcolm~A. Faers, and Paul Bartlett.
\newblock Ageing and collapse in gels with long-range attractions.
\newblock {\em Soft Matter}, 7(4):1341--1351, 2011.

\bibitem{klotsa2011}
Daphne Klotsa and Robert~L. Jack.
\newblock Predicting the self-assembly of a model colloidal crystal.
\newblock {\em Soft Matter}, 7(13):6294, 2011.

\bibitem{whitelam2015}
Stephen Whitelam and Robert~L. Jack.
\newblock The {{Statistical Mechanics}} of {{Dynamic Pathways}} to
  {{Self}}-{{Assembly}}.
\newblock {\em Annu. Rev. Phys. Chem.}, 66(1):143--163, April 2015.

\bibitem{ferreirocordova2020}
Claudia {Ferreiro-C{\'o}rdova}, C.~Patrick Royall, and Jeroen~S. {van
  Duijneveldt}.
\newblock Anisotropic viscoelastic phase separation in polydisperse hard rods
  leads to nonsticky gelation.
\newblock {\em Proc Natl Acad Sci USA}, 117(7):3415--3420, February 2020.

\bibitem{russel}
W.~Russel, D.~Saville, and W.~Schowalter.
\newblock {\em Russel, {{W}}.; {{Saville}}, {{D}}. \& {{Schowalter}}, {{W}}.
  {{Colloidal Dispersions}}}.
\newblock {Cambridge Univ. Press}, {Cambridge}, 1989.

\bibitem{weitz1984}
David~A. Weitz and J.M Oliveira.
\newblock Fractal {{Structures Formed}} by {{Kinetic Aggregation}} of {{Aqueous
  Gold Colloids}}.
\newblock {\em Phys Rev Lett}, 52:1433--1438, 1984.

\bibitem{asakura1954}
{Asakura, S. {and} Ooswawa, F.}
\newblock On {{Interaction}} between {{Two Bodies Immersed}} in a {{Solution}}
  of {{Macromolecules}}.
\newblock {\em J Chem Phys}, 22:1255--1256, 1954.

\bibitem{long1973}
J.~A. Long, D.~W.~J. Osmond, and B.~Vincent.
\newblock The equilibrium aspects of weak flocculation.
\newblock {\em J Coll Interf Sci}, 42:545--553, 1976.

\bibitem{appell1998}
Jacqueline Appell, Gr{\'e}goire Porte, and Michel Rawiso.
\newblock Interactions between {{Nonionic Surfactant Micelles Introduced}} by a
  {{Telechelic Polymer}}. {{A Small Angle Neutron Scattering Study}}.
\newblock {\em Langmuir}, 14(16):4409--4414, August 1998.

\bibitem{gao2015}
Yongxiang Gao.
\newblock Microdynamics and arrest of coarsening during spinodal decomposition
  in thermoreversible colloidal gels.
\newblock {\em Soft Matter}, page~12, 2015.

\bibitem{fussell2019}
S.~L. Fussell, K.~Bayliss, C.~Coops, L.~Matthews, W.~Li, W.~H. Briscoe, M.~A.
  Faers, C.~P. Royall, and J.~S. {van Duijneveldt}.
\newblock Reversible temperature-controlled gelation in mixtures of {{pNIPAM}}
  microgels and non-ionic polymer surfactant.
\newblock {\em Soft Matter}, 15(42):8578--8588, 2019.

\bibitem{veen2012}
Sandra~J. Veen, Oleg Antoniuk, Bart Weber, Marco A.~C. Potenza, Stefano
  Mazzoni, Peter Schall, and Gerard~H. Wegdam.
\newblock Colloidal {{Aggregation}} in {{Microgravity}} by {{Critical Casimir
  Forces}}.
\newblock {\em Phys. Rev. Lett.}, 109(24):248302, December 2012.

\bibitem{hertlein2008}
C.~Hertlein, L.~Helden, A.~Gambassi, S.~Dietrich, and C.~Bechinger.
\newblock Direct measurement of critical {{Casimir}} forces.
\newblock {\em Nature}, 451(7175):172--175, January 2008.

\bibitem{bonn2009}
Daniel Bonn, Jakub Otwinowski, Stefano Sacanna, Hua Guo, Gerard Wegdam, and
  Peter Schall.
\newblock Direct {{Observation}} of {{Colloidal Aggregation}} by {{Critical
  Casimir Forces}}.
\newblock {\em Phys. Rev. Lett.}, page~4, 2009.

\bibitem{rouwhorst2020ncomms}
Joep Rouwhorst, Christopher Ness, Simeon Stoyanov, Alessio Zaccone, and Peter
  Schall.
\newblock Nonequilibrium continuous phase transition in colloidal gelation with
  short-range attraction.
\newblock {\em Nat Commun}, 11(1):3558, December 2020.

\bibitem{rouwhorst2020pre}
Joep Rouwhorst, Peter Schall, Christopher Ness, Theo Blijdenstein, and Alessio
  Zaccone.
\newblock Nonequilibrium master kinetic equation modeling of colloidal
  gelation.
\newblock {\em Phys. Rev. E}, 102(2):022602, August 2020.

\bibitem{decamp2015}
Stephen~J. DeCamp, Gabriel~S. Redner, Aparna Baskaran, Michael~F. Hagan, and
  Zvonimir Dogic.
\newblock Orientational order of motile defects in active nematics.
\newblock {\em Nature Mater}, 14(11):1110--1115, November 2015.

\bibitem{sakai2020}
Nariaki Saka{\"i} and C.~Patrick Royall.
\newblock Active {{Dipolar Colloids}} in {{Three Dimensions}}: {{Strings}},
  {{Sheets}}, {{Labyrinthine Textures}} and {{Crystals}}.
\newblock {\em ArXiv201003925 Cond-Mat Physicsphysics}, October 2020.

\bibitem{ivlev}
{Ivlev, A. {and} Lowen, H {and} Morfill, G. E. {and} Royall, C. P.}
\newblock {\em Complex {{Plasmas}} and {{Colloidal Dispersions}}:
  {{Particle}}-Resolved {{Studies}} of {{Classical Liquids}} and {{Solids}}}.
\newblock {World Scientific Publishing Co., Singapore Scientific}, {Singapore},
  2012.

\bibitem{peterquote}
``{{The}} perceived wisdom of materials science is that the microscopic
  structure determines the dynamics and macroscopic behaviour of the
  material.'' {{Attributted}} to {{Peter Harrowell}}.

\bibitem{baxter1968}
R.~Baxter.
\newblock Percus-{{Yevick Equation}} for {{Hard Spheres}} with {{Surface
  Adhesion}}.
\newblock {\em J. Chem. Phys.}, 49:2770--2774, 1968.

\bibitem{miller2003}
Mark~A. Miller and Daan Frenkel.
\newblock Competition of {{Percolation}} and {{Phase Separation}} in a
  {{Fluid}} of {{Adhesive Hard Spheres}}.
\newblock {\em Phys. Rev. Lett.}, 90(13):135702, April 2003.

\bibitem{pham2002}
K.~N. Pham.
\newblock Multiple {{Glassy States}} in a {{Simple Model System}}.
\newblock {\em Science}, 296(5565):104--106, April 2002.

\bibitem{royall2018jcp}
C.~Patrick Royall, Stephen~R. Williams, and Hajime Tanaka.
\newblock Vitrification and gelation in sticky spheres.
\newblock {\em J. Chem. Phys.}, 148(4):044501, January 2018.

\bibitem{foffi2002}
G.~Foffi, K.~A. Dawson, S.~V. Buldyrev, F.~Sciortino, E.~Zaccarelli, and
  P.~Tartaglia.
\newblock Evidence for an unusual dynamical-arrest scenario in short-ranged
  colloidal systems.
\newblock {\em Phys. Rev. E}, 65(5):050802, May 2002.

\bibitem{bergenholtz2003}
J.~Bergenholtz, W.~C.~K. Poon, and M.~Fuchs.
\newblock Gelation in {{Model Colloid}}-{{Polymer Mixtures}}.
\newblock {\em Langmuir}, 19(10):4493--4503, May 2003.

\bibitem{chen2004}
Yeng-Long Chen and Kenneth~S. Schweizer.
\newblock Microscopic theory of gelation and elasticity in polymer\textendash
  particle suspensions.
\newblock {\em The Journal of Chemical Physics}, 120(15):7212--7222, April
  2004.

\bibitem{shah2003}
S~A Shah, Y-L Chen, S~Ramakrishnan, K~S Schweizer, and C~F Zukoski.
\newblock Microstructure of dense colloid\textendash polymer suspensions and
  gels.
\newblock {\em J. Phys.: Condens. Matter}, 15(27):4751--4778, July 2003.

\bibitem{buscall1987}
Richard Buscall and Lee~R. White.
\newblock The consolidation of concentrated suspensions. {{Part}}
  1.\textemdash{{The}} theory of sedimentation.
\newblock {\em J. Chem. Soc., Faraday Trans. 1}, 83(3):873, 1987.

\bibitem{zaccone2009}
Alessio Zaccone, Hua Wu, and Emanuela~Del Gado.
\newblock Elasticity of {{Arrested Short}}-{{Ranged Attractive Colloids}}:
  {{Homogeneous}} and {{Heterogeneous Glasses}}.
\newblock {\em Phys. Rev. Lett.}, page~4, 2009.

\bibitem{puertas2004}
Antonio~M. Puertas, Matthias Fuchs, and Michael~E. Cates.
\newblock Dynamical heterogeneities close to a colloidal gel.
\newblock {\em J. Chem. Phys.}, 121(6):2813, 2004.

\bibitem{padmanabhan2018}
Poornima Padmanabhan and Roseanna Zia.
\newblock Gravitational collapse of colloidal gels: Non-equilibrium phase
  separation driven by osmotic pressure.
\newblock {\em Soft Matter}, 14(17):3265--3287, 2018.

\bibitem{delgado2003}
E.~Del Gado, A~Fierro, L.~de~Arcangelis, and A~Coniglio.
\newblock A unifying model for chemical and colloidal gels.
\newblock {\em Europhys. Lett.}, 63(1):1--7, July 2003.

\bibitem{berthier2011}
Ludovic Berthier and Giulio Biroli.
\newblock Theoretical perspective on the glass transition and amorphous
  materials.
\newblock {\em Rev. Mod. Phys.}, 83(2):587--645, June 2011.

\bibitem{manley2005time}
S.~Manley, Benny Davidovitch, Neil~R. Davies, L.~Cipelletti, A.~E. Bailey,
  R.~J. Christianson, U.~Gasser, V.~Prasad, P.~N. Segre, M.~P. Doherty,
  S.~Sankaran, A.~L. Jankovsky, B.~Shiley, J.~Bowen, J.~Eggers, C.~Kurta,
  T.~Lorik, and D.~A. Weitz.
\newblock Time-{{Dependent Strength}} of {{Colloidal Gels}}.
\newblock {\em Phys. Rev. Lett.}, 95(4):048302, July 2005.

\bibitem{krishnareddy2012}
Naveen Krishna~Reddy, Zhenkun Zhang, M.~Paul~Lettinga, Jan K.~G. Dhont, and Jan
  Vermant.
\newblock Probing structure in colloidal gels of thermoreversible rodlike virus
  particles: {{Rheology}} and scattering.
\newblock {\em Journal of Rheology}, 56(5):1153--1174, September 2012.

\bibitem{faers2006}
Malcolm~A. Faers, Tahsin~H. Choudhury, Bobby Lau, Kevin McAllister, and Paul~F.
  Luckham.
\newblock Syneresis and rheology of weak colloidal particle gels.
\newblock {\em Colloids and Surfaces A: Physicochemical and Engineering
  Aspects}, 288(1-3):170--179, October 2006.

\bibitem{starrs2002}
L~Starrs, W~C~K Poon, D~J Hibberd, and M~M Robins.
\newblock Collapse of transient gels in colloid-polymer mixtures.
\newblock {\em J. Phys.: Condens. Matter}, 14(10):2485--2505, March 2002.

\bibitem{bartlett2012}
Paul Bartlett, Lisa~J. Teece, and Malcolm~A. Faers.
\newblock Sudden collapse of a colloidal gel.
\newblock {\em Phys. Rev. E}, 85(2):021404, February 2012.

\bibitem{carpineti1992}
M.~Carpineti and M.~Giglio.
\newblock Spinodal--{{Type Dynamics}} in {{Fractal Aggregation}} of {{Collodal
  Clusters}}.
\newblock {\em Phys. Rev. Lett.}, 68:3327--3330, 1992.

\bibitem{cipelletti2000}
Luca Cipelletti, S.~Manley, R.~C. Ball, and D.~A. Weitz.
\newblock Universal {{Aging Features}} in the {{Restructuring}} of {{Fractal
  Colloidal Gels}}.
\newblock {\em Phys. Rev. Lett.}, 84(10):2275--2278, March 2000.

\bibitem{verhaegh1999}
Nynke A~M Verhaegh, Daniela Asnaghi, and Henk N~W Lekkerkerker.
\newblock Transient gels in colloid\textendash polymer mixtures studied with
  uorescence confocal scanning laser microscopy.
\newblock {\em Phys. A}, page~11, 1999.

\bibitem{chaikin}
{Chaikin, P. M. {and} Lubeneksy, T. C.}
\newblock {\em Principles of {{Consdensed Matter Physics}}}.
\newblock {Cambridge University Press}, 1995.

\bibitem{cahn1959}
{Cahn, J, {and} Hilliard, J.}
\newblock Free {{Energy}} of a {{Nonuniform System III}}. {{Nucleation}} in a
  {{Two}}--{{Component Incompressible Fluid}}.
\newblock {\em J. Chem. Phys.}, 31:688--699, 1959.

\bibitem{cavagna2009}
Andrea Cavagna.
\newblock Supercooled liquids for pedestrians.
\newblock {\em Physics Reports}, 476(4-6):51--124, June 2009.

\bibitem{royall2015physrep}
C.~Patrick Royall and Stephen~R. Williams.
\newblock The role of local structure in dynamical arrest.
\newblock {\em Physics Reports}, 560:1--75, February 2015.

\bibitem{royall2018jpcm}
C~Patrick Royall, Francesco Turci, Soichi Tatsumi, John Russo, and Joshua
  Robinson.
\newblock The race to the bottom: Approaching the ideal glass?
\newblock {\em J. Phys.: Condens. Matter}, 30(36):363001, September 2018.

\bibitem{berthier2016}
Ludovic Berthier and Mark~D. Ediger.
\newblock Facets of glass physics.
\newblock {\em Physics Today}, 69(1):40--46, January 2016.

\bibitem{biroli2013}
Giulio Biroli and Juan~P. Garrahan.
\newblock Perspective: {{The}} glass transition.
\newblock {\em The Journal of Chemical Physics}, 138(12):12A301, March 2013.

\bibitem{debenedetti2001}
Pablo~G. Debenedetti and Frank~H Stillinger.
\newblock Debenedetti, {{P}}. \& {{Stillinger}}, {{F}}. {{Supercooled}} liquids
  and the glass transition.
\newblock {\em Nature}, 410:259--267, 2001.

\bibitem{royall2020}
C.~Patrick Royall, Francesco Turci, and Thomas Speck.
\newblock Dynamical phase transitions and their relation to structural and
  thermodynamic aspects of glass physics.
\newblock {\em J. Chem. Phys.}, 153(9):090901, September 2020.

\bibitem{charbonneau2017}
Patrick Charbonneau, Jorge Kurchan, Giorgio Parisi, Pierfrancesco Urbani, and
  Francesco Zamponi.
\newblock Glass and {{Jamming Transitions}}: {{From Exact Results}} to
  {{Finite}}-{{Dimensional Descriptions}}.
\newblock {\em Annu. Rev. Condens. Matter Phys.}, 8(1):265--288, March 2017.

\bibitem{manley2005spinodal}
S.~Manley, H.~M. Wyss, K.~Miyazaki, J.~C. Conrad, V.~Trappe, L.~J. Kaufman,
  D.~R. Reichman, and D.~A. Weitz.
\newblock Glasslike {{Arrest}} in {{Spinodal Decomposition}} as a {{Route}} to
  {{Colloidal Gelation}}.
\newblock {\em Phys. Rev. Lett.}, 95(23):238302, December 2005.

\bibitem{lu2008}
Peter~J. Lu, Emanuela Zaccarelli, Fabio Ciulla, Andrew~B. Schofield, Francesco
  Sciortino, and David~A. Weitz.
\newblock Gelation of particles with short-range attraction.
\newblock {\em Nature}, 453(7194):499--503, May 2008.

\bibitem{zaccarelli2008}
Emanuela Zaccarelli, Peter~J Lu, Fabio Ciulla, David~A Weitz, and Francesco
  Sciortino.
\newblock Gelation as arrested phase separation in short-ranged attractive
  colloid\textendash polymer mixtures.
\newblock {\em J. Phys.: Condens. Matter}, 20(49):494242, December 2008.

\bibitem{griffiths2017}
Samuel Griffiths, Francesco Turci, and C.~Patrick Royall.
\newblock Local structure of percolating gels at very low volume fractions.
\newblock {\em The Journal of Chemical Physics}, 146(1):014905, January 2017.

\bibitem{jamie2012}
E.~A.~G. Jamie, R.~P.~A. Dullens, and D.~G. A.~L. Aarts.
\newblock Spinodal decomposition of a confined colloid-polymer system.
\newblock {\em The Journal of Chemical Physics}, 137(20):204902, November 2012.

\bibitem{fullerton2020}
Christopher~J. Fullerton and Ludovic Berthier.
\newblock Glassy behaviour of sticky spheres: {{What}} lies beyond experimental
  timescales?
\newblock {\em ArXiv200714165 Cond-Mat}, July 2020.

\bibitem{gao2007}
Y.~Gao and M.~L. Kilfoil.
\newblock Direct {{Imaging}} of {{Dynamical Heterogeneities}} near the
  {{Colloid}}-{{Gel Transition}}.
\newblock {\em Phys. Rev. Lett.}, 99(7):078301, August 2007.

\bibitem{dibble2008}
Clare~J Dibble, Michael Kogan, and Michael~J Solomon.
\newblock Structural origins of dynamical heterogeneity in colloidal gels.
\newblock {\em Phys. Rev. E}, 77:050401R, 2008.

\bibitem{royall2008naturemater}
C.~Patrick~Royall, Stephen~R. Williams, Takehiro Ohtsuka, and Hajime Tanaka.
\newblock Direct observation of a local structural mechanism for dynamic
  arrest.
\newblock {\em Nature Mater}, 7(7):556--561, July 2008.

\bibitem{zhang2013}
Isla Zhang, C.~Patrick Royall, Malcolm~A. Faers, and Paul Bartlett.
\newblock Phase separation dynamics in colloid\textendash polymer mixtures: The
  effect of interaction range.
\newblock {\em Soft Matter}, 9(6):2076, 2013.

\bibitem{vanblaaderen1995}
A.~{van Blaaderen} and P.~Wiltzius.
\newblock Real-{{Space Structure}} of {{Colloidal Hard}}-{{Sphere Glasses}}.
\newblock {\em Science}, 270(5239):1177--1179, November 1995.

\bibitem{weeks2000}
Eric~R Weeks, J~C Crocker, Andrew~C Levitt, Andrew Schofield, and D~A Weitz.
\newblock Three-{{Dimensional Direct Imaging}} of {{Structural Relaxation
  Near}} the {{Colloidal Glass Transition}}.
\newblock {\em Science}, 287:627, 2000.

\bibitem{hallett2020}
James~E Hallett, Francesco Turci, and C~Patrick Royall.
\newblock The devil is in the details: Pentagonal bipyramids and dynamic
  arrest.
\newblock {\em J. Stat. Mech.}, 2020(1):014001, January 2020.

\bibitem{kuhn}
T.~S. Kuhn.
\newblock {\em The {{Structure}} of {{Scientific Revolutions}}}.
\newblock 1962.

\bibitem{goetze}
W.~Goetze.
\newblock {\em Complex {{Dynamics}} of {{Glass}}-{{Forming Liquids}}: {{A
  Mode}}-{{Coupling Theory}}}.
\newblock {Oxford Univ. Press}, 2009.

\bibitem{reichman2005}
David~R Reichman and Patrick Charbonneau.
\newblock Mode-coupling theory.
\newblock {\em J. Stat. Mech.}, 2005(05):P05013, May 2005.

\bibitem{brambilla2009}
G.~Brambilla, D.~El~Masri, M.~Pierno, L.~Berthier, L.~Cipelletti, G.~Petekidis,
  and A.~B. Schofield.
\newblock Probing the {{Equilibrium Dynamics}} of {{Colloidal Hard Spheres}}
  above the {{Mode}}-{{Coupling Glass Transition}}.
\newblock {\em Phys. Rev. Lett.}, 102(8):085703, February 2009.

\bibitem{hallett2018}
James~E. Hallett, Francesco Turci, and C.~Patrick Royall.
\newblock Local structure in deeply supercooled liquids exhibits growing
  lengthscales and dynamical correlations.
\newblock {\em Nat Commun}, 9(1):3272, December 2018.

\bibitem{janssen2015}
Liesbeth M.~C. Janssen and David~R. Reichman.
\newblock Microscopic {{Dynamics}} of {{Supercooled Liquids}} from {{First
  Principles}}.
\newblock {\em Phys. Rev. Lett.}, 115(20):205701, November 2015.

\bibitem{adam1965}
{Adam, G. {and} Gibbs, J.}
\newblock On the temperature dependence of relaxation phenomena in
  glass-forming liquids.
\newblock {\em J. Chem. Phys.}, 43:139--146, 1965.

\bibitem{lubchenko2007}
Vassiliy Lubchenko and Peter~G. Wolynes.
\newblock Theory of {{Structural Glasses}} and {{Supercooled Liquids}}.
\newblock {\em Annu. Rev. Phys. Chem.}, 58(1):235--266, May 2007.

\bibitem{parisi2010}
Giorgio Parisi and Francesco Zamponi.
\newblock Mean-field theory of hard sphere glasses and jamming.
\newblock {\em Rev Mod Phys}, 82(1):57, 2010.

\bibitem{tarjus2005}
G~Tarjus, S~A Kivelson, Z~Nussinov, and P~Viot.
\newblock The frustration-based approach of supercooled liquids and the glass
  transition: A review and critical assessment.
\newblock {\em J. Phys.: Condens. Matter}, 17(50):R1143--R1182, December 2005.

\bibitem{chandler2010}
David Chandler and Juan~P. Garrahan.
\newblock Dynamics on the {{Way}} to {{Forming Glass}}: {{Bubbles}} in
  {{Space}}-{{Time}}.
\newblock {\em Annu. Rev. Phys. Chem.}, 61(1):191--217, March 2010.

\bibitem{turci2017}
Francesco Turci, C.~Patrick Royall, and Thomas Speck.
\newblock Nonequilibrium {{Phase Transition}} in an {{Atomistic Glassformer}}:
  {{The Connection}} to {{Thermodynamics}}.
\newblock {\em Phys. Rev. X}, 7(3):031028, August 2017.

\bibitem{hecksher2008}
Tina Hecksher, Albena~I. Nielsen, Niels~Boye Olsen, and Jeppe~C. Dyre.
\newblock Little evidence for dynamic divergences in ultraviscous molecular
  liquids.
\newblock {\em Nature Phys}, 4(9):737--741, September 2008.

\bibitem{ozawa2019}
Misaki Ozawa, Camille Scalliet, Andrea Ninarello, and Ludovic Berthier.
\newblock Does the {{Adam}}-{{Gibbs}} relation hold in simulated supercooled
  liquids?
\newblock {\em J. Chem. Phys.}, 151(8):084504, August 2019.

\bibitem{cates2004}
M~E Cates, M~Fuchs, K~Kroy, W~C~K Poon, and A~M Puertas.
\newblock Theory and simulation of gelation, arrest and yielding in attracting
  colloids.
\newblock {\em J. Phys.: Condens. Matter}, 16(42):S4861--S4875, October 2004.

\bibitem{ball1987}
R.~C. Ball, D.~A. Weitz, T.~A. Witten, and F.~Leyvraz.
\newblock Universal kinetics in reaction-limited aggregation.
\newblock {\em Phys. Rev. Lett.}, 58:274--277, 1987.

\bibitem{kroy2004}
K.~Kroy, M.~E. Cates, and W.~C.~K. Poon.
\newblock Cluster {{Mode}}-{{Coupling Approach}} to {{Weak Gelation}} in
  {{Attractive Colloids}}.
\newblock {\em Phys. Rev. Lett.}, 92(14):148302, April 2004.

\bibitem{eberle2011}
A.~P.~R. Eberle, N.~J. Wagner, and R.~Castaneda-Priego.
\newblock Dynamical arrest transition in nanoparticle dispersions with
  short-range interactions.
\newblock {\em Phys. Rev. Lett.}, 106:105704, 2011.

\bibitem{laurati2009}
M.~Laurati, G.~Petekidis, N.~Koumakis, F.~Cardinaux, A.~B. Schofield, J.~M.
  Brader, M.~Fuchs, and S.~U. Egelhaaf.
\newblock Structure, dynamics, and rheology of colloid-polymer mixtures:
  {{From}} liquids to gels.
\newblock {\em J. Chem. Phys.}, 130(13):134907, April 2009.

\bibitem{gopalakrishnan2006}
V~Gopalakrishnan, K~S Schweizer, and C~F Zukoski.
\newblock Linking single particle rearrangements to delayed collapse times in
  transient depletion gels.
\newblock {\em J. Phys.: Condens. Matter}, 18(50):11531--11550, December 2006.

\bibitem{mewis2009}
N.~J. Mewis, J. \&~Wagner.
\newblock Thixotropy.
\newblock {\em Adv. Coll. Interf. Sci.,}, 147--148:214--227, 2009.

\bibitem{fielding2014}
S~M Fielding.
\newblock Shear banding in soft glassy materials.
\newblock {\em Rep. Prog. Phys.}, 77(10):102601, October 2014.

\bibitem{bonn2017}
Daniel Bonn, Morton~M. Denn, Ludovic Berthier, Thibaut Divoux, and
  S{\'e}bastien Manneville.
\newblock Yield stress materials in soft condensed matter.
\newblock {\em Rev. Mod. Phys.}, 89(3):035005, August 2017.

\bibitem{nicolas2018}
Alexandre Nicolas, Ezequiel~E. Ferrero, Kirsten Martens, and Jean-Louis Barrat.
\newblock Deformation and flow of amorphous solids: {{Insights}} from
  elastoplastic models.
\newblock {\em Rev. Mod. Phys.}, 90(4):045006, December 2018.

\bibitem{dijkstra1999}
Marjolein Dijkstra, Joseph~M Brader, and Robert Evans.
\newblock Phase behaviour and structure of model colloid-polymer mixtures.
\newblock {\em J. Phys.: Condens. Matter}, 11(50):10079--10106, December 1999.

\bibitem{dijkstra2000}
Marjolein Dijkstra, Rene {van Roij}, and Robert Evans.
\newblock Effective interactions, structure, and isothermal compressibility of
  colloidal suspensions.
\newblock {\em J Chem Phys}, 113(11):9, 2000.

\bibitem{taffs2010jpcm}
Jade Taffs, Alex Malins, Stephen~R Williams, and C~Patrick Royall.
\newblock A structural comparison of models of colloid\textendash polymer
  mixtures.
\newblock {\em J. Phys.: Condens. Matter}, 22(10):104119, March 2010.

\bibitem{dinsmore2002}
A~D Dinsmore and D~A Weitz.
\newblock Direct imaging of three-dimensional structure and topology of
  colloidal gels.
\newblock {\em J. Phys.: Condens. Matter}, 14(33):7581--7597, August 2002.

\bibitem{hunter2012}
G~L Hunter and E~R Weeks.
\newblock The physics of the colloidal glass transition.
\newblock {\em Rep Prog Phys}, page~31, 2012.

\bibitem{lu2013}
Peter~J Lu and David~A. Weitz.
\newblock Colloidal {{Particles}}: {{Crystals}}, {{Glasses}}, and {{Gels}}.
\newblock {\em Annu. Rev. Condens. Matter Phys.}, 4(1):217--233, April 2013.

\bibitem{yunker2014}
Peter~J Yunker, Ke~Chen, Matthew~D Gratale, Matthew~A Lohr, Tim Still, and A~G
  Yodh.
\newblock Physics in ordered and disordered colloidal matter composed of poly(
  {{{\emph{N}}}} -isopropylacrylamide) microgel particles.
\newblock {\em Rep. Prog. Phys.}, 77(5):056601, May 2014.

\bibitem{vanblaaderen1992}
Alfons {van Blaaderen}, Arnout Imhof, W.~Hage, and A.~Vrij.
\newblock Van {{Blaaderen}}, {{A}}.; {{Imhof}}, {{A}}.; {{Hage}}, {{W}}. \&
  {{Vrij}}, {{A}}. {{Three}}-{{Dimensional Imaging}} of {{Submicrometer
  Colloidal Particles}} in {{Concentrated Suspensions Using Confocal Scanning
  Laser Microscopy}}.
\newblock {\em Langmuir}, 8:1514--1517, 1992.

\bibitem{gasser2001}
U.~Gasser.
\newblock Real-{{Space Imaging}} of {{Nucleation}} and {{Growth}} in
  {{Colloidal Crystallization}}.
\newblock {\em Science}, 292(5515):258--262, April 2001.

\bibitem{taffs2013}
Jade Taffs, Stephen~R. Williams, Hajime Tanaka, and C.~Patrick Royall.
\newblock Structure and kinetics in the freezing of nearly hard spheres.
\newblock {\em Soft Matter}, 9:297--305, 2013.

\bibitem{crocker1996}
John~C. Crocker and David~G. Grier.
\newblock Methods of {{Digital Video Microscopy}} for {{Colloidal Studies}}.
\newblock {\em Journal of Colloid and Interface Science}, 179(1):298--310,
  April 1996.

\bibitem{leocmach2013}
Mathieu Leocmach and Hajime Tanaka.
\newblock A novel particle tracking method with individual particle size
  measurement and its application to ordering in glassy hard sphere colloids.
\newblock {\em Soft Matter}, 9(5):1447--1457, 2013.

\bibitem{gao2009}
Yongxiang Gao and Maria~L. Kilfoil.
\newblock Accurate detection and complete tracking of large populations of
  features in three dimensions.
\newblock {\em Opt. Express}, 17(6):4685, March 2009.

\bibitem{bierbaum2017}
Matthew Bierbaum, Brian~D. Leahy, Alexander~A. Alemi, Itai Cohen, and James~P.
  Sethna.
\newblock Light {{Microscopy}} at {{Maximal Precision}}.
\newblock {\em Phys. Rev. X}, 7(4):041007, October 2017.

\bibitem{kurita2012}
Rei Kurita and Eric~R. Weeks.
\newblock Measuring every particle's size from three-dimensional imaging
  experiments.
\newblock {\em Nat Commun}, 3(1):1127, January 2012.

\bibitem{koenderink1999}
G.~H. Koenderink, G.~A. Vliegenthart, S.~G. J.~M. Kluijtmans, A.~{van
  Blaaderen}, A.~P. Philipse, and H.~N.~W. Lekkerkerker.
\newblock Depletion-{{Induced Crystallization}} in {{Colloidal Rod}}-{{Sphere
  Mixtures}}.
\newblock {\em Langmuir}, 15(14):4693--4696, July 1999.

\bibitem{tolpekin2004}
V.~A. Tolpekin, M.~H.~G. Duits, D.~{van den Ende}, and J.~Mellema.
\newblock Aggregation and {{Breakup}} of {{Colloidal Particle Aggregates}} in
  {{Shear Flow}}, {{Studied}} with {{Video Microscopy}}.
\newblock {\em Langmuir}, 20(7):2614--2627, March 2004.

\bibitem{dehoog2001}
E.~H.~A. {de Hoog}, W.~K. Kegel, A.~{van Blaaderen}, and H.~N.~W. Lekkerkerker.
\newblock Direct observation of crystallization and aggregation in a
  phase-separating colloid-polymer suspension.
\newblock {\em Phys. Rev. E}, 64(2):021407, July 2001.

\bibitem{pusey1986}
P.~N. Pusey and {van Megen, W.}
\newblock Phase behaviour of concentrated suspensions of nearly hard colloidal
  spheres.
\newblock {\em Nature}, 320:340--342, 1986.

\bibitem{poon2012}
Wilson C~K Poon, Eric~R Weeks, and C~Patrick Royall.
\newblock On measuring colloidal volume fractions.
\newblock {\em Soft Matter}, 8:21--30, 2012.

\bibitem{royall2013myth}
C.~Patrick Royall, Wilson C.~K. Poon, and Eric~R. Weeks.
\newblock In search of colloidal hard spheres.
\newblock {\em Soft Matter}, 9(1):17--27, 2013.

\bibitem{royall2005sedimentation}
C~P Royall, R~van Roij, and A~van Blaaderen.
\newblock Extended sedimentation profiles in charged colloids: The
  gravitational length, entropy, and electrostatics.
\newblock {\em J. Phys.: Condens. Matter}, 17(15):2315--2326, April 2005.

\bibitem{royall2003}
C~P Royall, M~E Leunissen, and A~van Blaaderen.
\newblock A new colloidal model system to study long-range interactions
  quantitatively in real space.
\newblock {\em J. Phys.: Condens. Matter}, 15(48):S3581--S3596, December 2003.

\bibitem{klix2010}
Christian~L. Klix, C.~Patrick Royall, and Hajime Tanaka.
\newblock Structural and {{Dynamical Features}} of {{Multiple Metastable Glassy
  States}} in a {{Colloidal System}} with {{Competing Interactions}}.
\newblock {\em Phys. Rev. Lett.}, 104(16):165702, April 2010.

\bibitem{royall2018mermaid}
C.~Patrick Royall.
\newblock Hunting mermaids in real space: Known knowns, known unknowns and
  unknown unknowns.
\newblock {\em Soft Matter}, 14(20):4020--4028, 2018.

\bibitem{elmasri2012}
Djamel El~Masri, Teun Vissers, Stephane Badaire, Johan C.~P. Stiefelhagen,
  Hanumantha~Rao Vutukuri, Peter Helfferich, Tian~Hui Zhang, Willem~K. Kegel,
  Arnout Imhof, and Alfons {van Blaaderen}.
\newblock A qualitative confocal microscopy study on a range of colloidal
  processes by simulating microgravity conditions through slow rotations.
\newblock {\em Soft Matter}, 8(26):6979, 2012.

\bibitem{sedgwick2004}
H~Sedgwick, S~U Egelhaaf, and W~C~K Poon.
\newblock Clusters and gels in systems of sticky particles.
\newblock {\em J. Phys.: Condens. Matter}, 16(42):S4913--S4922, October 2004.

\bibitem{campbell2005}
Andrew~I. Campbell, Valerie~J. Anderson, Jeroen~S. {van Duijneveldt}, and Paul
  Bartlett.
\newblock Dynamical {{Arrest}} in {{Attractive Colloids}}: {{The Effect}} of
  {{Long}}-{{Range Repulsion}}.
\newblock {\em Phys. Rev. Lett.}, 94(20):208301, May 2005.

\bibitem{yethiraj2003}
Anand Yethiraj and Alfons {van Blaaderen}.
\newblock A colloidal model system with an interaction tunable from hard sphere
  to soft and dipolar.
\newblock {\em Nature}, 421(6922):513--517, January 2003.

\bibitem{riosdeanda2015}
I.~{Rios de Anda}, A.~Statt, F.~Turci, and C.P. Royall.
\newblock Low-{{Density Crystals}} in {{Charged Colloids}}: {{Comparison}} with
  {{Yukawa Theory}}: {{Low}}-{{Density Crystals}} in {{Charged Colloids}}:
  {{Comparison}} with {{Yukawa Theory}}.
\newblock {\em Contrib. Plasma Phys.}, 55(2-3):172--179, February 2015.

\bibitem{leunissen2007}
M.~E. Leunissen, A.~{van Blaaderen}, A.~D. Hollingsworth, M.~T. Sullivan, and
  P.~M. Chaikin.
\newblock Electrostatics at the oil-water interface, stability, and order in
  emulsions and colloids.
\newblock {\em Proceedings of the National Academy of Sciences},
  104(8):2585--2590, February 2007.

\bibitem{dinsmore2006}
A.~D. Dinsmore, V.~Prasad, I.~Y. Wong, and D.~A. Weitz.
\newblock Microscopic {{Structure}} and {{Elasticity}} of {{Weakly Aggregated
  Colloidal Gels}}.
\newblock {\em Phys. Rev. Lett.}, 96(18):185502, May 2006.

\bibitem{zhao2007}
Kun Zhao and Thomas~G. Mason.
\newblock Directing {{Colloidal Self}}-{{Assembly}} through
  {{Roughness}}-{{Controlled Depletion Attractions}}.
\newblock {\em Phys. Rev. Lett.}, 99(26):268301, December 2007.

\bibitem{lu2006}
Peter~J. Lu, Jacinta~C. Conrad, Hans~M. Wyss, Andrew~B. Schofield, and David~A.
  Weitz.
\newblock Fluids of {{Clusters}} in {{Attractive Colloids}}.
\newblock {\em Phys. Rev. Lett.}, 96(2):028306, January 2006.

\bibitem{leunissenthesis}
Mirjam~E. Leunissen.
\newblock {\em Manipulating {{Colloids}} with {{Charge}} and {{Electric
  Fields}}}.
\newblock PhD thesis, Utrecht, 2006.

\bibitem{royall2005competing}
C~Patrick Royall, Dirk G A~L Aarts, and Hajime Tanaka.
\newblock Fluid structure in colloid\textendash polymer mixtures: The
  competition between electrostatics and depletion.
\newblock {\em J. Phys.: Condens. Matter}, 17(45):S3401--S3408, November 2005.

\bibitem{groenewold2001}
Jan Groenewold and Willem~K. Kegel.
\newblock Anomalously {{Large Equilibrium Clusters}} of {{Colloids}}
  {\textsuperscript{\textdagger}}.
\newblock {\em J. Phys. Chem. B}, 105(47):11702--11709, November 2001.

\bibitem{stradner2004}
Anna Stradner, Helen Sedgwick, Fr{\'e}d{\'e}ric Cardinaux, Wilson C.~K. Poon,
  Stefan~U. Egelhaaf, and Peter Schurtenberger.
\newblock Equilibrium cluster formation in concentrated protein solutions and
  colloids.
\newblock {\em Nature}, 432(7016):492--495, November 2004.

\bibitem{stradner2020}
Anna Stradner.
\newblock Potential and limits of a colloid approach to protein solutions.
\newblock {\em Soft Matter}, page~17, 2020.

\bibitem{Bucciarelli2015}
Saskia Bucciarelli, Luc{\'i}a {Casal-Dujat}, Cristiano De~Michele, Francesco
  Sciortino, Jan Dhont, Johan Bergenholtz, Bela Farago, Peter Schurtenberger,
  and Anna Stradner.
\newblock Unusual {{Dynamics}} of {{Concentration Fluctuations}} in
  {{Solutions}} of {{Weakly Attractive Globular Proteins}}.
\newblock {\em J. Phys. Chem. Lett.}, 6(22):4470--4474, November 2015.

\bibitem{bucciarelli2016}
Saskia Bucciarelli, Jin~Suk Myung, Bela Farago, Shibananda Das, Gerard~A.
  Vliegenthart, Olaf Holderer, Roland~G. Winkler, Peter Schurtenberger, Gerhard
  Gompper, and Anna Stradner.
\newblock Dramatic influence of patchy attractions on short-time protein
  diffusion under crowded conditions.
\newblock {\em Sci. Adv.}, 2(12):e1601432, December 2016.

\bibitem{myung2018}
Jin~Suk Myung, Felix {Roosen-Runge}, Roland~G. Winkler, Gerhard Gompper, Peter
  Schurtenberger, and Anna Stradner.
\newblock Weak {{Shape Anisotropy Leads}} to a {{Nonmonotonic Contribution}} to
  {{Crowding}}, {{Impacting Protein Dynamics}} under {{Physiologically Relevant
  Conditions}}.
\newblock {\em J. Phys. Chem. B}, 122(51):12396--12402, December 2018.

\bibitem{shukla2008}
Anuj Shukla, Efstratios Mylonas, Emanuela Di~Cola, Stephanie Finet, Peter
  Timmins, Theyencheri Narayanan, and Dmitri~I. Svergun.
\newblock Absence of equilibrium cluster phase in concentrated lysozyme
  solutions.
\newblock {\em PNAS}, 105(13):5075--5080, April 2008.

\bibitem{zhang2008}
F.~Zhang, M.~W.~A. Skoda, R.~M.~J. Jacobs, S.~Zorn, R.~A. Martin, C.~M. Martin,
  G.~F. Clark, S.~Weggler, A.~Hildebrandt, O.~Kohlbacher, and F.~Schreiber.
\newblock Reentrant {{Condensation}} of {{Proteins}} in {{Solution Induced}} by
  {{Multivalent Counterions}}.
\newblock {\em Phys. Rev. Lett.}, 101(14):148101, September 2008.

\bibitem{vanschooneveld2009}
Matti~M. {van Schooneveld}, Volkert W.~A. {de Villeneuve}, Roel P.~A Dullens,
  Dirk G. A.~L. Aarts, Mirjam~E. Leunissen, and Willem~K. Kegel.
\newblock Structure, {{Stability}}, and {{Formation Pathways}} of {{Colloidal
  Gels}} in {{Systems}} with {{Short}}-{{Range Attraction}} and
  {{Long}}-{{Range Repulsion}}.
\newblock {\em J. Phys. Chem. B}, 113(14):4560--4564, April 2009.

\bibitem{klix2013}
Christian~L. Klix, Ken-ichiro Murata, Hajime Tanaka, Stephen~R. Williams, Alex
  Malins, and C.~Patrick Royall.
\newblock Novel kinetic trapping in charged colloidal clusters due to
  self-induced surface charge organization.
\newblock {\em Sci Rep}, 3(1):2072, December 2013.

\bibitem{capellmann2016}
Ronja~F. Capellmann, N{\'e}stor~E. {Valadez-P{\'e}rez}, Benedikt Simon,
  Stefan~U. Egelhaaf, Marco Laurati, and Ram{\'o}n {Casta{\~n}eda-Priego}.
\newblock Structure of colloidal gels at intermediate concentrations: The role
  of competing interactions.
\newblock {\em Soft Matter}, 12(46):9303--9313, 2016.

\bibitem{ohtsuka2008}
Takehiro Ohtsuka, C.~Patrick Royall, and Hajime Tanaka.
\newblock Local structure and dynamics in colloidal fluids and gels.
\newblock {\em Europhys. Lett.}, 84(4):46002, November 2008.

\bibitem{sciortino2005}
F.~Sciortino, P.~Tartaglia, and E.~Zaccarelli.
\newblock One-{{Dimensional Cluster Growth}} and {{Branching Gels}} in
  {{Colloidal Systems}} with {{Short}}-{{Range Depletion Attraction}} and
  {{Screened Electrostatic Repulsion}}.
\newblock {\em J. Phys. Chem. B}, 109(46):21942--21953, November 2005.

\bibitem{kohl2016}
M.~Kohl, R.~F. Capellmann, M.~Laurati, S.~U. Egelhaaf, and M.~Schmiedeberg.
\newblock Directed percolation identified as equilibrium pre-transition towards
  non-equilibrium arrested gel states.
\newblock {\em Nat Commun}, 7(1):11817, September 2016.

\bibitem{kim2013}
Jung~Min Kim, Jun Fang, Aaron P.~R. Eberle, Ram{\'o}n {Casta{\~n}eda-Priego},
  and Norman~J. Wagner.
\newblock Gel {{Transition}} in {{Adhesive Hard}}-{{Sphere Colloidal
  Dispersions}}: {{The Role}} of {{Gravitational Effects}}.
\newblock {\em Phys. Rev. Lett.}, 110(20):208302, May 2013.

\bibitem{richard2018jcp}
David Richard, C.~Patrick Royall, and Thomas Speck.
\newblock Communication: {{Is}} directed percolation in colloid-polymer
  mixtures linked to dynamic arrest?
\newblock {\em The Journal of Chemical Physics}, 148(24):241101, June 2018.

\bibitem{noro2000}
Massimo~G. Noro and Daan Frenkel.
\newblock Extended corresponding-states behavior for particles with variable
  range attractions.
\newblock {\em The Journal of Chemical Physics}, 113(8):2941--2944, August
  2000.

\bibitem{vanmegen1998}
W.~{van Megen}, T.~C. Mortensen, S.~R. Williams, and J.~M{\"u}ller.
\newblock Measurement of the self-intermediate scattering function of
  suspensions of hard spherical particles near the glass transition.
\newblock {\em Phys. Rev. E}, 58(5):6073--6085, November 1998.

\bibitem{frank1952}
F.~Charles Frank.
\newblock Supercooling of {{Liquids}}.
\newblock {\em Proc. R. Soc. A.}, 215:43--46, 1952.

\bibitem{steinhardt1983}
P.~J. Steinhardt, David~R. Nelson, and M.~Ronchetti.
\newblock Bond-{{Orientational Order In Liquids And Glasses}}.
\newblock {\em Phys. Rev. B}, 28:784--805, 1983.

\bibitem{robinson2019}
Joshua~F. Robinson, Francesco Turci, Roland Roth, and C.~Patrick Royall.
\newblock Morphometric {{Approach}} to {{Many}}-{{Body Correlations}} in {{Hard
  Spheres}}.
\newblock {\em Phys. Rev. Lett.}, 122(6):068004, February 2019.

\bibitem{meng2010}
G.~Meng, N.~Arkus, M.~P. Brenner, and V.~N. Manoharan.
\newblock The {{Free}}-{{Energy Landscape}} of {{Clusters}} of {{Attractive
  Hard Spheres}}.
\newblock {\em Science}, 327(5965):560--563, January 2010.

\bibitem{malins2009}
Alex Malins, Stephen~R Williams, Jens Eggers, Hajime Tanaka, and C~Patrick
  Royall.
\newblock Geometric frustration in small colloidal clusters.
\newblock {\em J. Phys.: Condens. Matter}, 21(42):425103, October 2009.

\bibitem{miller1999}
Mark~A. Miller, Jonathan P.~K. Doye, and David~J. Wales.
\newblock Structural relaxation in {{Morse}} clusters: {{Energy}} landscapes.
\newblock {\em The Journal of Chemical Physics}, 110(1):328--334, January 1999.

\bibitem{doye1995}
Jonathan P.~K. Doye, David~J. Wales, and R.~Stephen Berry.
\newblock The effect of the range of the potential on the structures of
  clusters.
\newblock {\em The Journal of Chemical Physics}, 103(10):4234--4249, September
  1995.

\bibitem{mossa2003}
S.~Mossa and G.~Tarjus.
\newblock Locally preferred structure in simple atomic liquids.
\newblock {\em The Journal of Chemical Physics}, 119(15):8069--8074, October
  2003.

\bibitem{taffs2010jcp}
Jade Taffs, Alex Malins, Stephen~R. Williams, and C.~Patrick Royall.
\newblock The effect of attractions on the local structure of liquids and
  colloidal fluids.
\newblock {\em The Journal of Chemical Physics}, 133(24):244901, December 2010.

\bibitem{royall2015jnonxtalsol}
C.~Patrick Royall, Alex Malins, Andrew~J. Dunleavy, and Rhiannon Pinney.
\newblock Strong geometric frustration in model glassformers.
\newblock {\em Journal of Non-Crystalline Solids}, 407:34--43, January 2015.

\bibitem{malins2013jcp}
Alex Malins, Jens Eggers, C.~Patrick Royall, Stephen~R. Williams, and Hajime
  Tanaka.
\newblock Identification of long-lived clusters and their link to slow dynamics
  in a model glass former.
\newblock {\em The Journal of Chemical Physics}, 138(12):12A535, March 2013.

\bibitem{malins2013fara}
Alex Malins, Jens Eggers, Hajime Tanaka, and C.~Patrick Royall.
\newblock Lifetimes and lengthscales of structural motifs in a model
  glassformer.
\newblock {\em Faraday Discuss.}, 167:405, 2013.

\bibitem{malins2013tcc}
Alex Malins, Stephen~R Williams, Jens Eggers, and C~Patrick Royall.
\newblock Identification of structure in condensed matter with the topological
  cluster classification.
\newblock {\em J Chem Phys}, page~21, 2013.

\bibitem{royall2008aip}
C.~Patrick Royall, Stephen~R. Williams, Takehiro Ohtsuka, Hajime Tanaka, Michio
  Tokuyama, Irwin Oppenheim, and Hideya Nishiyama.
\newblock Direct {{Observation}} of {{Low}}-{{Energy Clusters}} in a
  {{Colloidal Gel}}.
\newblock In {\em {{AIP Conference Proceedings}}}, volume 982, pages 97--101.
  {AIP}, 2008.

\bibitem{hsiao2012}
L.~C. Hsiao, R.~S. Newman, S.~C. Glotzer, and M.~J. Solomon.
\newblock Role of isostaticity and load-bearing microstructure in the
  elasticity of yielded colloidal gels.
\newblock {\em Proceedings of the National Academy of Sciences},
  109(40):16029--16034, October 2012.

\bibitem{tsurusawa2018}
Hideyo Tsurusawa, Mathieu Leocmach, John Russo, and Hajime Tanaka.
\newblock Gelation as condensation frustrated by hydrodynamics and mechanical
  isostaticity.
\newblock {\em ArXiv180404370 Cond-Mat}, April 2018.

\bibitem{whitaker2019}
Kathryn~A. Whitaker, Zsigmond Varga, Lilian~C. Hsiao, Michael~J. Solomon,
  James~W. Swan, and Eric~M. Furst.
\newblock Colloidal gel elasticity arises from the packing of locally glassy
  clusters.
\newblock {\em Nat Commun}, 10(1):2237, December 2019.

\bibitem{richard2018softmatter}
David Richard, James Hallett, Thomas Speck, and C.~Patrick Royall.
\newblock Coupling between criticality and gelation in ``sticky'' spheres: A
  structural analysis.
\newblock {\em Soft Matter}, 14(27):5554--5564, 2018.

\bibitem{royall2007naturephys}
C.~Patrick~Royall, Dirk G. A.~L. Aarts, and Hajime Tanaka.
\newblock Bridging length scales in colloidal liquids and interfaces from
  near-critical divergence to single particles.
\newblock {\em Nature Phys}, 3(9):636--640, September 2007.

\bibitem{royall2007jcp}
C.~Patrick Royall, Ard~A. Louis, and Hajime Tanaka.
\newblock Measuring colloidal interactions with confocal microscopy.
\newblock {\em The Journal of Chemical Physics}, 127(4):044507, July 2007.

\bibitem{hurley1995}
{Hurley, M. M. {and} Harrowell, P.}
\newblock Kinetic structure of a two-dimensional liquid.
\newblock {\em Phys. Rev. E.}, 52:1694--1698, 1995.

\bibitem{royall2015prl}
C.~Patrick Royall, Jens Eggers, Akira Furukawa, and Hajime Tanaka.
\newblock Probing {{Colloidal Gels}} at {{Multiple Length Scales}}: {{The
  Role}} of {{Hydrodynamics}}.
\newblock {\em Phys. Rev. Lett.}, 114(25):258302, June 2015.

\bibitem{prasad2003}
V~Prasad, V~Trappe, A~D Dinsmore, P~N Segre, L~Cipelletti, and D~A Weitz.
\newblock Universal features of the fluid to solid transition for attractive
  colloidal particles.
\newblock {\em Faraday Discuss}, page~12, 2003.

\bibitem{vandoorn2018}
Jan~Maarten {van Doorn}, Joanne~E. Verweij, Joris Sprakel, and Jasper {van der
  Gucht}.
\newblock Strand {{Plasticity Governs Fatigue}} in {{Colloidal Gels}}.
\newblock {\em Phys. Rev. Lett.}, 120(20):208005, May 2018.

\bibitem{verweij2019}
Joanne E.~Verweij, Frans~A. M.~Leermakers, Joris Sprakel, and Jasper van~der
  Gucht.
\newblock Plasticity in colloidal gel strands.
\newblock {\em Soft Matter}, 15(32):6447--6454, 2019.

\bibitem{furukawa2010}
Akira Furukawa and Hajime Tanaka.
\newblock Key {{Role}} of {{Hydrodynamic Interactions}} in {{Colloidal
  Gelation}}.
\newblock {\em Phys. Rev. Lett.}, page~4, 2010.

\bibitem{tsurusawa2019}
Hideyo Tsurusawa, Mathieu Leocmach, John Russo, and Hajime Tanaka.
\newblock Direct link between mechanical stability in gels and percolation of
  isostatic particles.
\newblock {\em Sci. Adv.}, 5(5):eaav6090, May 2019.

\bibitem{manley2004}
S.~Manley, L.~Cipelletti, V.~Trappe, A.~E. Bailey, R.~J. Christianson,
  U.~Gasser, V.~Prasad, P.~N. Segre, M.~P. Doherty, S.~Sankaran, A.~L.
  Jankovsky, B.~Shiley, J.~Bowen, J.~Eggers, C.~Kurta, T.~Lorik, and D.~A.
  Weitz.
\newblock Limits to {{Gelation}} in {{Colloidal Aggregation}}.
\newblock {\em Phys. Rev. Lett.}, 93(10):108302, September 2004.

\bibitem{tanaka2005}
Hajime Tanaka, Yuya Nishikawa, and Takehito Koyama.
\newblock Network-forming phase separation of colloidal suspensions.
\newblock {\em J. Phys.: Condens. Matter}, 17(15):L143--L153, April 2005.

\bibitem{zhang2012}
Tian~Hui Zhang, Jan Klok, R.~Hans~Tromp, Jan Groenewold, and Willem~K. Kegel.
\newblock Non-equilibrium cluster states in colloids with competing
  interactions.
\newblock {\em Soft Matter}, 8(3):667--672, 2012.

\bibitem{royall2018molphys}
C.~Patrick~Royall.
\newblock Kinetic crystallisation instability in liquids with short-ranged
  attractions.
\newblock {\em Molecular Physics}, 116(21-22):3076--3084, November 2018.

\bibitem{royall2012}
C.~P. Royall and A.~Malins.
\newblock The role of quench rate in colloidal gels.
\newblock {\em Faraday Discuss.}, 158:301–311, 2012.

\bibitem{sollich2010}
Peter Sollich and Nigel~B. Wilding.
\newblock Crystalline {{Phases}} of {{Polydisperse Spheres}}.
\newblock {\em Phys. Rev. Lett.}, 104(11):118302, March 2010.

\bibitem{tenwolde1997}
Pieter~Rein ten Wolde and Daan Frenkel.
\newblock Enhancement of {{Protein Crystal Nucleation}} by {{Critical Density
  Fluctuations}}.
\newblock {\em Science}, 277(5334):1975--1978, September 1997.

\bibitem{savage2009}
J.~R. Savage and A.~D. Dinsmore.
\newblock Experimental {{Evidence}} for {{Two}}-{{Step Nucleation}} in
  {{Colloidal Crystallization}}.
\newblock {\em Phys. Rev. Lett.}, 102(19):198302, May 2009.

\bibitem{taylor2012}
Shelley~L Taylor, Robert Evans, and C~Patrick~Royall.
\newblock Temperature as an external field for colloid\textendash polymer
  mixtures: `quenching' by heating and `melting' by cooling.
\newblock {\em J. Phys.: Condens. Matter}, 24(46):464128, November 2012.

\bibitem{tsurusawa2017}
Hideyo Tsurusawa, John Russo, Mathieu Leocmach, and Hajime Tanaka.
\newblock Formation of porous crystals via viscoelastic phase separation.
\newblock {\em Nat. Mater.}, 16(10):1022--1028, October 2017.

\bibitem{poon1999}
W~C~K Poon and M~M Robins.
\newblock Delayed sedimentation of transient gels in colloid\textendash polymer
  mixtures : Dark-\"ueld observation, rheology and dynamic light scattering
  studies.
\newblock {\em Faraday Discuss}, page~12, 1999.

\bibitem{buscall2009}
Richard Buscall, Tahsin~H. Choudhury, Malcolm~A. Faers, James~W. Goodwin,
  Paul~A. Luckham, and Susan~J. Partridge.
\newblock Towards rationalising collapse times for the delayed sedimentation of
  weakly-aggregated colloidal gels.
\newblock {\em Soft Matter}, 5(7):1345, 2009.

\bibitem{teece2014}
Lisa~J. Teece, James~M. Hart, Kerry Yen~Ni Hsu, Stephen Gilligan, Malcolm~A.
  Faers, and Paul Bartlett.
\newblock Gels under stress: {{The}} origins of delayed collapse.
\newblock {\em Colloids and Surfaces A: Physicochemical and Engineering
  Aspects}, 458:126--133, September 2014.

\bibitem{piazza2012}
Roberto Piazza, Stefano Buzzaccaro, and Eleonora Secchi.
\newblock The unbearable heaviness of colloids: Facts, surprises, and puzzles
  in sedimentation.
\newblock {\em J. Phys.: Condens. Matter}, 24(28):284109, July 2012.

\bibitem{piazza2014}
Roberto Piazza.
\newblock Settled and unsettled issues in particle settling.
\newblock {\em Rep. Prog. Phys.}, 77(5):056602, May 2014.

\bibitem{razali2017}
Azaima Razali, Christopher~J. Fullerton, Francesco Turci, James~E. Hallett,
  Robert~L. Jack, and C.~Patrick Royall.
\newblock Effects of vertical confinement on gelation and sedimentation of
  colloids.
\newblock {\em Soft Matter}, 13(17):3230--3239, 2017.

\bibitem{kamp2009}
Stephen~W. Kamp and Maria~L. Kilfoil.
\newblock Universal behaviour in the mechanical properties of weakly aggregated
  colloidal particles.
\newblock {\em Soft Matter}, 5(12):2438, 2009.

\bibitem{faers2003}
Malcolm~A. Faers.
\newblock The importance of the interfacial stabilising layer on the
  macroscopic flow properties of suspensions dispersed in non-adsorbing polymer
  solution.
\newblock {\em Advances in Colloid and Interface Science}, 106(1-3):23--54,
  December 2003.

\bibitem{laurati2011}
M.~Laurati, S.~U. Egelhaaf, and G.~Petekidis.
\newblock Nonlinear rheology of colloidal gels with intermediate volume
  fraction.
\newblock {\em Journal of Rheology}, 55(3):673--706, April 2011.

\bibitem{koumakis2015}
Nick Koumakis, Esmaeel Moghimi, Rut Besseling, Wilson~C. K.~Poon, John
  F.~Brady, and George Petekidis.
\newblock Tuning colloidal gels by shear.
\newblock {\em Soft Matter}, 11(23):4640--4648, 2015.

\bibitem{smith2007}
P.~A. Smith, G.~Petekidis, S.~U. Egelhaaf, and W.~C.~K. Poon.
\newblock Yielding and crystallization of colloidal gels under oscillatory
  shear.
\newblock {\em Phys. Rev. E}, 76(4):041402, October 2007.

\bibitem{lin2014}
Neil Y.~C. Lin, Jonathan~H. McCoy, Xiang Cheng, Brian Leahy, Jacob~N.
  Israelachvili, and Itai Cohen.
\newblock A multi-axis confocal rheoscope for studying shear flow of structured
  fluids.
\newblock {\em Review of Scientific Instruments}, 85(3):033905, March 2014.

\bibitem{isa2007}
Lucio Isa, Rut Besseling, and Wilson C~K Poon.
\newblock Shear {{Zones}} and {{Wall Slip}} in the {{Capillary Flow}} of
  {{Concentrated Colloidal Suspensions}}.
\newblock {\em Phys. Rev. Lett.}, 98(19):198305, May 2007.

\bibitem{pandey2014}
Rahul Pandey, Melissa Spannuth, and Jacinta~C. Conrad.
\newblock Confocal {{Imaging}} of {{Confined Quiescent}} and {{Flowing
  Colloid}}-polymer {{Mixtures}}.
\newblock {\em JoVE J. Vis. Exp.}, (87):e51461, May 2014.

\bibitem{kohl2017}
Matthias Kohl and Michael Schmiedeberg.
\newblock Shear-induced slab-like domains in a directed percolated colloidal
  gel.
\newblock {\em Eur. Phys. J. E}, 40(8):71, August 2017.

\bibitem{schwen2020}
Eric~M. Schwen, Meera Ramaswamy, Chieh-Min Cheng, Linda Jan, and Itai Cohen.
\newblock Embedding orthogonal memories in a colloidal gel through oscillatory
  shear.
\newblock {\em Soft Matter}, 16(15):3746--3752, April 2020.

\bibitem{moller2008}
P.~C.~F. M{\o}ller, S.~Rodts, M.~A.~J. Michels, and Daniel Bonn.
\newblock Shear banding and yield stress in soft glassy materials.
\newblock {\em Phys. Rev. E}, 77(4):041507, April 2008.

\bibitem{fall2010}
Abdoulaye Fall, Jose Paredes, and Daniel Bonn.
\newblock Yielding and {{Shear Banding}} in {{Soft Glassy Materials}}.
\newblock {\em Phys. Rev. Lett.}, 105(22):225502, November 2010.

\bibitem{sprakel2011}
Joris Sprakel, Stefan~B. Lindstr{\"o}m, Thomas~E. Kodger, and David~A. Weitz.
\newblock Stress {{Enhancement}} in the {{Delayed Yielding}} of {{Colloidal
  Gels}}.
\newblock {\em Phys. Rev. Lett.}, 106(24):248303, June 2011.

\bibitem{han2019}
Ming Han, Jonathan~K. Whitmer, and Erik Luijten.
\newblock Dynamics and structure of colloidal aggregates under microchannel
  flow.
\newblock {\em Soft Matter}, 15(4):744--751, January 2019.

\bibitem{conrad2008}
Jacinta~C. Conrad and Jennifer~A. Lewis.
\newblock Structure of {{Colloidal Gels}} during {{Microchannel Flow}}.
\newblock {\em Langmuir}, 24(15):7628--7634, August 2008.

\bibitem{conrad2010}
Jacinta~C. Conrad and Jennifer~A. Lewis.
\newblock Structural {{Evolution}} of {{Colloidal Gels During Constricted
  Microchannel Flow}}.
\newblock {\em Langmuir}, 26(9):6102--6107, May 2010.

\bibitem{solomon2010}
Michael~J. Solomon and Patrick~T. Spicer.
\newblock Microstructural regimes of colloidal rod suspensions, gels, and
  glasses.
\newblock {\em Soft Matter}, 6(7):1391, 2010.

\bibitem{schilling2007}
T.~Schilling, S.~Jungblut, and Mark~A. Miller.
\newblock Depletion-{{Induced Percolation}} in {{Networks}} of {{Nanorods}}.
\newblock {\em Phys. Rev. Lett.}, 98(10):108303, March 2007.

\bibitem{kyrylyuk2011}
Andriy~V. Kyrylyuk, Marie~Claire Hermant, Tanja Schilling, Bert Klumperman,
  Cor~E. Koning, and Paul {van der Schoot}.
\newblock Controlling electrical percolation in multicomponent carbon nanotube
  dispersions.
\newblock {\em Nature Nanotech}, 6(6):364--369, June 2011.

\bibitem{lekkerkerker}
H~N~W Lekkerkerker and Remco Tuinier.
\newblock {\em Colloids and the {{Depletion Interaction}}}, volume 833 of {\em
  Lecture {{Notes}} in {{Physics}}}.
\newblock 2011.

\bibitem{hsiao2015}
Lilian~C. Hsiao, Benjamin~A. Schultz, Jens Glaser, Michael Engel, Megan~E.
  Szakasits, Sharon~C. Glotzer, and Michael~J. Solomon.
\newblock Metastable orientational order of colloidal discoids.
\newblock {\em Nat Commun}, 6(1):8507, December 2015.

\bibitem{wilkins2009}
Georgina M.~H. Wilkins, Patrick~T. Spicer, and Michael~J. Solomon.
\newblock Colloidal {{System To Explore Structural}} and {{Dynamical
  Transitions}} in {{Rod Networks}}, {{Gels}}, and {{Glasses}}.
\newblock {\em Langmuir}, 25(16):8951--8959, August 2009.

\bibitem{rice2012}
Rebecca Rice, Roland Roth, and C.~Patrick Royall.
\newblock Polyhedral colloidal `rocks': Low-dimensional networks.
\newblock {\em Soft Matter}, 8(4):1163--1167, 2012.

\bibitem{zhang2018}
Isla Zhang, Rattachai Pinchaipat, Nigel~B Wilding, Malcolm~A Faers, Paul
  Bartlett, Robert Evans, and C~Patrick Royall.
\newblock Composition inversion in mixtures of binary colloids and polymer.
\newblock {\em J Chem Phys}, page~11, 2018.

\bibitem{cates2008}
Michael~E. Cates and Paul~S. Clegg.
\newblock Bijels: A new class of soft materials.
\newblock {\em Soft Matter}, 4(11):2132, 2008.

\bibitem{varrato2012}
F.~Varrato, L.~Di~Michele, M.~Belushkin, N.~Dorsaz, S.~H. Nathan, E.~Eiser, and
  G.~Foffi.
\newblock Arrested demixing opens route to bigels.
\newblock {\em Proceedings of the National Academy of Sciences},
  109(47):19155--19160, November 2012.

\bibitem{herzig2007}
E.~M. Herzig, K.~A. White, A.~B. Schofield, W.~C.~K. Poon, and P.~S. Clegg.
\newblock Bicontinuous emulsions stabilized solely by colloidal particles.
\newblock {\em Nature Mater}, 6(12):966--971, December 2007.

\bibitem{filali2001}
Mohammed Filali, Mohamed~Jamil Ouazzani, Eric Michel, Raymond Aznar,
  Gr{\'e}goire Porte, and Jacqueline Appell.
\newblock Robust {{Phase Behavior}} of {{Model Transient Networks}}.
\newblock {\em J. Phys. Chem. B}, 105(43):10528--10535, November 2001.

\bibitem{fussell2021}
S~L Fussell, C~P Royall, and J~S {van Duijneveldt}.
\newblock Controlling phase separation in microgel-polymeric micelle mixtures
  using variable quench rates.
\newblock {\em Soft Matter}, page~12, 2021.

\bibitem{blumlein2015}
Alice Blumlein and Jennifer~J. McManus.
\newblock Bigels formed via spinodal decomposition of unfolded protein.
\newblock {\em J. Mater. Chem. B}, 3(17):3429--3435, 2015.

\bibitem{cheng2021}
R.~Cheng, I.~Rios~de Anda, T.~W.~C. Taylor, M.~A. Faers, J.~L.~R. Anderson,
  A.~M. Seddon, and C.~P. Royall.
\newblock Protein--polymer mixtures in the colloid limit: Aggregation,
  sedimentation.
\newblock {\em submitted to J. Chem. Phys.}, 2021.

\bibitem{bechinger2016}
Clemens Bechinger, Roberto Di~Leonardo, Hartmut L{\"o}wen, Charles Reichhardt,
  Giorgio Volpe, and Giovanni Volpe.
\newblock Active {{Particles}} in {{Complex}} and {{Crowded Environments}}.
\newblock {\em Rev. Mod. Phys.}, 88(4):045006, November 2016.

\bibitem{marchetti2013}
M.~C. Marchetti, J.~F. Joanny, S.~Ramaswamy, T.~B. Liverpool, J.~Prost, Madan
  Rao, and R.~Aditi Simha.
\newblock Hydrodynamics of soft active matter.
\newblock {\em Rev. Mod. Phys.}, 85(3):1143--1189, July 2013.

\bibitem{dassanayake2000}
U~Dassanayake, S~Fraden, and A~{van Blaaderen}.
\newblock Structure of electrorheological fluids.
\newblock {\em J. Chem. Phys.}, 112:3851--3858, 2000.

\bibitem{aime2018}
Stefano Aime, Laurence Ramos, and Luca Cipelletti.
\newblock Microscopic dynamics and failure precursors of a gel under mechanical
  load.
\newblock {\em PNAS}, 115(14):3587--3592, April 2018.

\bibitem{harich2016}
R.~Harich, T.~W. Blythe, M.~Hermes, E.~Zaccarelli, A.~J. Sederman, L.~F.
  Gladden, and W.~C.~K. Poon.
\newblock Gravitational collapse of depletion-induced colloidal gels.
\newblock {\em Soft Matter}, 12(19):4300--4308, 2016.

\bibitem{varga2016}
Zsigmond Varga and James Swan.
\newblock Hydrodynamic interactions enhance gelation in dispersions of colloids
  with short-ranged attraction and long-ranged repulsion.
\newblock {\em Soft Matter}, 12(36):7670--7681, 2016.

\bibitem{allen2009}
R.~J. Allen, C.~Valeriani, and P.~Rein ten Wolde.
\newblock Forward flux sampling for rare event simulations.
\newblock {\em J. Phys. Cond. Matt.}, 21:463102, 2009.

\bibitem{charbonneau2014}
Patrick Charbonneau, Jorge Kurchan, Giorgio Parisi, Pierfrancesco Urbani, and
  Francesco Zamponi.
\newblock Fractal free energy landscapes in structural glasses.
\newblock {\em Nat Commun}, 5(1):3725, September 2014.

\bibitem{klongvessa2019}
Natsuda Klongvessa, F{\'e}lix Ginot, Christophe Ybert, C{\'e}cile
  {Cottin-Bizonne}, and Mathieu Leocmach.
\newblock Active {{Glass}}: {{Ergodicity Breaking Dramatically Affects
  Response}} to {{Self}}-{{Propulsion}}.
\newblock {\em Phys. Rev. Lett.}, 123(24):248004, December 2019.

\end{thebibliography}
\end{document}